\documentclass{jfm}

\usepackage[dvipsnames]{xcolor}
\usepackage{graphicx}
\usepackage{newtxtext}
\usepackage{newtxmath}
\usepackage{natbib}
\usepackage{hyperref}
\hypersetup{
    colorlinks = true,
    urlcolor   = blue,
    citecolor  = black,
}

\newcommand{\RomanNumeralCaps}[1]
\usepackage{subfig}
\usepackage{amsmath}
\usepackage{epstopdf}
\usepackage{multicol}
\usepackage{amssymb}
\usepackage{textcomp}
\usepackage{color}

 \newcommand\nb{\boldsymbol{n}}
 \newcommand\xb{\boldsymbol{x}}
 \newcommand\xo{\xb_o}
 \newcommand\ubar{\overline{U}_e}

 \newcommand\gamtil{\widetilde{\gamma}}
 \newcommand\alf{\alpha}
 
 \newcommand\ub{\boldsymbol{u}}

 \newcommand\uinf{U_{\infty}}

\newcommand{\rmd}{\mathrm{d}}
\newcommand{\rmi}{\mathrm{i}}
\newcommand{\rme}{\mathrm{e}}

\newcommand{\bx}{\boldsymbol{x}}
\newcommand{\bu}{\boldsymbol{u}}

\newcommand{\bn}{\boldsymbol{n}}

\title[]{On the stability of an in-line formation of hydrodynamically interacting flapping plates}
\author{Monika Nitsche\aff{1}
  \corresp{\email{nitsche@unm.edu}},
  Anand U. Oza\aff{2}\corresp{\email{oza@njit.edu}}
 \and Michael Siegel\aff{2}}

\affiliation{\aff{1}Department of Mathematics and Statistics, University of New Mexico, Albuquerque, NM 87131, USA
\aff{2}Department of Mathematics \& Center for Applied Mathematics and Statistics, New Jersey Institute of Technology, Newark, NJ 07102, USA}

\begin{document}
\maketitle

\abstract{
The motion of several plates in an inviscid and incompressible fluid is studied numerically using a vortex sheet model. Two to four plates are initially placed in-line, separated by a specified distance, and actuated in the vertical direction with a prescribed oscillatory heaving motion. The vertical motion induces the plates' horizontal acceleration due to their self-induced thrust and fluid drag forces. In certain parameter regimes, the plates adopt equilibrium ``schooling modes," wherein they translate at a steady horizontal velocity while maintaining a constant separation distance between them. The separation distances are found to be quantized on the flapping wavelength. As either the number of plates increases or the flapping amplitude decreases, the schooling modes destabilize via oscillations that propagate downstream from the leader and cause collisions between the plates, an instability that is similar to that observed in recent experiments on flapping wings in a water tank~\citep{Newbolt_2024}. A simple control mechanism is implemented, wherein each plate accelerates or decelerates according to its velocity relative to the plate directly ahead by modulating its own flapping amplitude. This mechanism is shown to successfully stabilize the schooling modes, with remarkable impact on the regularity of the vortex pattern in the wake. Several phenomena observed in the simulations are obtained by a reduced model based on linear thin-airfoil theory. 
}
%The numerical method accounts for vortex shedding, thrust and drag forces acting on the plates, and accurate evaluation of near-singular integrals. , the general motivation being to understand the hydrodynamic interactions in schooling and flocking behavior in animal collectives

\section{Introduction}

There has been considerable interest recently in understanding the hydrodynamic interactions between flapping wings in moderate to high Reynolds number flows, for which inertial effects are relevant. Apart from scientific curiosity, this interest is motivated by numerous biological and engineering applications. Specifically, understanding why fish swim in schools~\citep{Pavlov_Review} and birds fly in flocks~\citep{Bajec_Review} has been of long-standing interest to biologists and physicists studying collective behavior in animal groups. While many studies have investigated how social interactions between fish may mediate schooling behavior~\citep{Jolles1,Swain1,Tunstrom1}, the so-called {\it Lighthill conjecture} posits that orderly formations in schools may arise passively due to hydrodynamic interactions, instead of requiring active control or decision making~\citep{Lighthill_Book}. The underlying mechanisms behind flow-induced collective behavior could have applications to the next generation of biomimetic underwater vehicles, which self-propel using mechanisms inspired from fish locomotion~\citep{Triantafyllou_Review, Geder_2017}.

Field observations of the formations adopted by fish schools are diverse and typically vary between species~\citep{Pavlov_Review}. For instance, some fish species (e.g. minnow, bream, saithe, herring) tend to adopt lattice-like schooling formations, while others (e.g. cod) adopt more disordered configurations~\citep{Pitcher_Minnow,Partridge_3D}. Recent laboratory experiments have revealed that red nose tetra fish may adopt different lattice-like formations when they school~\citep{AshrafINT,AshrafPNAS}. %While fish have been observed to school in a number of different formations, The simplest formation is
Our paper is concerned with the simplest of these formations, 
the so-called {\it in-line} formation, in which the line connecting the fish is in the swimming direction. Indeed, fish have been observed to swim in linear chains~\citep{Gudger}, and pairs of goldfish have recently been observed to school in an in-line formation in which the follower beats its tail to synchronize with the oncoming vortices shed by the leader~\citep{Li2020}. Flying ibises have also been observed to flock in an in-line formation, a statistical analysis of which reveals certain preferred spatial phase relationships between the wingtip paths of successive birds~\citep{Portugal2014}.

A number of recent papers have investigated the Lighthill conjecture numerically, by simulating the coupled dynamics of bodies with a prescribed time-periodic flapping kinematics (e.g. heaving, pitching or undulating) and the fluid that surrounds them. The motion of the fluid is determined by solving the Navier-Stokes equations with no-slip boundary conditions on the body, and the dynamics of the bodies are evolved through the influence of drag and thrust forces on them. Simulations of pairs of elastic filaments~\citep{Zhu_2014,Dai_2018,Park_2018} and hydrofoils~\citep{Lin_2020} in two dimensions (2D) have shown that, when the bodies are free to evolve in the streamwise ($x$) direction but constrained in the lateral ($y$) direction, they tend to adopt certain {\it schooling modes}, in which their cycle-averaged velocities and separation distance is constant in time. Simulations have also shown that up to eight flexible plates in 2D~\citep{Peng_2018} and two flexible plates in 3D~\citep{Arranz_2022} may execute a variety of stable in-line schooling modes.

The drawback of numerical methods that solve the Navier-Stokes equations directly is that they are typically limited to moderate Reynolds numbers, $\mathrm{Re} = O(10^2)-O(10^3)$, due to the presence of boundary layers near the bodies that require increasingly refined grids and thus computational time as the Reynolds number is increased progressively. However, schooling fish often operate in relatively high Reynolds number regimes, $\mathrm{Re} = O(10^4)$ or higher~\citep{Triantafyllou_Review}, simulations of which have remained elusive. A tractable alternative is a vortex sheet method, in which fluid viscosity is neglected and the Euler equations are solved directly: the bodies are represented as bound vortex sheets, and free vortex sheets are shed from the bodies' sharp trailing edges~\citep{Alben_Flexible,Alben_Street,Huang2016,Fang2017}. Vortex sheet simulations showed that pairs of rigid plates undergoing either heaving or pitching kinematics can adopt stable in-line schooling modes~\citep{Fang_Thesis,Heydari_2021}. A recent study of larger in-line formations found that the number of wings that could stably school together was limited by hydrodynamic interactions~\citep{Heydari_2024}. That is, the number of wings in an in-line school could be increased up to a point, after which the last wing would fall behind and be unable to keep up with the school.

Experimental platforms in which collectives of hydrofoils are actuated in a water tank offer a valuable testbed for investigating the hydrodynamic mechanisms that underpin schooling, as they are more controllable and reproducible than biological systems~\citep{Becker}. A study showed that a pair of hydrofoils that execute heaving kinematics adopt stable steady schooling modes that are quantized on the so-called {\it schooling number}, $S = d/\lambda$, where $d$ is the tail-to-tip distance between the wings and $\lambda = U/f$ is the wavelength of the flapping motion, $U$ being the translational velocity and $f$ the flapping frequency. The schooling number measures the spatial phase coherence of the two swimmers: integer values of the schooling number, which were observed in the experiment, correspond to in-phase states in which the pair traces out the same path through space, while half-integer values indicate out-of-phase states. A recent experimental study on larger collectives of hydrofoils showed that increasingly large chains are unstable to flow-induced instabilities termed {\it flonons}, wherein the horizontal positions of downstream wings begin to oscillate with increasingly large amplitude down the group and thus cause collisions between the trailing members~\citep{Newbolt_2024}. That study suggests that, while hydrodynamic interactions can bind wings into a schooling mode, they can also lead to an instability that breaks the bound mode.

In this paper, we conduct vortex sheet simulations of collectives of in-line heaving plates, with a view to investigating the bound states and flow-induced instabilities that may arise. Care is taken to evaluate the near-singular integrals that arise when a vortex sheet is near a body, which is essential to prevent the vortex sheets from unphysically crossing the bodies~\citep{Nitsche_2021}. In summary, we observe that the collectives adopt a number of schooling modes, which become more stable as the flapping amplitude is increased progressively. The separation distances between the plates are quantized on the schooling number, in agreement with experiments on heaving hydrofoils in a water tank~\citep{Sophie,Newbolt_2022,Newbolt_2024}. Moreover, as the number of wings is increased, the collective tends to destabilize via an oscillatory instability that propagates downstream from the leader and causes collisions between the followers, a phenomenon that is reminiscent of the flonons observed in recent experiments~\citep{Newbolt_2024}. We rationalize these results qualitatively by proposing a reduced model for the dynamics of interacting plates based on the seminal thin-airfoil theory of~\citet{Wu1961Swimmingwavingplate}. We also demonstrate that a simple control mechanism, wherein each plate adjusts its flapping amplitude according to its velocity relative to the plate directly ahead, mitigates the flow-induced oscillatory instability.

The paper is organized as follows. In \S~\ref{sec:model} we present the vortex sheet model (\S~\ref{ssec:vs}), formulae for the thrust and drag forces (\S~\ref{ssec:thrust}), and the numerical method used to solve the model equations (\S~\ref{ssec:numerical}). The numerical results are presented in \S~\ref{sec:results}: specifically, the steady state motion for one plate (\S~\ref{ssec:single}); sample results for more than one plate (\S~\ref{ssec:several}); the evolution towards steady schooling states in multi-plate configurations (\S~\ref{ssec:steadypositions}); the set of all equilibria and their loss of stability (\S~\ref{ssec:loss}); our proposed stabilization mechanism (\S~\ref{ssec:stabilization}); and the reduced model based on thin-airfoil theory (\S~\ref{ssec:reduced}). Concluding remarks are given in \S~\ref{sec:conc}.

%%%%% sec02_model.tex %%%%%

\section{Numerical Approach}\label{sec:model}

%\newpage 
\begin{figure}
%\centering
\centerline{\includegraphics[trim=0 0 0 35, clip, width=0.99\textwidth]{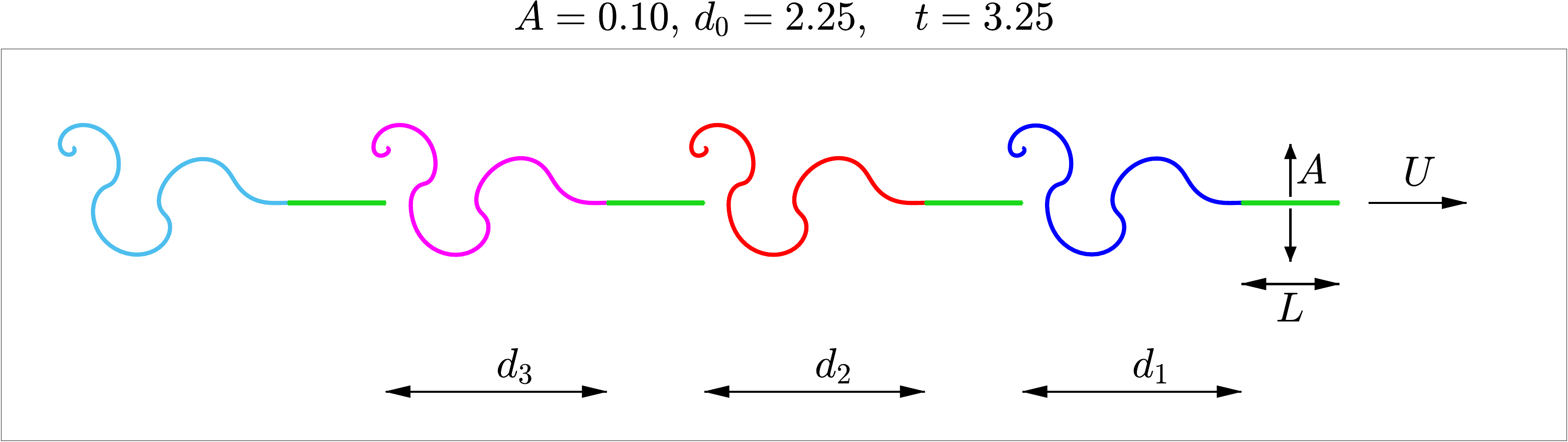}}
\caption{Schematic of the vortex sheet model. The plates (green), each of length $L$, are in an in-line formation and heave periodically in the vertical direction with amplitude $A$ and frequency $f$ while translating to the right with a horizontal velocity $U$. The vortex sheet shed by the first, second, third or fourth plate is colored blue, red, pink or cyan, respectively, a color scheme that will be repeated throughout the text. The figure shows a sample simulation at $t=3.25$ (which equals 1.625 flapping periods) of $n=4$ plates with $A = 0.1$ that were initially equispaced by a distance $d_0 = 2.25$. The tail-to-tip distances $d_i$ are also shown. 
}
\label{F:sketch}
\end{figure}

The model used to simulate $n$ flapping plates is illustrated in figure \ref{F:sketch}, for $n=4$. The fluid is assumed to be inviscid and incompressible, and also irrotational except for the wake behind each plate. The wake is modelled by a vortex sheet formed by shedding vorticity from the plates' trailing edges at each time step. The plates are modelled by bound vortex sheets, with vortex sheet strength determined to satisfy the no-penetration boundary condition, or the requirement that no fluid flows through the plates. The fluid velocity in turn is determined by the vorticity in the plates and the wakes. 

The plates are all of length $L$ and mass per unit span $M$, and are constrained to maintain an in-line formation, in that their cycle-averaged positions lie on a single horizontal line. They are initially separated by a distance $d_0$ from their nearest neighbors, and are given the same initial horizontal velocity $U_0$. The prescribed vertical heaving motion is oscillatory with vertical position $A\sin(2\pi f t)$, corresponding to frequency $f$ and amplitude $A$. This vertical motion induces a horizontal thrust that accelerates the plates, and is countered by a skin friction drag. Thrust and drag forces vary from plate to plate, causing distinct horizontal plate motion and thus distinct evolution of the separation distances $d_j(t)$ between the plates. 

%The flow is non-dimensionalized by the plate length $L$ and frequency $2f$.
While the flow is assumed to be inviscid, viscous effects are accounted for in two ways. First, they are necessary for vorticity to separate from the bodies. Second, they lead to a skin friction drag $F_{\mathrm{d}}$ on each plate, which we model using the Blasius boundary layer drag law, $F_{\mathrm{d}}=-C_{\mathrm{d}}\rho\sqrt{L\nu}U^{3/2}$, where $C_{\mathrm{d}}$ is the drag coefficient, $\nu$ the fluid kinematic viscosity, $\rho$ the fluid density and $U$ the instantaneous horizontal velocity of the plate. 
An alternative drag model used by \cite{Fang_Thesis} and \cite{Heydari_2021} more closely accounts for the velocity above and below the plate. %However, a comparison of these two models presented in Appendix~\ref{App:Drag} shows the differences between them to be minor. 
However, in Appendix~\ref{App:Drag} we present a comparison between these two models, which shows that the differences between them are minor. We thus opt to proceed with the simpler drag model herein.
The resulting governing equations of the fluid flow and plates' dynamics are non-dimensionalized by using the plate length $L$ as the length scale, the flapping half-period $1/(2f)$ as the time scale and $4\rho (Lf)^2$ as the pressure scale. In this paper, we fix the values of the dimensionless plate mass $\tilde{M}=M/\rho L^2$ and drag coefficient $\tilde{C}_{\mathrm{d}}=C_{\mathrm{d}}\sqrt{\nu/(2fL^2)}$, as will be discussed in \S~\ref{ssec:single}, and study the dependence of the plate motion on 
the initial velocity $\tilde{U}_0=U_0/(2Lf)$,
the initial separation distance $\tilde{d}_0=d_0/L$, 
the flapping amplitude $\tilde{A}=A/L$, 
and the number of plates $n$. From here on, all variables are non-dimensional and the tildes are dropped.

\subsection{The vortex sheet model}\label{ssec:vs}
The $n$ bound vortex sheets representing the flat plates have positions 
\begin{equation}
\xb_b^j(\alpha,t)=(x_b^j,y_b^j)(\alpha,t)\,,\quad
\alpha\in[0,\pi]~,\quad j=1,\dots,n\,,
\label{E:plate}
\end{equation}
and sheet strengths $\gamma^j(\alpha,t)$. The sheets are initially on the horizontal $x$-axis, with position
\begin{align}
x_b^j(\alpha,0)&=s(\alpha) - (j-1)(d_0+1),~~ s(\alpha)=\cos(\alpha)/2\,,\\
y_b^j(\alpha,0)&=0\,.
\label{E:init}
\end{align}
That is, each plate has unit length, with $\alf=0$ at the leading edge (right), $\alf=\pi$ at the trailing edge (left), and $d_0$ the tail-to-tip distance between plates.
The horizontal evolution of $x^j_b$ is determined by the displacement $X^j(t)$ of plate $j$ from its initial position, while the oscillatory vertical component $y^j_b(t)$ is prescribed:
\begin{subequations}
\begin{align}
x^j_b(\alpha,t)&=X^j(t) + x_b^j(\alf,0)\,,\quad X^j(0)=0\,,\\
y^j_b(\alpha,t)&=A\sin(\pi t)\,.
\label{E:prescribedY}
% CAREFUL: in code s(alf)=-cos(alf)/2 so leading and trailing are reversed
\end{align}
\end{subequations}
The plate position $X^j(t)$ evolves according to the induced thrust and drag forces acting on the plate (see \S~\ref{ssec:thrust}).
The plate vortex sheet strength is determined numerically by enforcing the no-penetration boundary condition on the plate walls.
%ensuring that the vertical fluid velocity at points on the plate equals the prescribed vertical plate velocity. 

The free vortex sheets representing the wakes are parameterized by the circulation parameter $\Gamma$, where $\Gamma_T^j$ is the total circulation around the $j$th sheet,
\begin{equation}
\xb_f^j(\Gamma,t)~,\quad \Gamma\in[0,\Gamma_T^j] ~,\quad j=1,\dots,n~.
\end{equation}
Together, the free and the bound vortex sheets induce the fluid velocity at any point $\xo$:
\begin{equation}
\ub(\xo,t) = \frac{1}{2\pi}\sum_{j=1}^n \left[
\int\limits_0^{\pi}
\frac{
%\kb\times(\xb_b^j(\alpha,t)-\xo)
(\xo-\xb_b^j(\alpha,t))^{\perp}
} {|\xo-\xb_b^j(\alpha,t)|^2}\gamma^j(\alpha,t)
s'(\alpha)\,\rmd\alpha
+%\sum_{j=1}^n
\int\limits_0^{\Gamma_T^j}
\frac{
%\kb\times(\xb_f^j(\Gamma,t)-\xo)
(\xo-\xb_f^j(\Gamma,t))^{\perp}
} {|\xo-\xb_f^j(\Gamma,t)|^2+\delta^2}\rmd\Gamma 
\right].\label{E:totalvel}
\end{equation}
where $(x,y)^{\perp}=(-y,x)$
Here, the parameter $\delta$ regularizes the flow for points $\xo$ near or on the free sheet, using the vortex blob method~\citep{Krasny1986}. If $\xo$ is a point on the plate, the integrals in~\eqref{E:totalvel} are evaluated in the principal value sense and yield the average of the velocity above and below the plate.

The free sheet evolves with the fluid velocity (\ref{E:totalvel}) as
\begin{equation}
\frac{\rmd\xb_f^j}{\rmd t}(\Gamma,t) =\ub\left(\xb_f^j(\Gamma,t),t\right) ~.
\label{E:velofree}
\end{equation}
Its circulation $\Gamma$ is determined using the circulation shedding algorithm of~\citet{Nitsche1994}. 
Vorticity is shed at time $t$  by releasing a particle from the plate's trailing edge, with vertical velocity equal to the  plate's vertical velocity and horizontal velocity equal to the average horizontal fluid velocity $u(\xb^j(\pi,t),t)=\ubar^j$ at the trailing edge, as obtained from (\ref{E:totalvel}).  The particle's circulation is such that the Kutta condition is satisfied, with
\begin{subequations}
\begin{align}
\frac{\rmd\Gamma_T^j}{\rmd t} &= -\frac{1}{2}[(u^j_+)^2-(u^j_-)^2]~,\label{E:kuttaA}\\
\ubar^j&=\frac{1}{2}(u^j_++u^j_-)\,,\label{E:kuttaB}\\
\gamma^j(\pi,t)&=u^j_--u^j_+\,.\label{E:kuttaC}
\end{align}
\label{E:kutta}
\end{subequations}
Here, $u^j_+$ and $u^j_-$ are the horizontal velocity components above and below the $j$th plate at the trailing edge,  obtained from $\gamma^j(\pi,t)$ and $\ubar^j$. 

\subsection{Thrust and Drag}\label{ssec:thrust}
The plates experience a thrust force due to the suction at their leading edges, which arises because the fluid velocity is singular there. Specifically, the thrust %force acting on the plates 
is obtained from the leading edge bound vortex sheet strength. The sheet strength is unbounded at the leading edge, with 
$\gamma^j(\alpha,t)\sim \Delta s^{-1/2}$ as $\Delta s\rightarrow 0$ where $\Delta s=1/2-s(\alpha)$ is the distance from the leading edge. However, the desingularized vortex sheet strength $\gamma^j(\alpha)s'(\alpha)$ remains bounded. The horizontal thrust force acting on the plate is given by \citep{Fang2017,EldredgeBook}
\begin{equation}
F^j_x=\frac{\pi}{4}\gamtil_j^2~, \quad \text{where}\quad \gamtil_j=\lim\limits_{s\to 1/2}\gamma^j(s)\sqrt{\frac{1}{4}-s^2}
\label{E:thrust}
\end{equation}
and, with our choice of $s(\alpha)=\cos\alpha/2$, 
% CAREFUL
%in code $s(\alpha)=-\cos\alpha/2$, 
\begin{equation}
\gamtil_j=-\lim\limits_{\alpha\to0}\gamma^j(\alpha)s'(\alpha)\,.\label{E:LEgamma}
\end{equation}

The relation (\ref{E:thrust}) holds for steady or unsteady plate motion. For steady motions, the result reduces to  $\gamtil=-2V$, where $V$ is the vertical velocity of the leading edge, which is used by~\citet{Alben_Street}. % and follows from a simpler argument \citep{Acheson}. 
For completeness, both the unsteady and steady results are derived here in Appendices~\ref{App:Thrust} and \ref{App:Steady}, respectively. Appendix~\ref{App:Steady} also shows a comparison between the flow field around a plate in a steady vertical flow, and the unsteady flow field around an impulsively started plate moving vertically upward with constant velocity, the latter being computed using the method described in \S~\ref{ssec:numerical}. We detail how the flow field (figure~\ref{F:streamlinesvx}) and circulation (figure~\ref{F:tipstrengthconstantvelo}) in the unsteady case approach that of the steady case in the long-time limit. % Steady and unsteady flow are then compared for an example. Consistency between the two results is shown for an unsteady flow of an impulsively started plate that approaches a steady flow

As discussed at the beginning of \S~\ref{sec:model}, the drag force acting on the plate is modelled to be proportional to $(U^j(t))^{3/2}$, where $U^j(t)$ is the plate velocity. The horizontal motion of the $n$ plates is thus governed by the system of ordinary differential equations
\begin{subequations}
\begin{eqnarray}
\frac{\rmd X^j}{\rmd t}&=&U^j ~,\quad X^j(0)=0,\label{E:pos}\\
M\frac{\rmd U^j}{\rmd t}&=&\frac{\pi}{4}\gamtil_j^2-C_{\mathrm{d}} (U^j)^{3/2}~,\quad U^j(0)=U_0\,.\label{E:vel}
\end{eqnarray}
\label{E:veloplate}
\end{subequations}
The flow evolution is determined by the system of ODEs (\ref{E:velofree}, \ref{E:kutta} and \ref{E:veloplate}), in addition to a Fredholm integral equation for $\gamtil$. Its numerical solution is described next.

\subsection{Numerical method}\label{ssec:numerical}
The plates (\ref{E:plate}) are each discretized by $N_b+1$ point vortices corresponding to $\alpha_k=k\pi/N_b$, $k=0,\dots,N_b$. The free vortex sheets are discretized by $N_f+1$ vortex blobs with total circulation $\Gamma_T^j$, created by shedding one blob at each timestep; thus, $N_f$ increases with time. Time is discretized by $t_m=m\Delta t(t)$ where typically $\Delta t$ is smaller at early times near the beginning of the motion and increases linearly from $\Delta t_1$ to a final timestep $\Delta t_2$ at a time $t_1$, after which it remains constant. The small initial timesteps are needed to resolve the large initial velocities about the trailing edge. The system of ODEs (\ref{E:velofree}, \ref{E:kutta} and \ref{E:veloplate}) is solved using the fourth order Runge-Kutta method. At each stage, the following steps are taken~\citep{Sheng2012}: 
\begin{itemize}
\item[(i)] Set the vertical plate velocity $V(t)=A\pi\cos(\pi t)$. 
\item[(ii)] For each plate, $j=1,\dots,n$, find the sheet strength $\gamma^j_k=\gamma^j(\alpha_k,t)$ such that the vertical fluid velocity  in~\eqref{E:totalvel} evaluated at the midpoints $\alpha_k^m=(\alpha_{k-1}+\alpha_k)/2$ equals the prescribed vertical plate velocity,
\begin{equation}
\ub\left(\xb^j_b(\alpha^m_k,t),t\right)\cdot \nb = V(t)~,\quad k=1,\dots,N_b
\end{equation}
where $\boldsymbol{n}=(0,1)$ is normal to the plate. Here, the singular integrals in the velocity~\eqref{E:totalvel} are computed with what is in essence the alternating point vortex method of~\citet{Shelley1992}. In addition, enforce zero total circulation around the plate and its wake. The resulting $(N_b+1)\times (N_b+1)$ linear system for $\gamma^j_k$, $k=0 \ldots N_b$, is inverted using a precomputed LU decomposition.
\item[(iii)] Set $\gamma^j(\pi,t)= \gamma^j_{N_b}$ and $\ubar^j= u\left(\xb_b^j\left(\alpha_{N_b}^m,t\right),t\right)$, and obtain $u_j^+$ and $u_j^-$ using~\eqref{E:kuttaB} and~\eqref{E:kuttaC}. Then shed a particle $\xb^j_{\mathrm{new}}$ from each trailing edge with velocity 
\begin{equation}
\frac{\rmd \xb^j_{\mathrm{new}}}{\rmd t}= (\ubar^j,V)
\label{E:veloshed}
\end{equation}
and circulation such that~\eqref{E:kuttaA} is satisfied. As explained by~\citet[\S2.4(iii)]{Nitsche1994}, only separated flow is considered when determining $u_j^+$ and $u_j^-$. 
\item[(iv)] Move all previously shed particles with velocity 
\begin{equation}
\frac{\rmd \xb^j_{f,k}}{\rmd t}= \ub\left(\xb^j_{f,k},t\right)~,\quad k=0,\dots,N^j_f
\label{E:velopoints}
\end{equation}
and add $\xb^j_{f,N^j_f+1}=\xb^j_{\mathrm{new}}$, then increase $N^j_f$ by one.  Here, the integral over the bound vortex sheets in (\ref{E:totalvel}) is near-singular when the target point is near a plate. The near-singular integrals are computed using the corrected trapezoid rule of~\citet{Nitsche_2021}.
\item[(v)] Determine the desingularized leading edge sheet strength $\gamtil_j$ using~\eqref{E:LEgamma} and move the plates by (\ref{E:veloplate}). 
\end{itemize}

\vspace{0.05in}
\noindent Typically, one new particle is shed per time step. In some cases a particle is shed every other timestep or more for the sake of computational efficiency.

%%%%% sec03_results.tex %%%%%

\section{Numerical Results}\label{sec:results}
This section presents numerical results for $n=1$, 2, 3 and 4 plates, with flapping amplitudes $A=0.1-0.4$, mass $M=1$ and drag coefficient $C_{\mathrm{d}}=0.1$. The numerical parameters are $\delta=0.2$, $N_b=100$, $t_1=1$, $\Delta t_2=0.008/A$ and $\Delta t_1\approx \Delta t_2/10$. Results were confirmed to remain basically unchanged under mesh refinement (decreasing $\Delta t$, increasing $N_b$) or smaller values of $\delta$.

%%%%% sec031_oneplate.tex %%%%%

%\newpage 
\begin{figure}
 \centering
\includegraphics[trim=0 0 0 20, clip, width=0.78\textwidth]{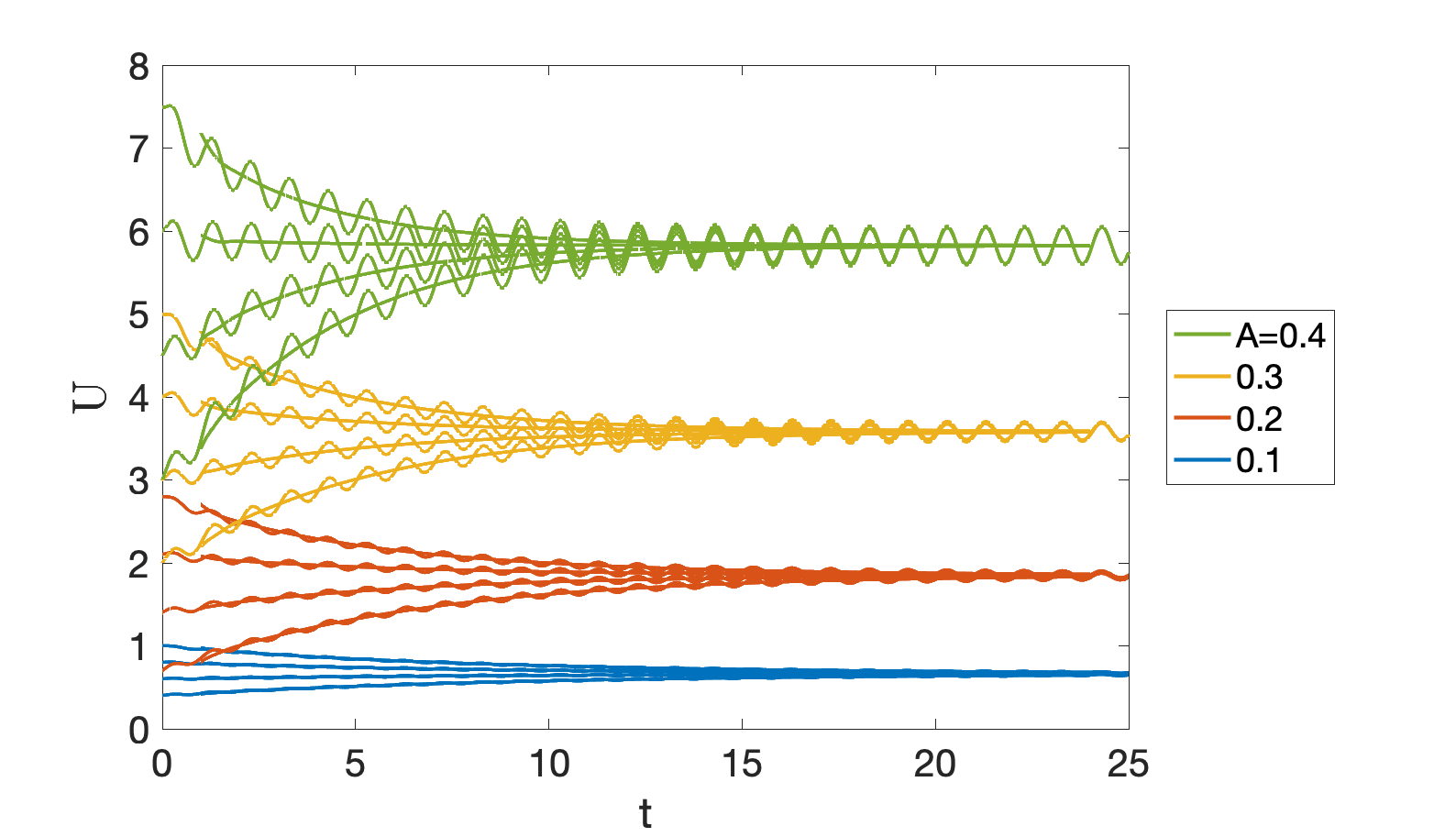}
\caption{Time evolution of the translation velocity $U$ of a single plate ($n=1$), for various initial velocities $U_0$ and the four amplitudes $A$ indicated in the legend. The oscillatory curves show the instantaneous velocity $U(t)$, upon which we have superimposed a cycle-averaged velocity obtained by averaging $U(t)$ over a moving window of size equal to the oscillation period.} 
\label{F:uinf}\end{figure}

\subsection{Single plate, $n=1$}\label{ssec:single} % Steady state velocity $\uinf$}%  ($n=1$)}
Here we consider one plate moving with a prescribed oscillatory heaving motion of amplitude $A$, and given an initial horizontal velocity $U_0$. The plate moves forward by the thrust force (\ref{E:thrust}), as vorticity is shed from its trailing edge. Figure \ref{F:uinf} shows the resulting horizontal plate velocity for a range of amplitudes $A$ and initial velocities $U_0$. The plate horizontal velocity oscillates with the same frequency as the heaving motion. For each case, the figure shows both the actual oscillatory velocity $U(t)$, and the cycle-averaged velocity obtained by averaging $U(t)$ over a moving window of size equal to the oscillation period. The figure shows that, in all cases, the cycle-averaged plate velocity approaches a steady state velocity $\uinf$ that depends solely on the amplitude $A$ and is independent of the initial velocity $U_0$. The steady state velocity $\uinf$ increases as $A$ increases, with values of
$\uinf=0.665$, 1.85, 3.58 and 5.83 for $A=0.1$, 0.2, 0.3 and 0.4, respectively. These values of $\uinf$ are roughly consistent with those obtained in experiments on heaving wings in a water tank~\citep{Sophie}, which motivates our choice of the drag coefficient $C_{\mathrm{d}}$.

\begin{figure}
 \centering
\includegraphics[trim=0 34 0 15, clip, width=0.995\textwidth]{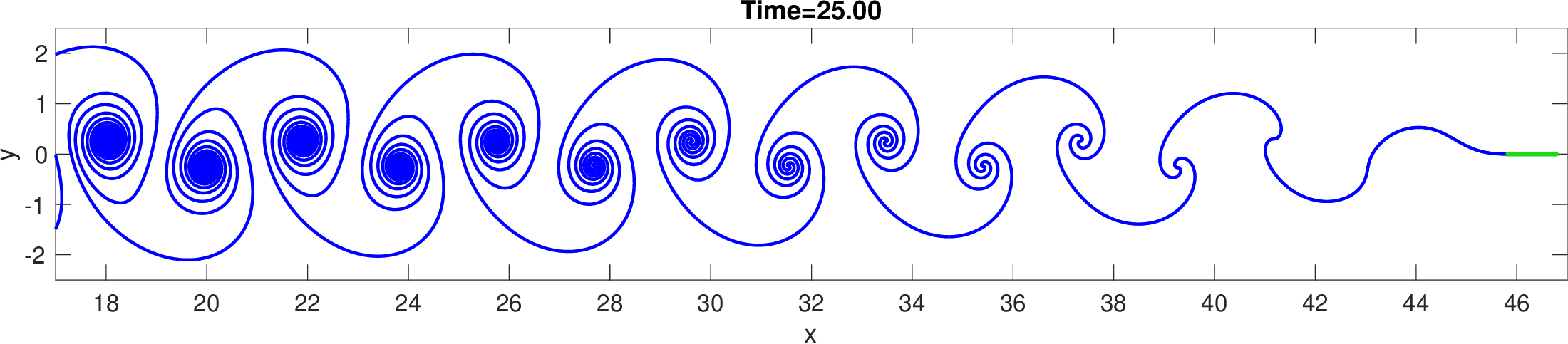}
\includegraphics[trim=0 34 0 15, clip, width=1.0\textwidth]{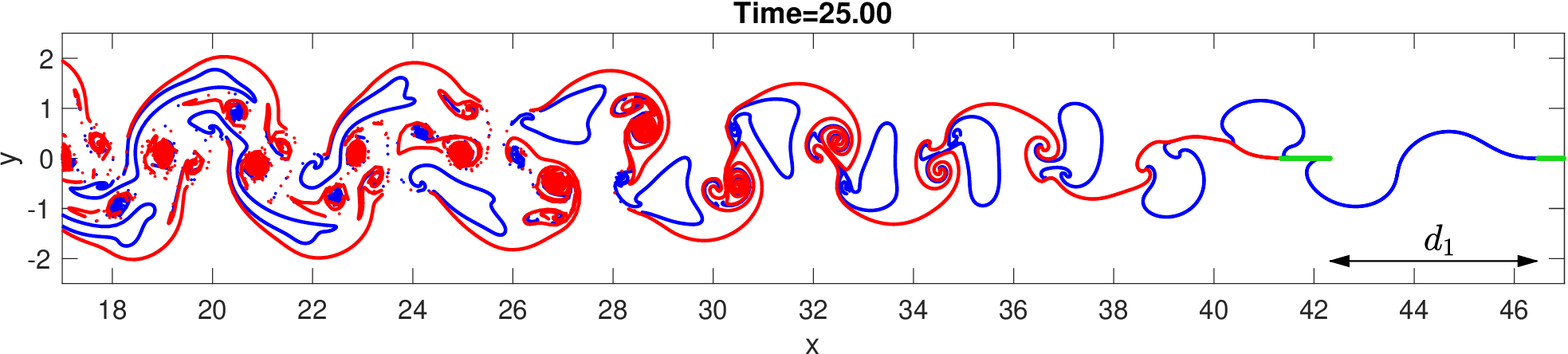}
\includegraphics[trim=115 110 110 105, clip, width=1.0\textwidth]{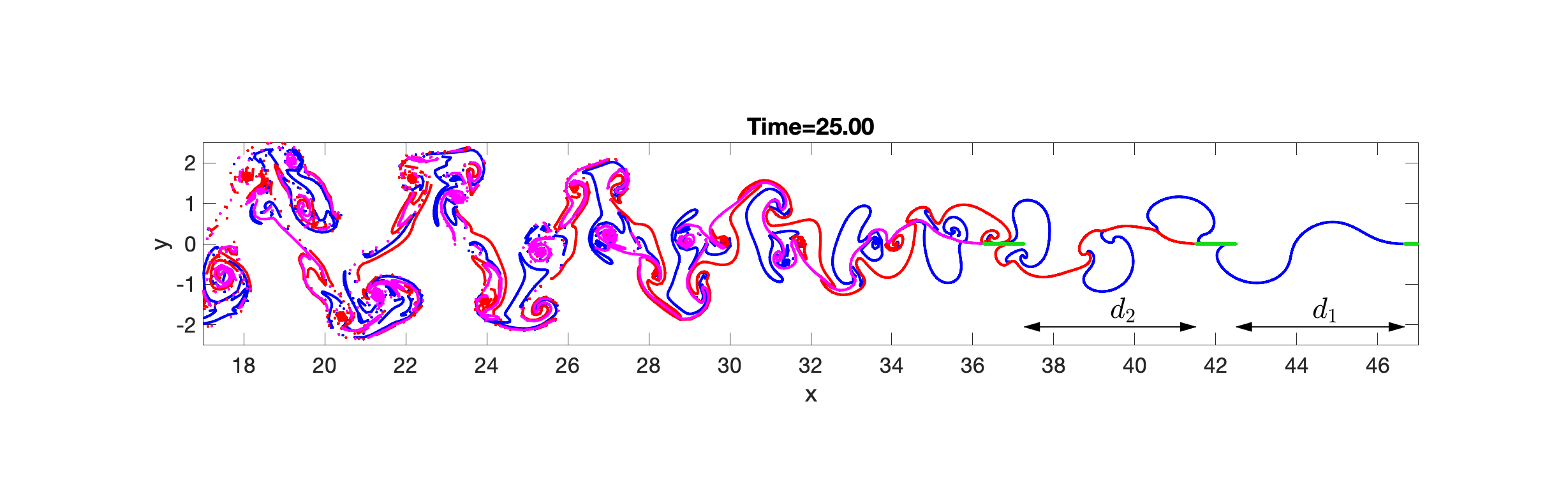}
\includegraphics[trim=115 75 110 105, clip, width=1.0\textwidth]{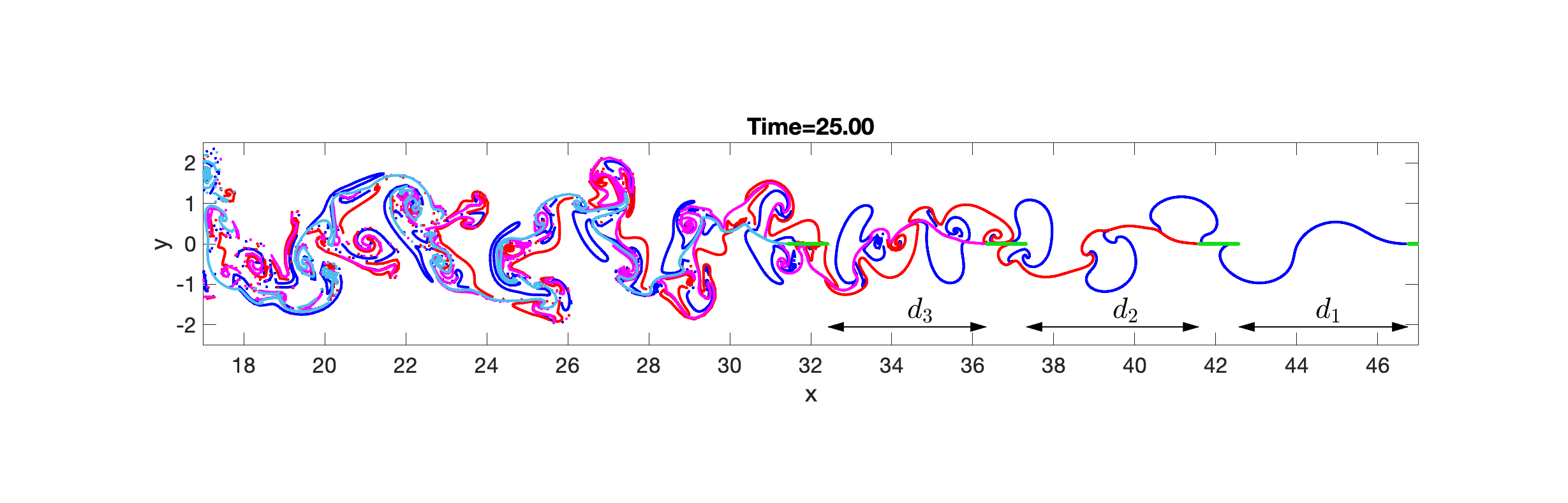}
\vspace*{-71ex}  % Tune this to the image height.
\begin{center}
\hspace{-5.4in}
{\normalsize (a)}
\end{center}
\vspace*{71ex}
\vspace*{-57ex}  % Tune this to the image height.
\begin{center}
\hspace{-5.4in}
{\normalsize (b)}
\end{center}
\vspace*{57ex}
\vspace*{-43ex}  % Tune this to the image height.
\begin{center}
\hspace{-5.4in}
{\normalsize (c)}
\end{center}
\vspace*{43ex}
\vspace*{-29ex}  % Tune this to the image height.
\begin{center}
\hspace{-5.4in}
{\normalsize (d)}
\end{center}
\vspace*{15ex}
\caption{Plate and vortex sheet position at $t=25$, for the flapping amplitude $A=0.2$ and $n=1$, 2, 3 and 4 plates (from top to bottom). All plates are initially equispaced with a distance near $d_0=4.2$.
%\colb{The vortex sheet in the wake is plotted using a linear interpolant of the shed point vortices if neighbouring points are sufficiently close to each other. As the sheet stretches far downstream, we simply plot the individual point vortex positions. }
}
\label{F:allsheet25}
\end{figure}

Figure \ref{F:allsheet25}(a) shows the shed vortex sheet behind one plate oscillating with amplitude $A=0.2$, at time $t=25$. The given initial velocity is that of the corresponding steady state, $U_0=1.85$. With each upward (downward) motion of the plate, positive (negative) vorticity is shed from the trailing edge, which leads to a sequence of counter-rotating vortex pairs shed during each  oscillation period. By taking the curl of~\eqref{E:totalvel}, we obtain the regularized vorticity corresponding to the shed vortex sheet, which is shown in figure \ref{F:allvort25}(a). Soon after being shed, the vorticity settles into a sequence of almost equally spaced vortices of alternating sign, with the positive (negative) vortex positioned slightly below (above) the centerline $y=0$. 

\subsection{Several plates, $n=2,3,4$}\label{ssec:several}
\begin{figure}
\centering
\includegraphics[trim=100 76 95 68, clip, width=1.0\textwidth]{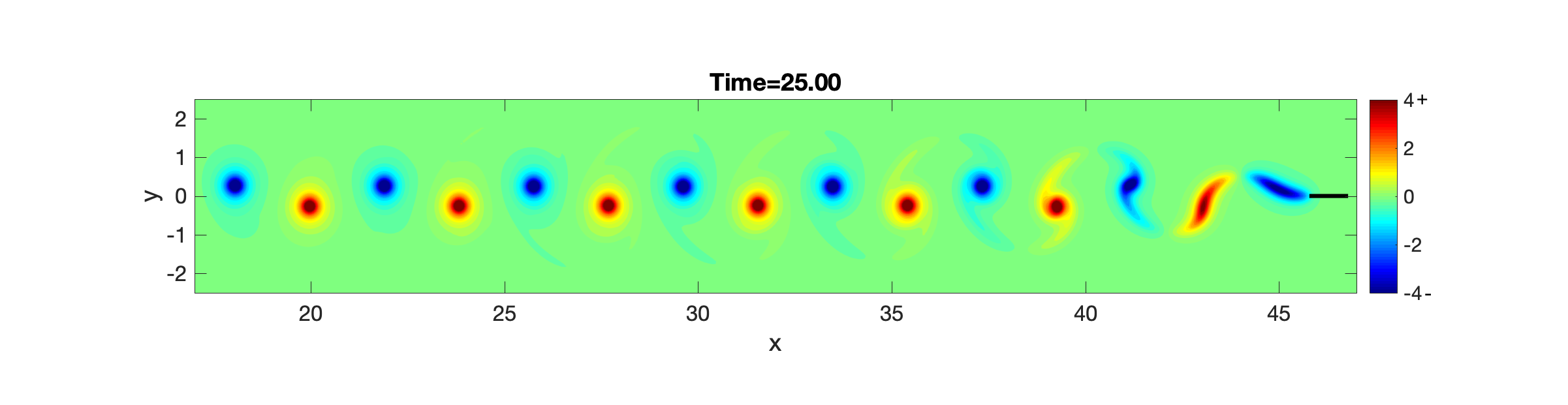}
\includegraphics[trim=100 76 95 68, clip, width=1.0\textwidth]{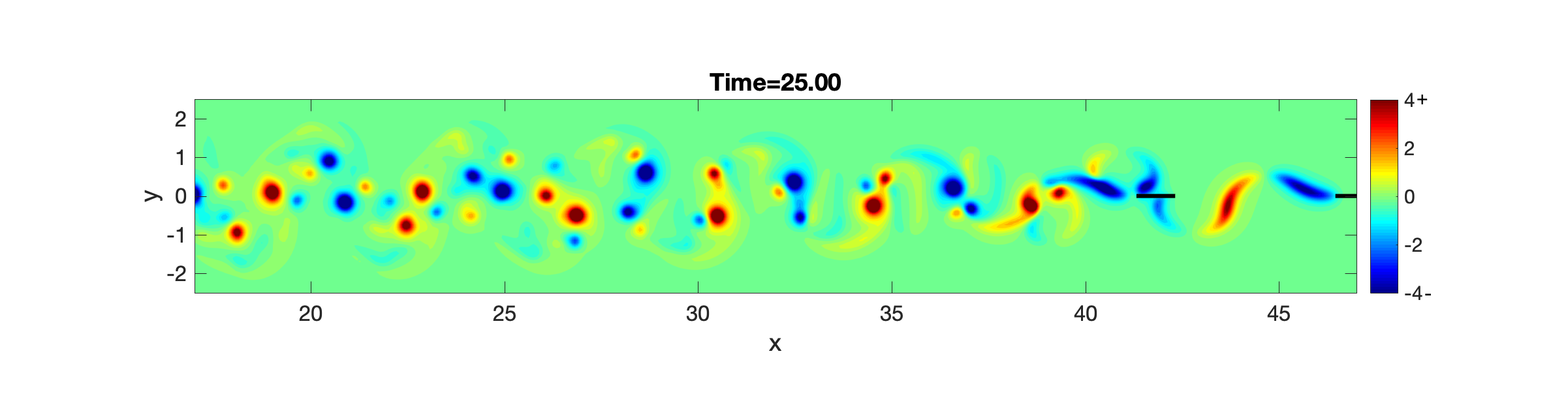}
\includegraphics[trim=100 76 95 68, clip, width=1.0\textwidth]{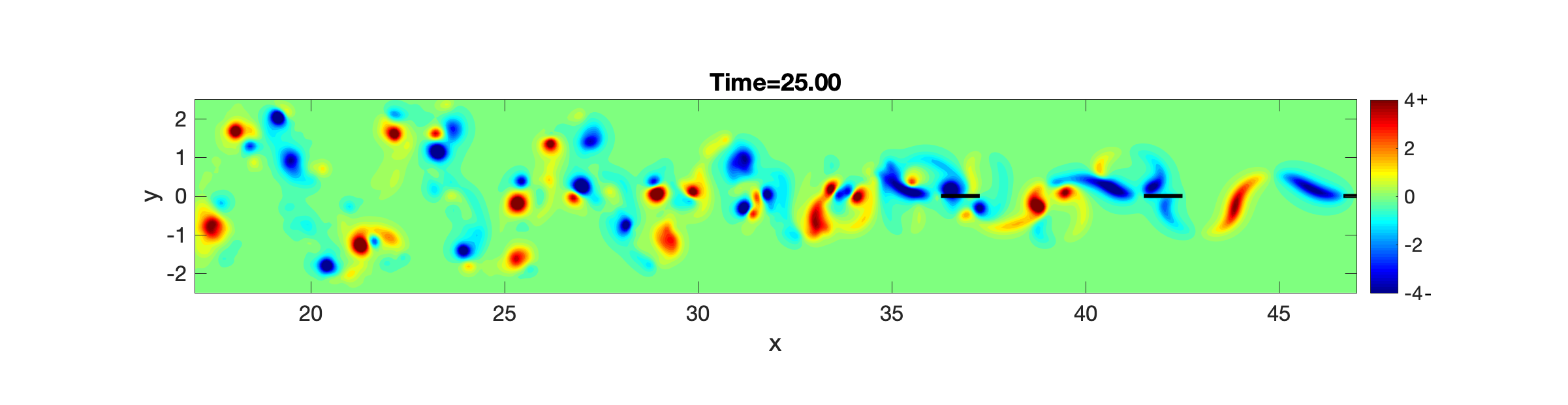}
\includegraphics[trim=100 40 95 68, clip, width=1.0\textwidth]{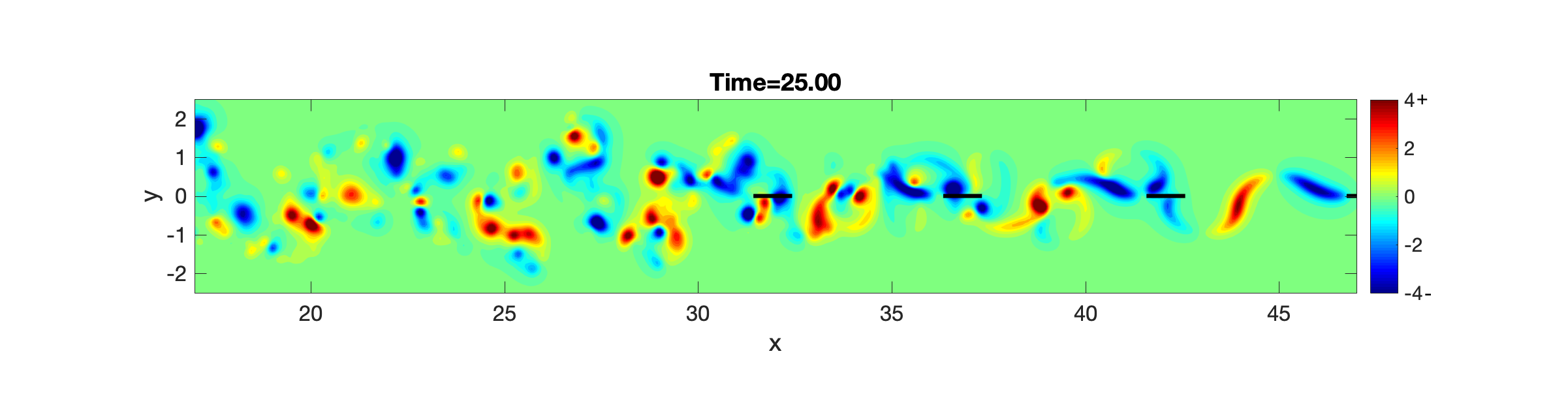}
\vspace*{-66.5ex}  % Tune this to the image height.
\begin{center}
\hspace{-5.4in}
{\normalsize (a)}
\end{center}
\vspace*{66.5ex}
\vspace*{-54ex}  % Tune this to the image height.
\begin{center}
\hspace{-5.4in}
{\normalsize (b)}
\end{center}
\vspace*{54ex}
\vspace*{-41ex}  % Tune this to the image height.
\begin{center}
\hspace{-5.4in}
{\normalsize (c)}
\end{center}
\vspace*{41ex}
\vspace*{-28ex}  % Tune this to the image height.
\begin{center}
\hspace{-5.4in}
{\normalsize (d)}
\end{center}
\vspace*{13ex}
\caption{Same as figure~\ref{F:allsheet25}, but the regularized vorticity is plotted instead of the vortex sheet. The plates are indicated in black. Absolute vorticity values larger than 4 are represented by the darkest blue and red colors.}
\label{F:allvort25}
\end{figure}

Figure \ref{F:allsheet25}(b) shows the two vortex sheets shed behind two plates, both heaving with amplitude $A=0.2$. The color scheme is the same as that in figure \ref{F:sketch}, with the vortex sheet shed by the leader (follower) plate indicated in blue (red).
There is a large amount of stretching far downstream of the plates as the two sheets interact. We plot the sheets using a linear interpolant of the shed point vortices if neighbouring points are sufficiently close to each other. Otherwise, after a sufficiently large amount of stretching, we simply plot the individual point vortex positions. 
The initial distance between the plates is $d_0=4.2$, which is, as will be seen later, near an equilibrium position. The two sheets are seen to roll up into four vortices per period, a phenomenon that is also evident in the top panel of supplementary movie 1. This is more clearly seen in figure \ref{F:allvort25}(b), which shows a sequence of two positive vortices followed by two negative vortices per period. This pattern repeats, although in an irregular fashion. Groups of four vortices are discernible, but their relative positions do not settle into a clear steady pattern. 

Figure \ref{F:allsheet25}(c) shows the three vortex sheets shed behind three plates heaving with amplitude $A=0.2$. The initial separation between all plates is as before, $d_0=4.2$. A pattern seems to appear farther downstream of the plates. Most notably, near $x=20$ the vortices have spread further from the midline $y=0$. Figure \ref{F:allvort25}(c) reveals more clearly a repeating pattern of groups of six vortices, three positive followed by three negative. Sufficiently far downstream, for $x<25$, the vorticity is clearly concentrated away from the midline, as it has spread significantly farther than in the $n=2$ case shown in figure~\ref{F:allvort25}(b). A simulation with $n=3$ plates is shown in supplementary movie 2.

Figures \ref{F:allsheet25}(d) and \ref{F:allvort25}(d) show the vortex sheets and corresponding vorticity shed behind $n=4$ moving plates. It is difficult to detect a pattern in the downstream vortex sheet positions. The vorticity is expected to consist of groups of eight vortices, four positive followed by  four negative, but the pattern is less clear and more random. Furthermore, the spread away from the midline is not as big as for $n=3$. A simulation with $n=4$ plates is shown in supplementary movie 3.

%\begin{figure}
% \centering
%\includegraphics[trim=100  76 95 68, clip, width=1.0\textwidth]%{figs/t25a20n1vort}
%\includegraphics[trim=100 76 95 68, clip, width=1.0\textwidth]{figs/t25a20n2vort}
%\includegraphics[trim=100 76 95 68, clip, width=1.0\textwidth]{figs/t25a20n3vort}
%\includegraphics[trim=100 40 95 68, clip, width=1.0\textwidth]{figs/t25a20n4vort}
%\caption{Same as figure~\ref{F:allsheet25}, but the regularized vorticity is plotted instead of the vortex sheet. The plates are indicated in black. Absolute vorticity values larger than 2 are represented by the darkest blue and red colors.}
%\label{F:allvort25}
%\end{figure}

The effect of the trailing plates on the leader is most clearly seen in figure~\ref{F:allsheet25}. In all four cases, the given initial velocity is $U_0=1.85$, and the initial distances between the plates are $d_j(0)= 4.2$. The shape of the sheet between the first and second plate is basically unchanged in all cases from the single plate case. That is, the second plate has little influence on the wake in front of it. Similarly, the wake behind the second and third plate is unchanged from the $n=2$ case, so the third plate also has little influence on the wake in front of it. %We observe 
It is evident that as the number of plates behind the front plate increases, the front plate has travelled slightly further by $t=25$. However, after an initial transient, we observe %the results show 
that $\uinf$ remains essentially unchanged, with only a small increase of less than 3\% as $n$ increases.  We attribute this increase in the leader's velocity, albeit small, to the fact that the leader's bound vorticity is modified by the downstream plates' bound and shed vorticity, as is evident from step (ii) of the numerical method presented in \S~\ref{ssec:numerical}. The leader's bound vorticity in turn affects its thrust, as given by Eq.~\eqref{E:thrust}.

%%%%% sec033_equilibria.tex %%%%%

\subsection{Steady state positions for $A=0.2$, $n=2$}\label{ssec:steadypositions}

\begin{figure}
 \centering
\includegraphics[trim=0 50 0 18, clip, width=0.699\textwidth]{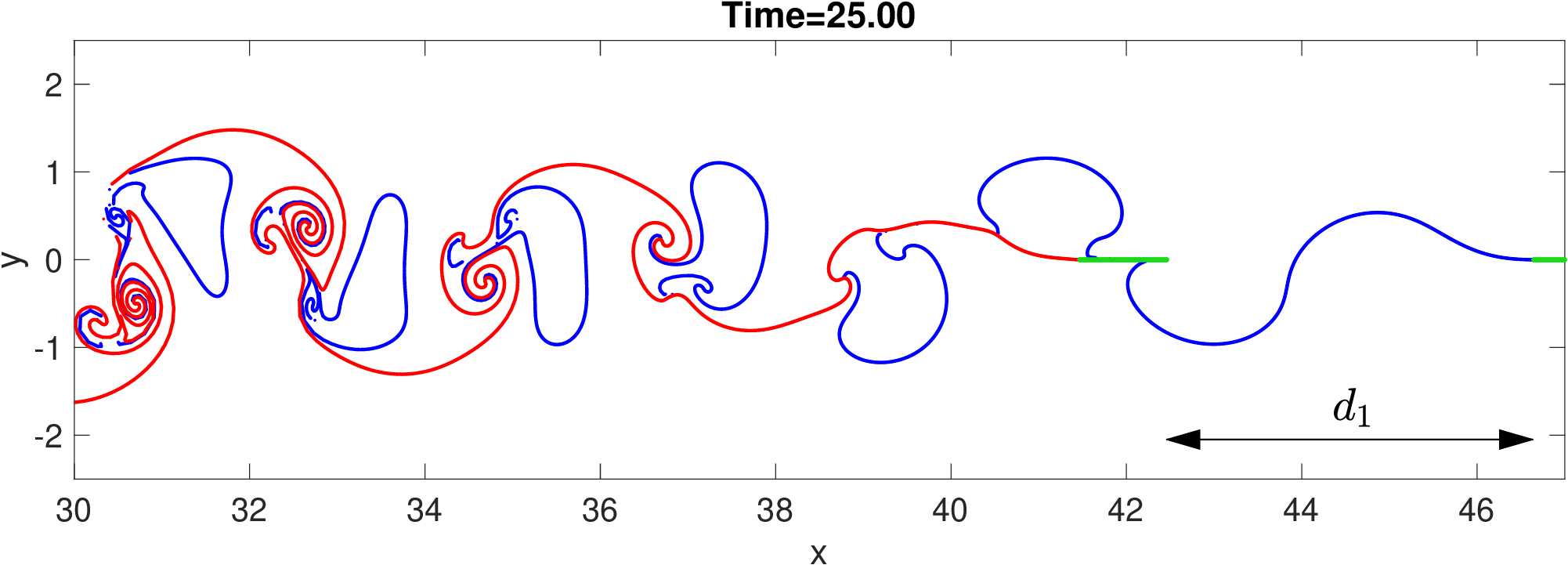}
\includegraphics[trim=0 50 0 18, clip, width=0.7\textwidth]{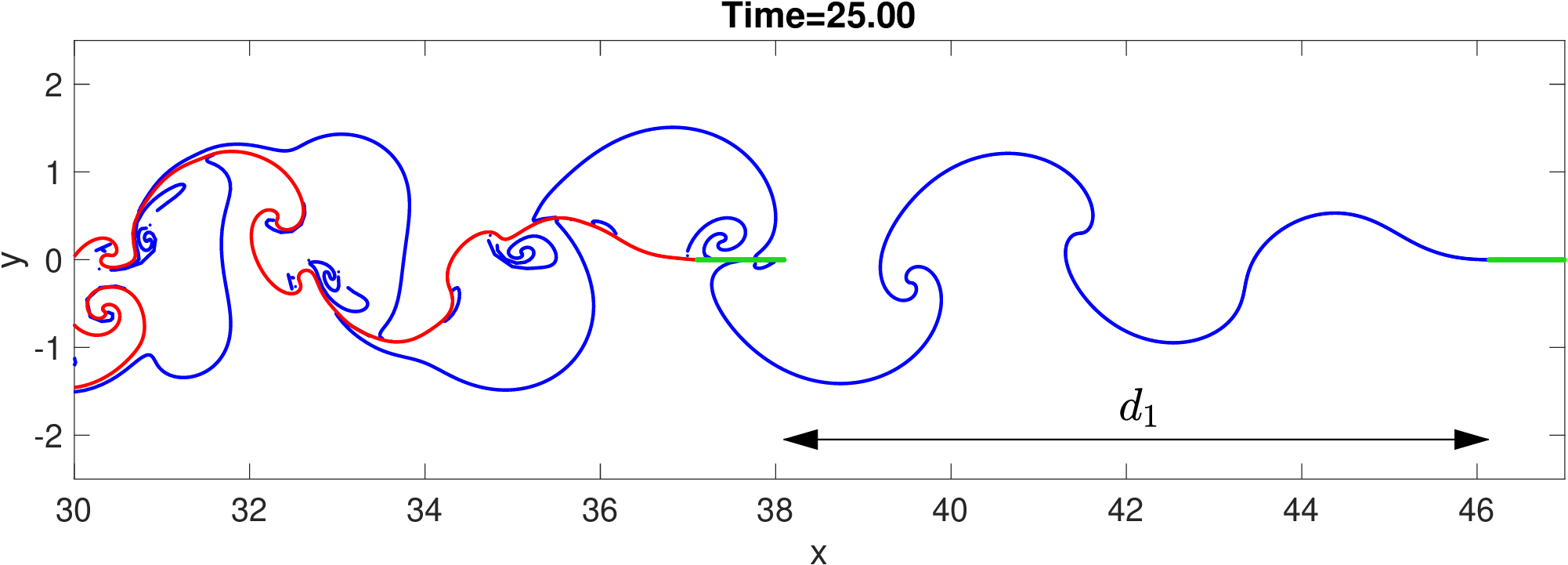}
\includegraphics[trim=0  0 0 18, clip, width=0.7\textwidth]{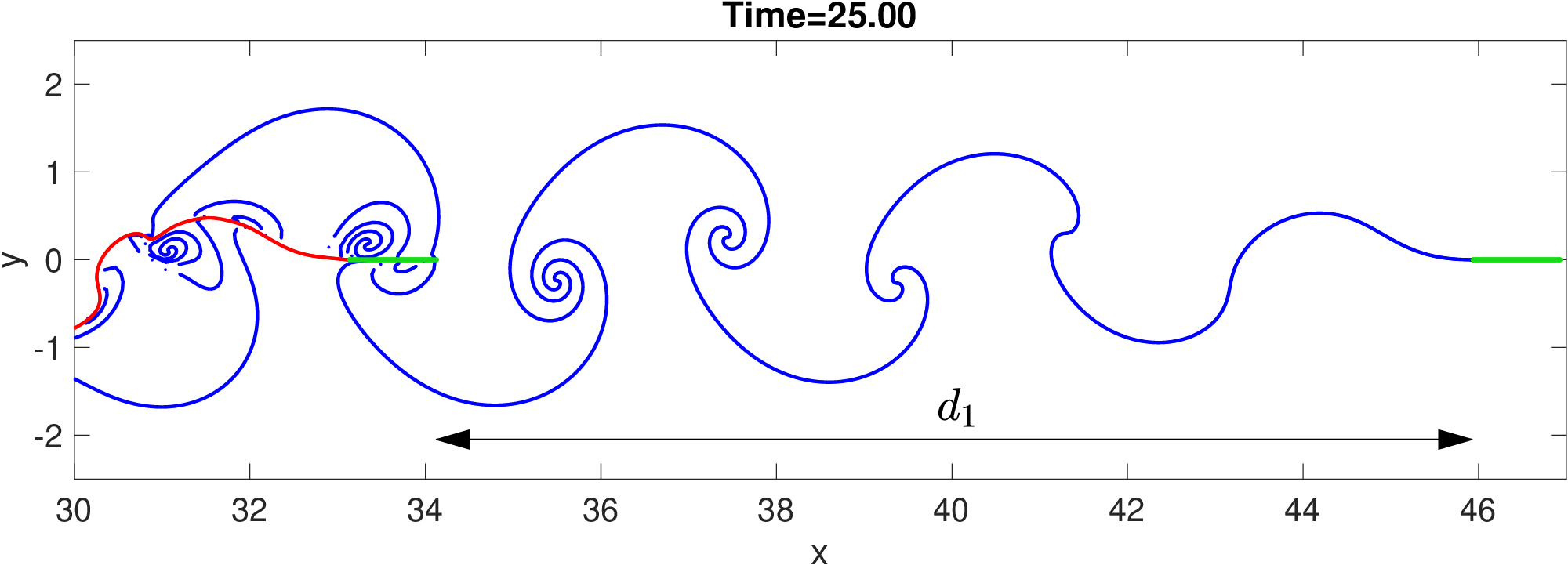}
\caption{
Snapshot at $t=25$ of three of the schooling modes obtained for a pair of plates ($n=2$) with flapping amplitude $A = 0.2$. The plates are initially located near the first, second and third equilibria, which, from top to bottom, correspond to the distances $d^{\infty,1}_1=4.18$, $d^{\infty,2}_1=8.01$ and $d^{\infty,3}_1=11.82$, respectively.
}
%dinf obtained from runs to t=25, havent quite reached steady state. using 1.85, 3 digits for all
%old values 4.17,8.0,11.8
\label{F:equilibria}
\end{figure}

The distances $d_j(t)$ between the plates evolve in time. We are interested in the long term behaviour of $d_j$ and the associated possible steady state configurations. 
This section presents results for a pair of wings ($n=2$) with flapping amplitude $A=0.2$, with a range of initial separation distances $d_0$. 
Since the steady state cycle-averaged plate velocity $\uinf$ is found to always be independent of $U_0$, from here on the initial horizontal velocity of each plate is taken to be the steady state cycle-averaged velocity of a single plate flapping with the same amplitude. %all plates are initialized with $U_0=\uinf$.
We observe that the two plates evolve towards a steady configuration in which the distance between them, after undergoing several oscillations, approaches a constant, but that this equilibrium distance $d_1^{\infty}$ depends on the initial distance $d_0$. Figure~\ref{F:equilibria} shows snapshots of simulations performed for three different values of $d_0$, at a time when the separation distance $d_1$ between the plates has reached the equilibrium values $d_1^{\infty,1}$, $d_1^{\infty,2}$ and $d_1^{\infty,3}$. Supplementary movie 1 also shows examples of simulations performed for different values of $d_0$, which illustrates how changing $d_0$ causes the pair of plates to attain different equilibria.
%Figure \ref{F:equilibria} shows the solution for $n=2$ plates, at a time when the distance $d_1$ between the plates has reached an equilibrium, for three different values of $d_0$. Figures (a,b,c) show three different equilibrium positions, with $d_1=d_1^{\infty,1}$, $d_1^{\infty,2}$, and$d_1^{\infty,3}$.

%figures made by twoswimmersmove/datout/plotclschool1.m

Note that in figure~\ref{F:equilibria}(a), the distance $d_1^{\infty,1}$ between the plates is approximately the distance travelled by the plate in one period of the heaving oscillation, as evident by the single upward and downward wave of the shed sheet. In figure~\ref{F:equilibria}(b), the equilibrium distance is approximately twice that of figure~\ref{F:equilibria}(a), and in figure~\ref{F:equilibria}(c) it is approximately 3 times. The distance traveled by a plate with velocity $\uinf$ in one oscillation period is the wavelength $\lambda=\uinf/f$, where $f=1/2$. Using the value $\uinf=1.85$ % [I didn't realize this: we use the single-wing steady state speed, rather than the steady state speed of the pair?] 
corresponding to $A=0.2$ we find that the equilibrium positions satisfy
%\begin{subequations}
%\begin{align}
%d^{\infty,1}_{1}\approx 4.17\approx &~ 1.12 \lambda \\
%d^{\infty,2}_{1}\approx 8.0 \approx &~ 2.12 \lambda\\
%d^{\infty,3}_{1}\approx 12 \approx &~ 3.12 \lambda
%\end{align}
%\end{subequations}
\begin{align}
d^{\infty,1}_{1}\approx 4.18\approx  1.13\lambda,\quad
d^{\infty,2}_{1}\approx 8.01 \approx  2.16\lambda\quad\text{and}\quad
d^{\infty,3}_{1}\approx 11.82 \approx  3.19\lambda.\label{E:n2d1}
\end{align}
Clearly, the plates have settled at equilibrium configurations near integer values of the non-dimensional \textit{schooling number}~\citep{Becker}
\begin{equation}
S_k=\frac{d^{\infty,k}_{1}}{\lambda}.
\label{E:schooling}
\end{equation}
Note that while the first equilibrium has schooling number $S_{1}=1.13$ slightly larger than 1, the following schooling numbers $S_2=2.16$ and $S_3=3.19$ increase approximately by integer values, $S_k\approx S_{k-1}+1$.

%%%%% sec034_lossofstab.tex %%%%%

\subsection{Equilibrium schooling states and their loss of stability}\label{ssec:loss} %: dependence on $n$ and $A$}
\begin{figure}
 \centering

\includegraphics[trim= 0 20 0  0, clip, width=0.335\textwidth]{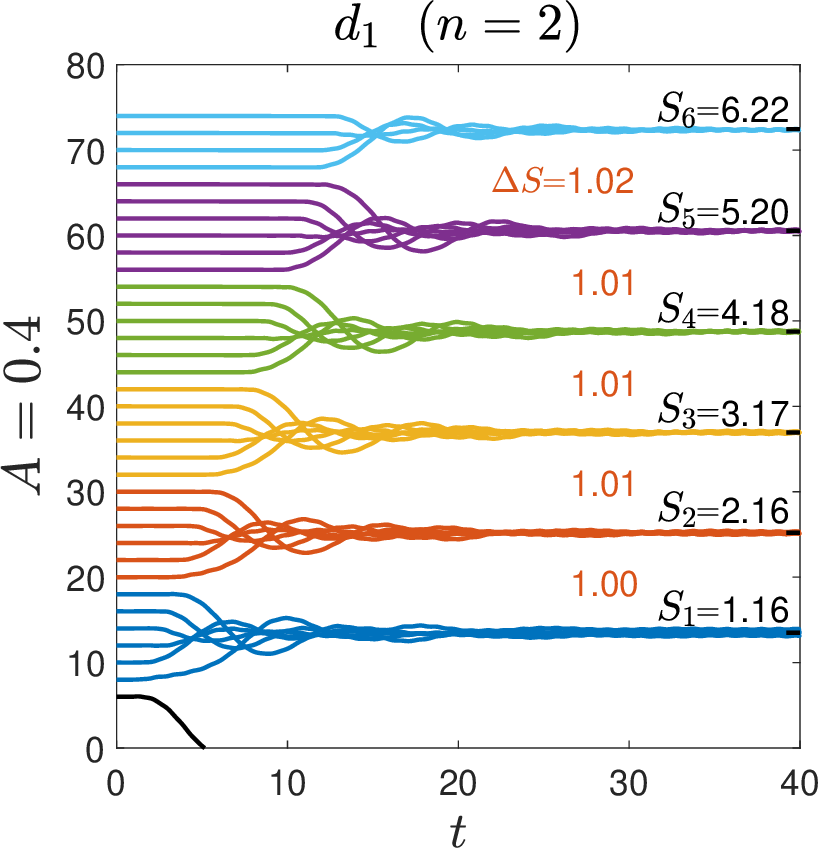}
\includegraphics[trim= 30 20 0  0, clip, width=0.31\textwidth]{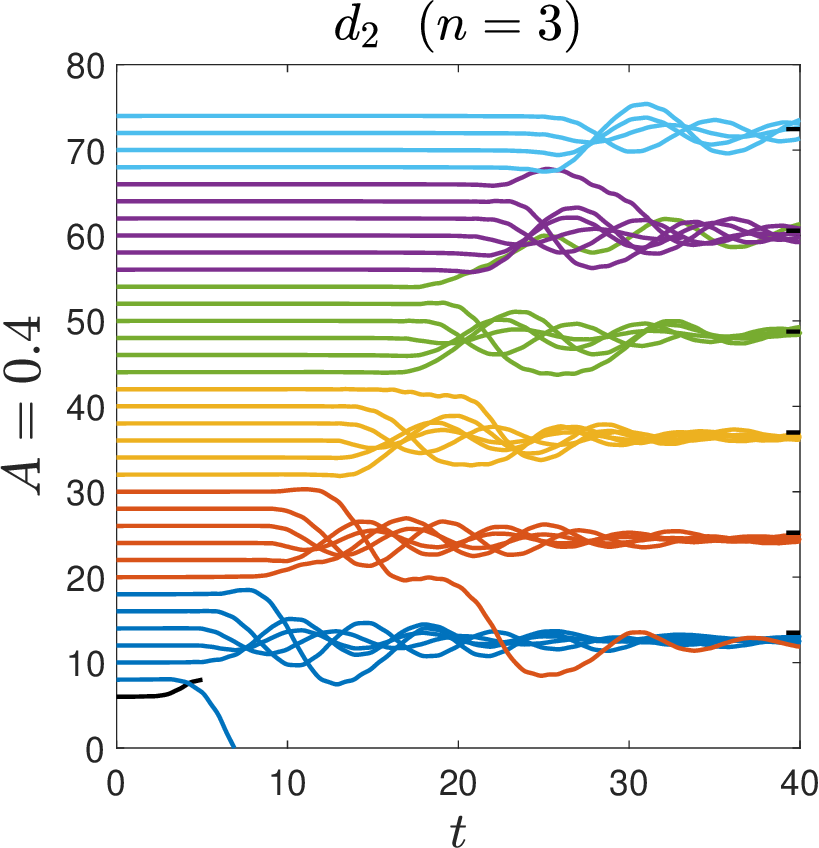}
\includegraphics[trim= 30 20 0  0, clip, width=0.31\textwidth]{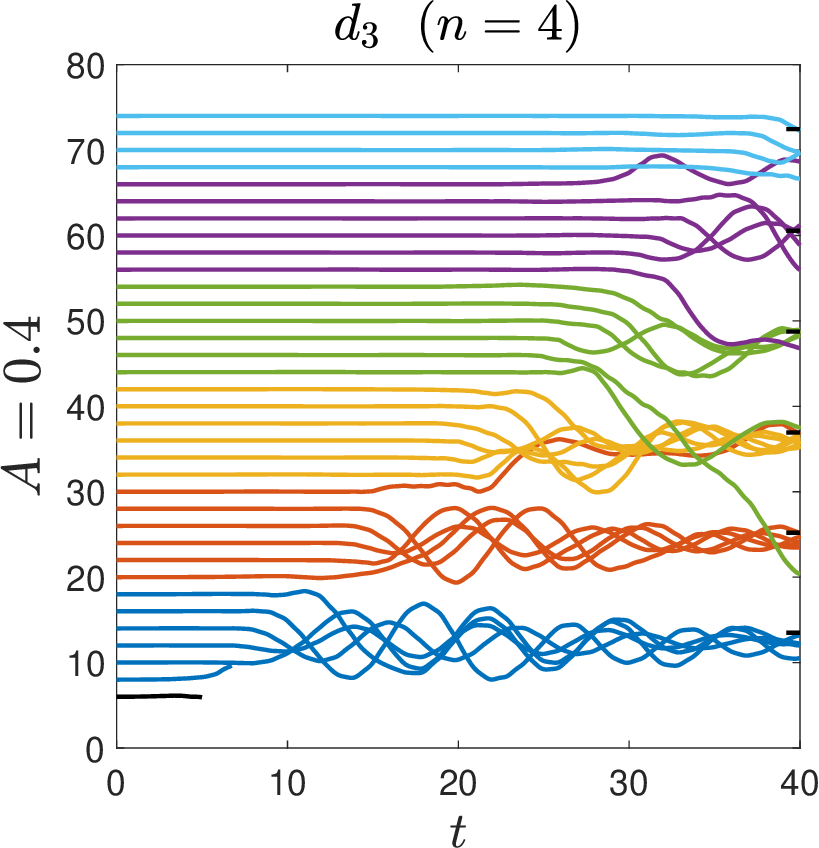}\\
\includegraphics[trim= 0 20 0 25, clip, width=0.335\textwidth]{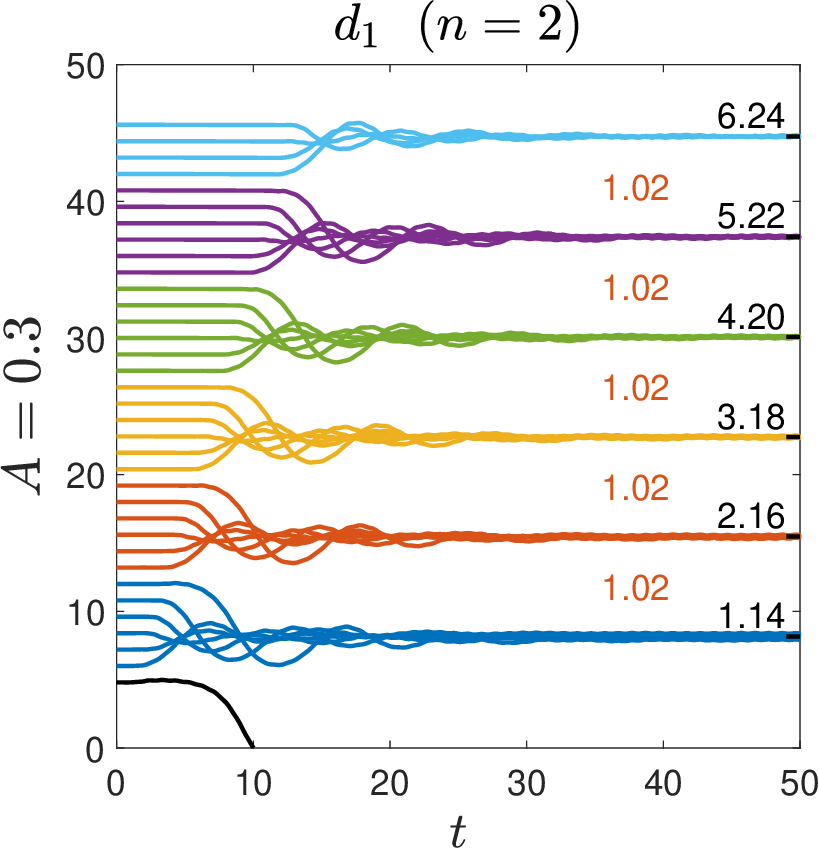}
\includegraphics[trim=30 20 0 25, clip, width=0.31\textwidth]{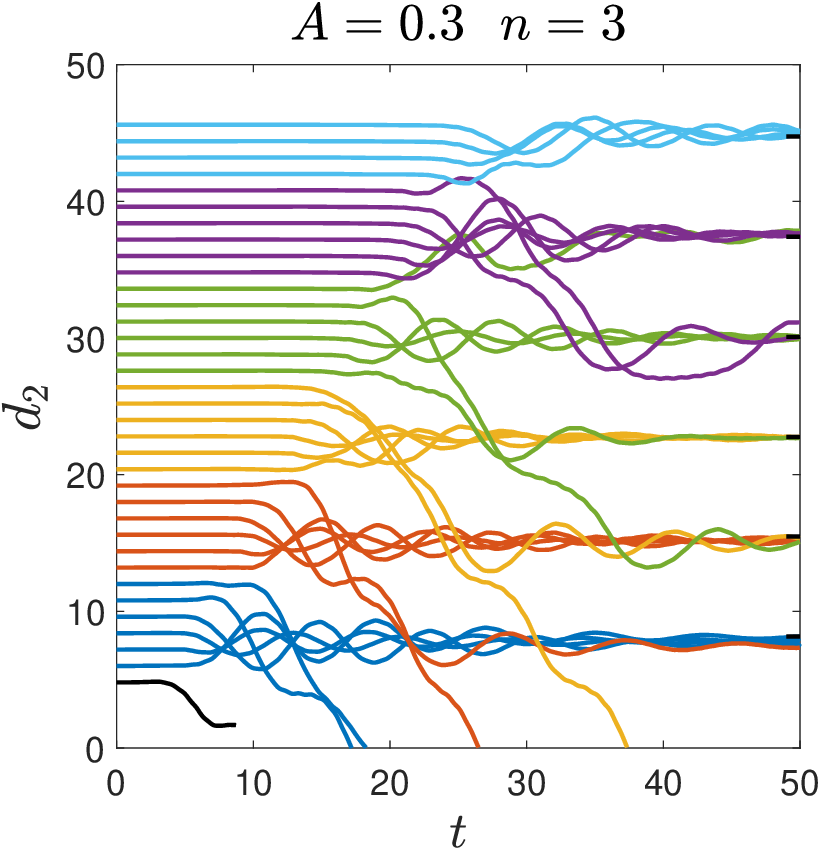}
\includegraphics[trim=30 20 0 25, clip, width=0.31\textwidth]{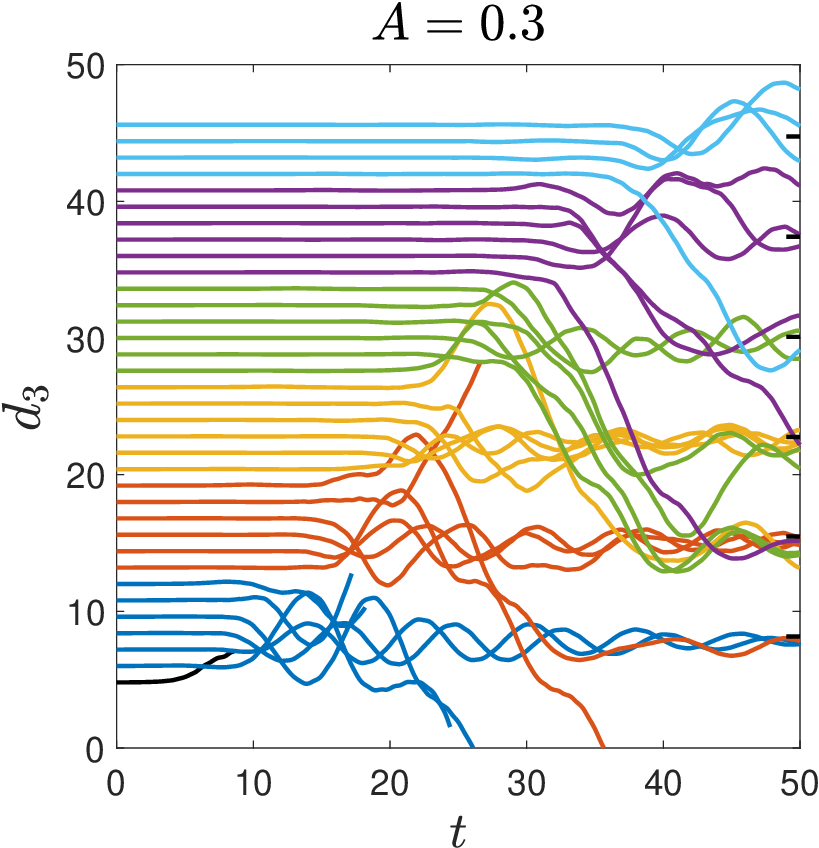}\\
\includegraphics[trim= 0  0 0 25, clip, width=0.335\textwidth]{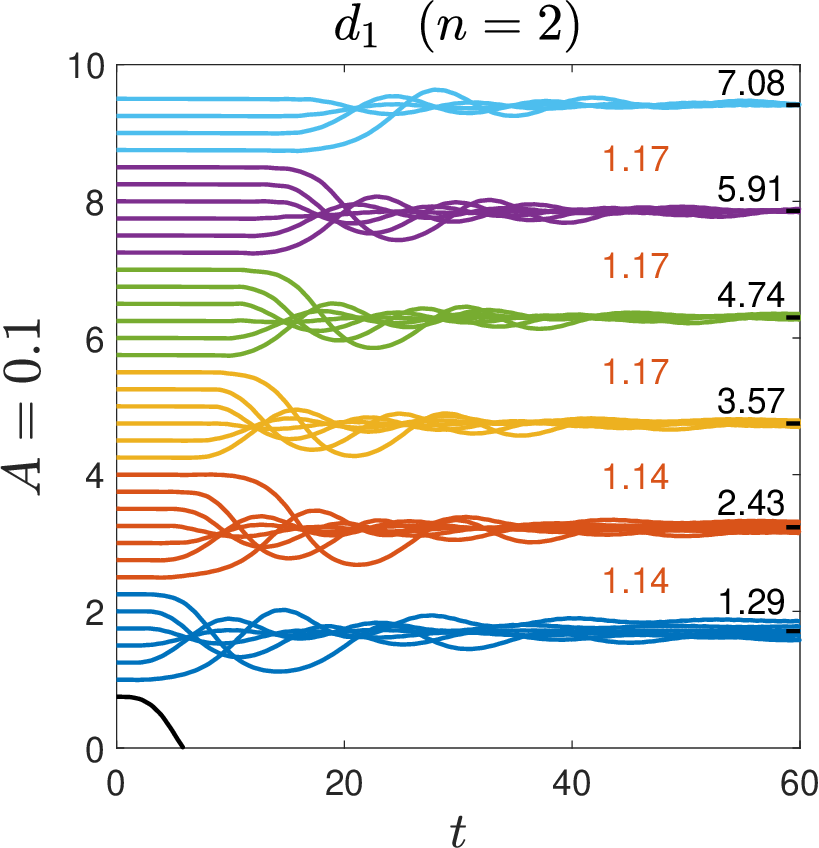}
\includegraphics[trim=30  0 0 25, clip, width=0.31\textwidth]{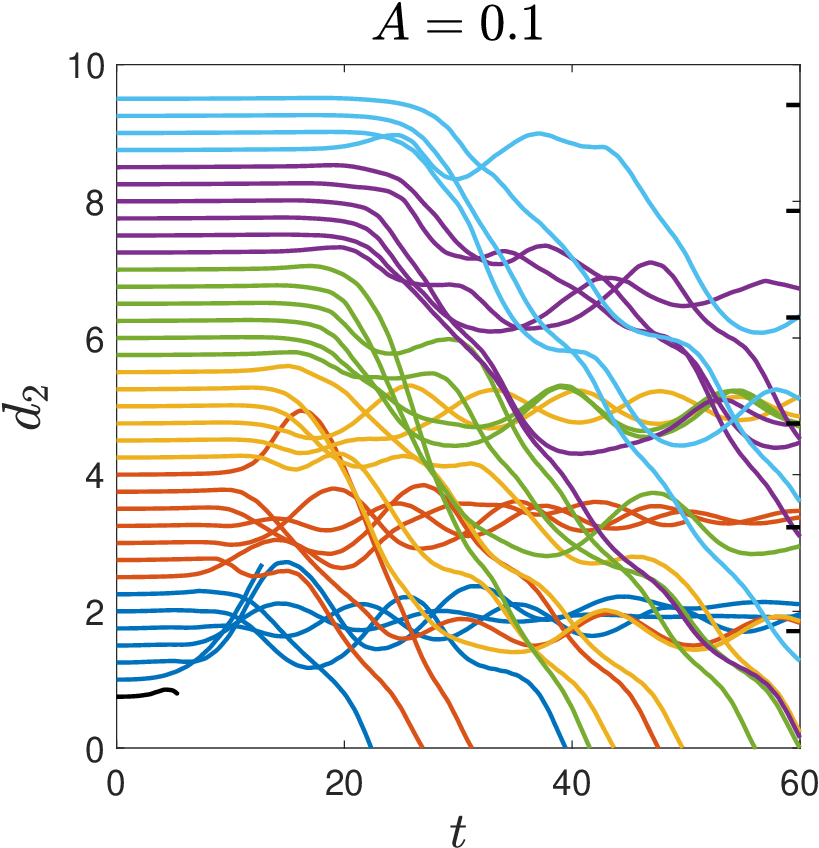}
\includegraphics[trim=30  0 0 24, clip, width=0.31\textwidth]{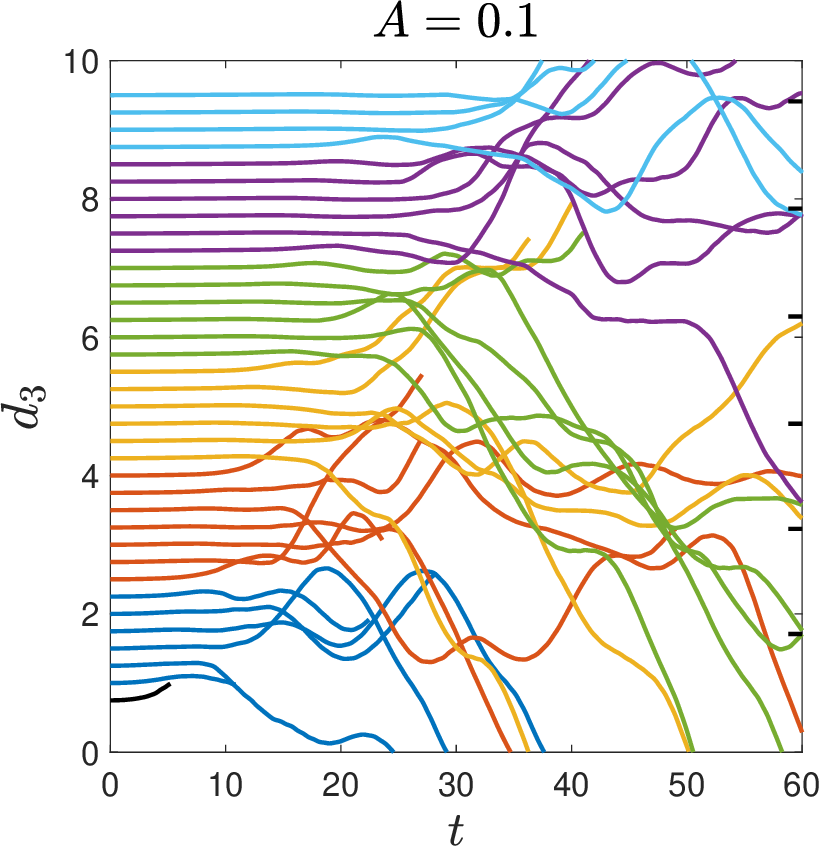}

\caption{Time evolution of the distances $d_1(t)$ (left column), $d_2(t)$ (middle column) and $d_3(t)$ (right column) for in-line formations of $n=2$, 3 and 4 plates, respectively. The plots in the top, middle and bottom rows correspond to the heaving amplitudes $A = 0.4$, 0.3 and 0.1, respectively. The different curves in each plot are obtained by varying the initial distances $d_j(0)$, and are color-coded according to the equilibrium schooling mode reached for two plates ($n=2$, left column). The schooling numbers $S_k=d^{\infty,k}_1/\lambda$ corresponding to each equilibrium distance $d^{\infty,k}_1$ are written in black in the first column, with the differences between them written in red.
Here $\lambda = 2U_{\infty}$ is computed using the steady-state $U_{\infty}$ for the pair of plates.}
\label{F:distall}
\end{figure}

Figure \ref{F:distall} shows the dependence of the distances $d_j(t)$ between plates on the initial separation distance $d_0$, for in-line formations of $n=2$, 3 and 4 wings (left to right column) and the three flapping amplitudes $A = 0.4$, 0.3 and 0.1 (top to bottom row). These flapping amplitudes complement the value $A=0.2$ considered in \S~\ref{ssec:steadypositions}. We proceed by discussing each column in turn.

The first column shows results for $n=2$ plates. It shows the distance $d_1(t)$ between the plates as a function of time, for 35 or 36 different initial separation distances $d_0$. During an initial time interval of length approximately $d_0/U_{\infty}$, the vorticity of the leader has not yet reached the follower and both plates travel with the same velocity, each without the influence of the other. During that time interval the distance between them stays constant. The larger the initial distance $d_0$, the longer this initial time interval is. Once the vortex wake of the leader reaches the follower, that vorticity changes the evolution of the follower and the distance between the plates changes. In all cases the distance $d_1(t)$ initially oscillates, and then the amplitude of these oscillations decreases in time as $d_1$ approaches a constant steady state value. Notice that the steady state distances $d_1^{\infty,k}$ decrease in proportion to $\uinf$ as $A$ decreases from 0.4 (top row) to 0.1 (bottom row). That is, as the heaving amplitude is decreased progressively, the plates are closer together in their steady configurations. Note also that the oscillations in $d_1$ are larger, relative to the value of $d_1$, for smaller values of $A$.

As already observed in figure \ref{F:equilibria}, there are several steady states, depending on the value of $d_0$. The steady states are measured by their schooling number (\ref{E:schooling}), using the value for $\uinf$ corresponding to the given value of $A$. The %35 
distinct values of $d_0$ yield six distinct corresponding steady states, $S_1,\dots,S_6$. The schooling numbers associated to each steady state are listed in black in the first column of figure~\ref{F:distall}, and the differences between them are in red. For all three values of $A$, $S_1$ is slightly larger than 1, with larger schooling numbers increasing by integer values as observed in \S~\ref{ssec:steadypositions}.

The curves are color-coded according to the steady state reached. The curve with smallest $d_0$ is black and reaches zero in finite time, which corresponds to collision of the two plates. While generally the leader travels without being influenced much by the follower, the follower may accelerate and collide with the leader if the initial separation distance is sufficiently small.

The second column in figure~\ref{F:distall} illustrates the case of $n=3$ plates, and shows the distance $d_2(t)$ between the second and third plate. As for $n=2$, the addition of an $n$th plate behind a group of $n-1$ plates leaves the motion of the first $n-1$ plates %almost 
mostly unchanged. That is, for $n=3$ the dynamics of $d_1(t)$ is roughly the same as that shown in the first column of figure~\ref{F:distall}, except in those cases of collision between the second and third plate after which the computation stops. The curves are colored by the equilibrium reached in the absence of the third plate, that is, using the same color for a given $d_0$ as in the first column of figure~\ref{F:distall}.

In the case of $n=3$ plates, the second and third plate move with the same velocity until the vortex wake from the first plate reaches the last plate, since, before then, they both are only affected by the (equal) wake from the plate directly ahead. The time interval of constant $d_2$ is therefore approximately $2d_0/U_{\infty}$, twice as long as %longer than 
the time interval of constant $d_1$. However, after this interval, the motion is more irregular than in the case of two plates, as is evident by comparing the first and second columns of figure~\ref{F:distall}. For $A=0.4$ (top row), one trajectory close to $S_2$ (red) jumps to $S_1$. For $A=0.3$ (middle row), there are more trajectories that move to a different, typically lower, equilibrium schooling number. There are also more collisions between the plates. The situation is noticeably worse for $A=0.1$ (bottom row). Only a few curves reach their corresponding equilibrium, with the rest either exhibiting collisions or being on the path to collision. We thus conclude that the equilibria $S_k$ appear to lose stability when $n$ is increased from 2 to 3, and when $A$ is decreased from 0.4 to 0.1. We note that some of the curves showing $d_2$, for example the black curves at the bottom of the panels in the second column, terminate abruptly despite staying strictly positive. This behavior indicates that the first two plates have collided while the third has not.

The case of $n=4$ plates, with $d_3$ plotted in the third column of figure~\ref{F:distall}, confirms this pattern. There are more irregular paths for $A=0.4$ (top row) than there were with $n=3$, with one, in green, possibly approaching collision. The results for $A=0.3$ (middle row) show many curves that end at a finite time, corresponding to collisions in $d_2$ that are shown in the middle column. In addition, there are more collisions and many more curves that leave their original equilibrium. The basins of attraction of these equilibrium schooling modes appear to have shrunk.
The results for $A=0.1$ are the most striking, as most of the curves either exhibit collisions or are on the path to collision and only a few reach the final time of $t=60$ (30 flapping periods). 

\begin{figure}
 \centering
\includegraphics[trim= 0 20 0  0, clip, width=0.335\textwidth]{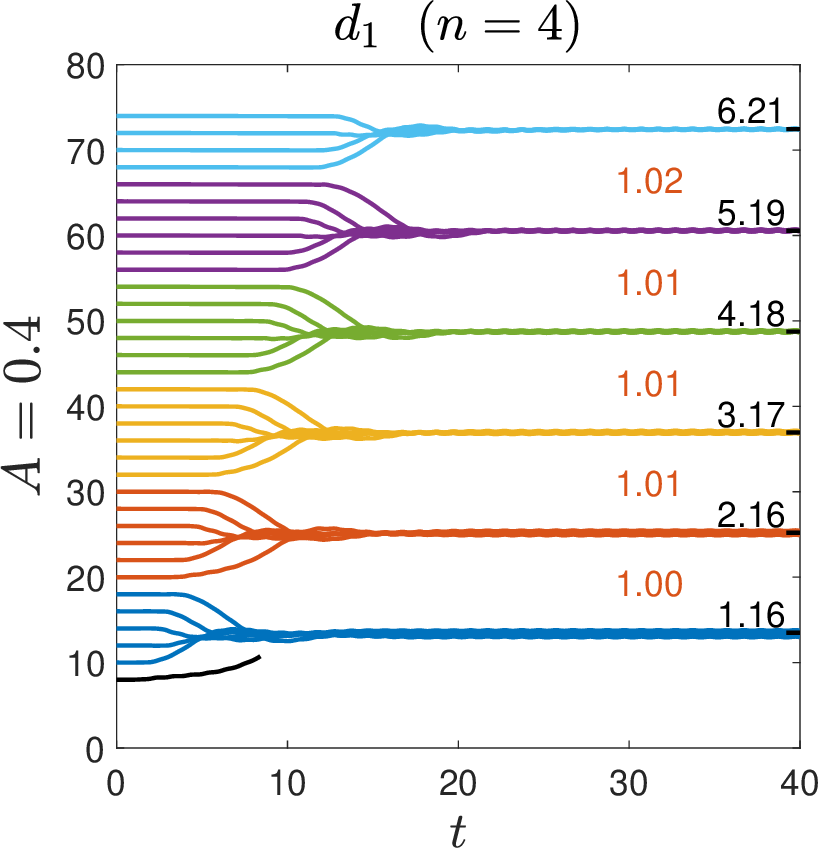}
\includegraphics[trim= 30 20 0  0, clip, width=0.31\textwidth]{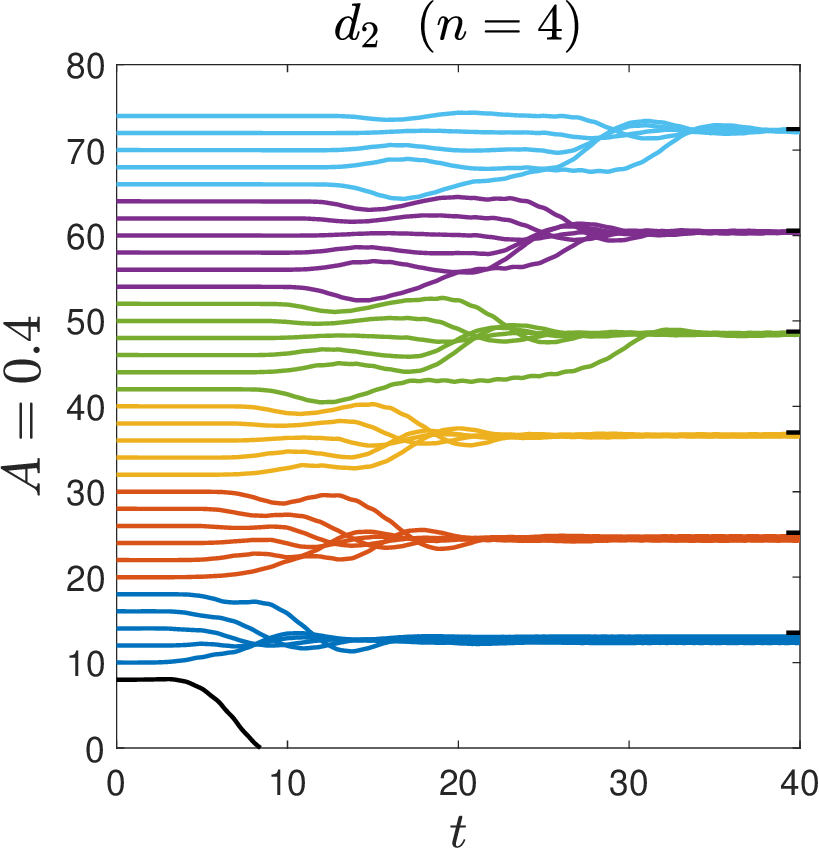}
\includegraphics[trim= 30 20 0  0, clip, width=0.31\textwidth]{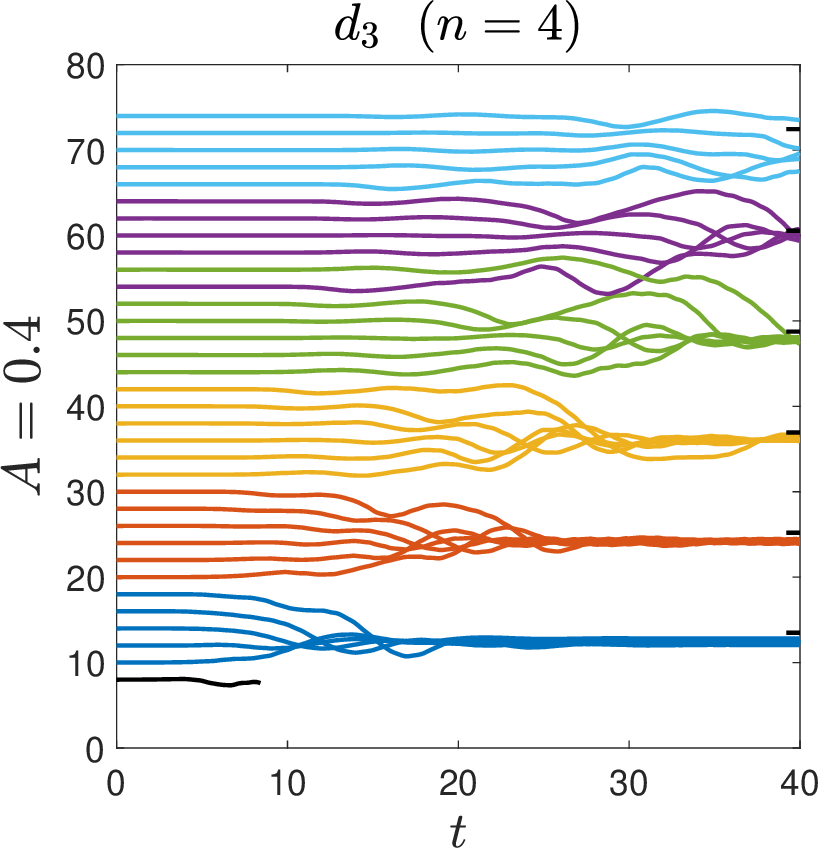}\\
\includegraphics[trim= 0 20 0 25, clip, width=0.335\textwidth]{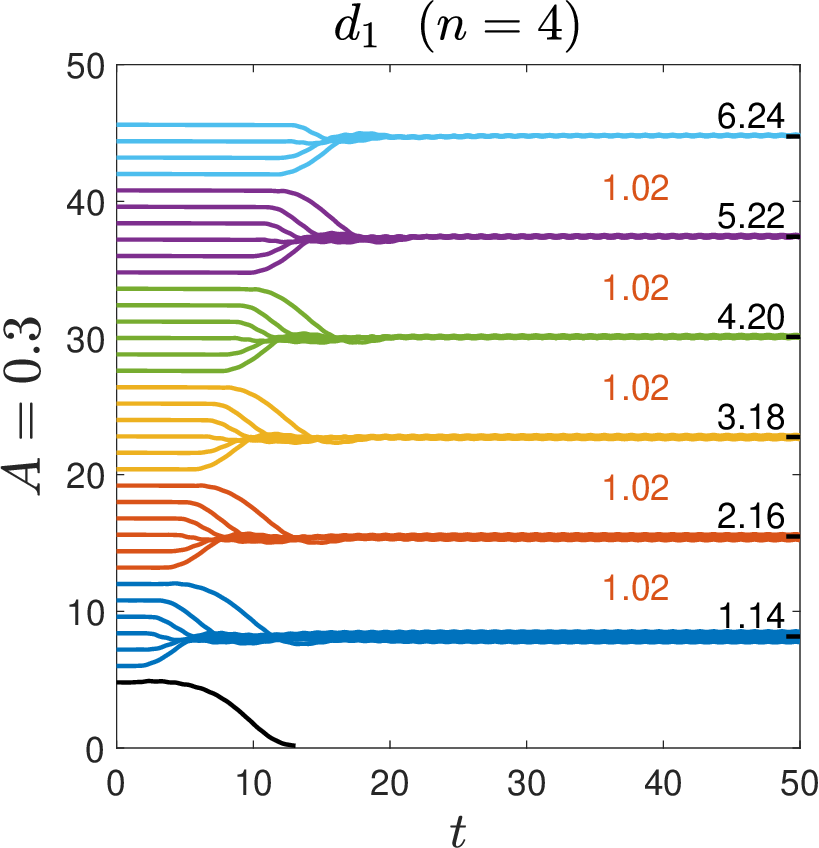}
\includegraphics[trim=30 20 0 25, clip, width=0.31\textwidth]{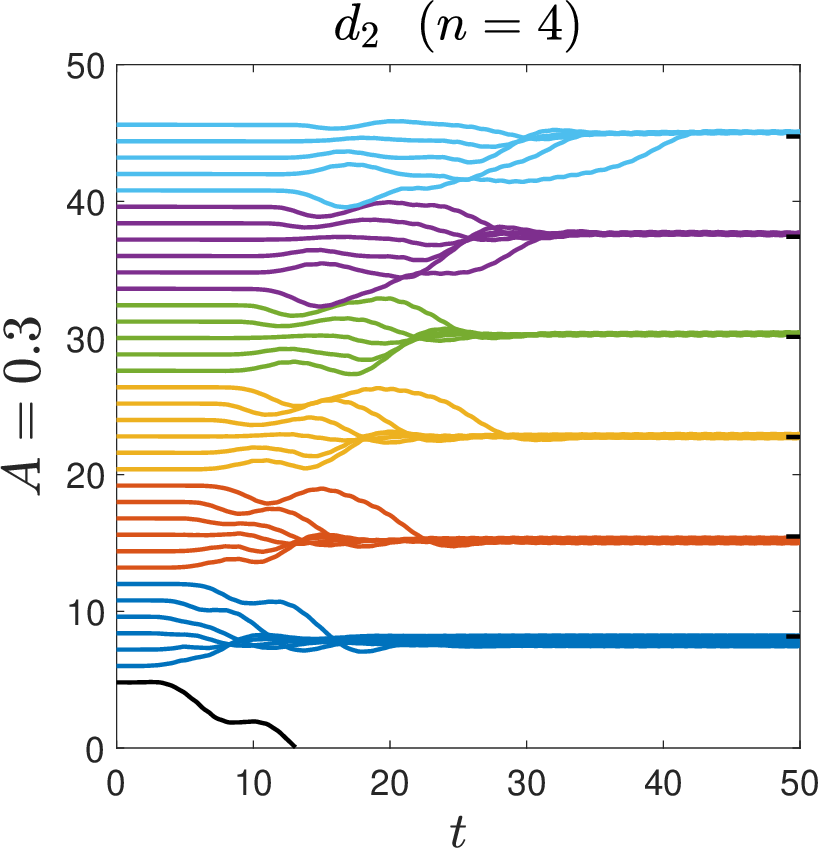}
\includegraphics[trim=30 20 0 25, clip, width=0.31\textwidth]{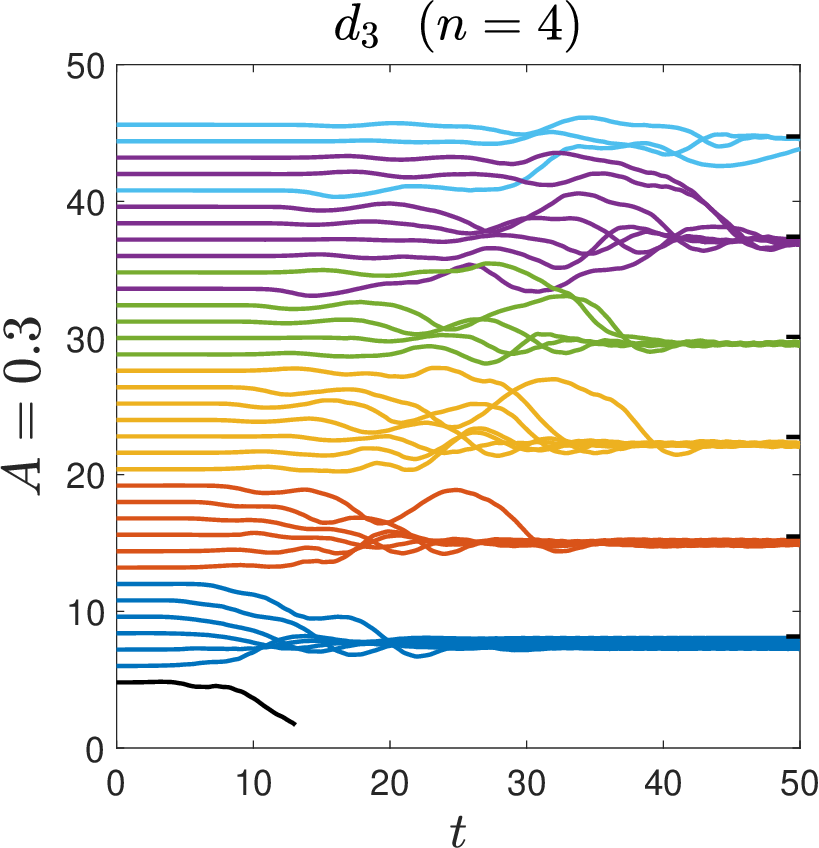}\\
\includegraphics[trim= 0 0 0 25, clip, width=0.336\textwidth]{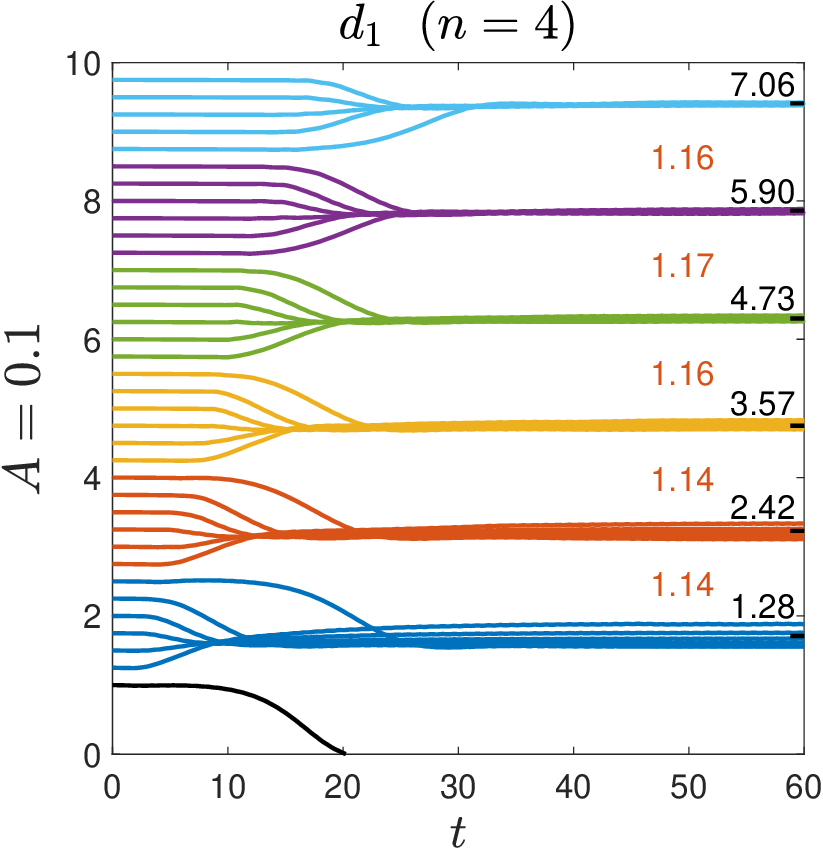}
\includegraphics[trim=30 0 0 25, clip, width=0.31\textwidth]{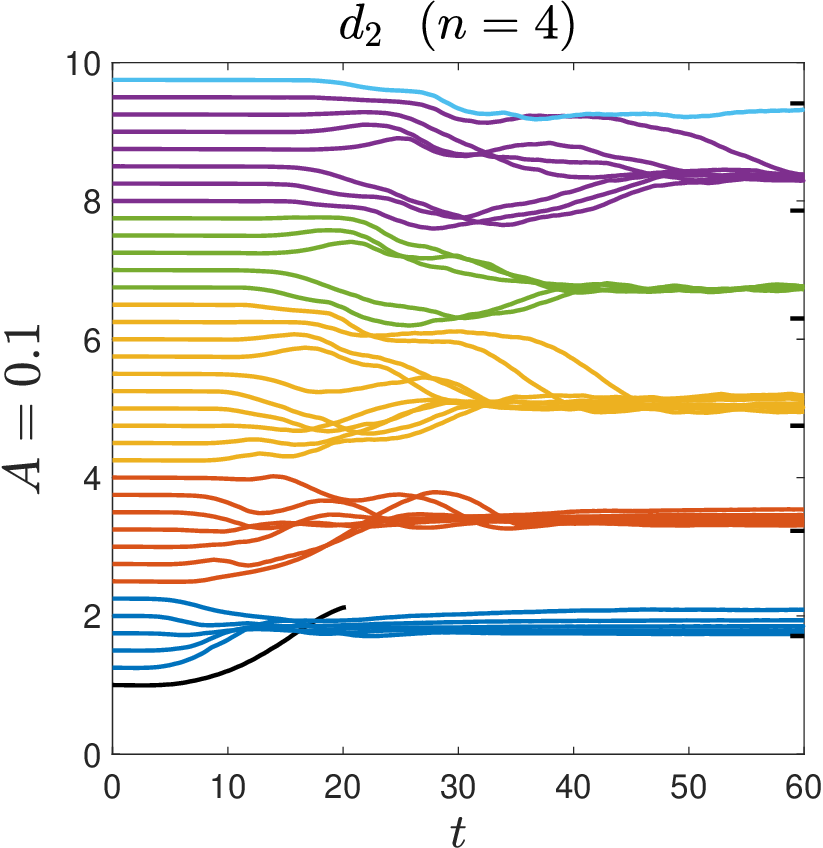}
\includegraphics[trim=30 0 0 25, clip, width=0.31\textwidth]{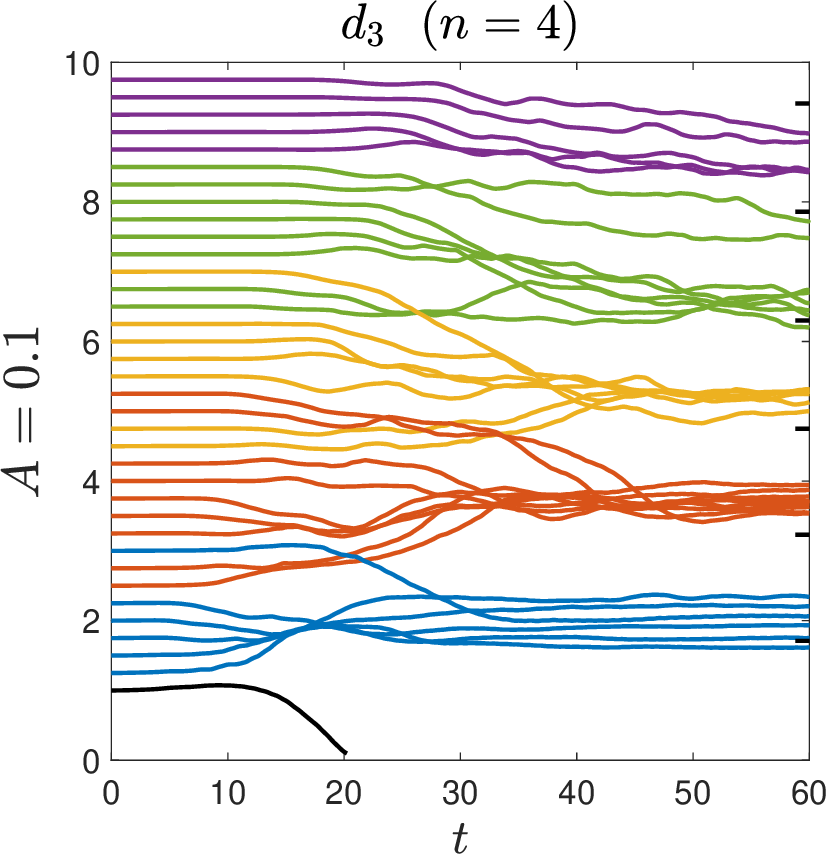}

\caption{Same as figure~\ref{F:distall}, but with the stabilization rule~\eqref{E:simplestab} and $n=4$ throughout. The stabilization factor $\beta=0.15$, 0.4, and 5.0 for the flapping amplitudes $A=0.4$, 0.3 and 0.1, respectively. %For numerical stability we choose $\beta\sim 1/(AU_{\infty})$.} 
Here the curves are color-coded simply by the equilibrium state they reach.
%Distance $d_1,d_2,d_3$ for $n=4$ plates, stabilized, for $A=0.4$ (top row), 0.3 (middle row), 0.1 (bottom row)..
}
\label{F:distall+stabilization}
\end{figure}

Taken together, the simulation results shown in figure~\ref{F:distall} demonstrate that both as the number $n$ of plates in an in-line formation increases, and as the heaving amplitude $A$ decreases, the basins of attraction of the equilibria shrink, with most of the stability being lost for $A=0.1$ and $n=4$. The loss of stability can be attributed to the increasing relative size of the oscillations in $d_k$ observed as $n$ increases and $A$ decreases.

%%%%% sec035_stabilization.tex %%%%%

\subsection{Stabilization of schooling modes}\label{ssec:stabilization}
To prevent the loss of stability and  collisions observed in figure 
\ref{F:distall},
we propose a simple stabilization algorithm:
if the distance decreases (increases) between a given plate and the one directly ahead of it, then slow it down (speed it up) by decreasing (increasing) its flapping amplitude. We implement this mechanism by allowing the heaving amplitude $A$ in (\ref{E:prescribedY}) to depend on each plate,
and evolving the amplitude $A^j$ of plate $j$ according to the rule
\begin{equation}
A^j(t)=A_0[1+\beta d_j'(t)],
\label{E:simplestab}
\end{equation}
where $A_0$ is the initial amplitude
and $d_j'=U^{j-1}-U^j$ is the rate of change of the distance between the plates.
%Equation (\ref{E:simplestab}) in turn is implemented by augmenting the system of dependent variables $U^j$ and $X^j$ by the variable $A^j$, in that the ODE 
%\begin{equation}
%\frac{\rmd A^j}{\rmd t}=A_0\beta \left( \frac{\rmd U^j}{\rmd t}-\frac{\rmd U^{j-1}}{\rmd t}\right)
%\end{equation}
%with initial condition $A^j(0)=A_0(1+\beta d_j'(0)$
%is added to (\ref{E:veloplate}ab). \colr{[Question (not important): why do we need to add an ODE? Can't we just impose $A^j$ at every timestep according to~\eqref{E:simplestab}?]} 
The updated values of $A^j$ are entered in step (i) of the numerical method (\S~\ref{ssec:numerical}), where $V$ is replaced by $V_j(t)=A^j(t)\pi\cos(\pi t)$, and thereby affect all remaining steps 
(ii-v) at each timestep.

The stabilization parameter $\beta$ determines how fast the amplitude changes occur as the distance between plates changes. Small values of $\beta$ may not be sufficient to prevent collisions in finite time. Large values of $\beta$ lead to fast adjustment, but can also lead to stiffness of the numerical scheme. For all runs shown below, we used $0.15\le\beta\le 5$. 

Figure \ref{F:distall+stabilization} shows the distances $d_1$, $d_2$ and $d_3$ between each pair of plates for a formation of $n=4$ wings, computed with the stabilization mechanism. %shows the distance between the two trailing plates for the same flapping amplitudes $A$ \colg{and} number of plates $n$ shown in figure \ref{F:distall}, but with the stabilization mechanism. \colr{[I think the labels at the top of figure~\ref{F:distall+stabilization} should read $n=2$, $n=3$ and $n=4$; otherwise this sentence is wrong.]} 
The figure shows that the stabilization either eliminates or significantly reduces the oscillations in $d_j$, which quickly approach a steady state. Apart from the smallest value of $d_0$ (black curves), all of the collisions observed in figure \ref{F:distall} are avoided, with steady state solutions approached even in the most unstable case, $n=4$ and $A=0.1$. We note that $d_1$ is not shown with 
$n=2$, nor $d_2$ with $n=3$, since, in the absence of collisions, the results are the same as with $n=4$. For the sake of clarity, the curves in figure~\ref{F:distall+stabilization} are color-coded by the equilibrium state they reach, as opposed to the equilibrium state reached by the corresponding wing pair ($n=2$) in figure~\ref{F:distall}. Supplementary movie 4 contains examples of simulations with $n=3$ plates, and shows how the stabilization algorithm prevents a collision between the two trailing plates.

\begin{figure}
\centering
%made in: twoswimmersmove/datout02x/plotclose15v2 with plotonepaperFig8 and then  fourswimmerscontrol/datout02ctl/plotxn2 with plotonepaperFig8: 
%n=2, t=25 stabilization alf=2 d=4.2
\includegraphics[trim=0 0 0 0, clip, width=0.99\textwidth]{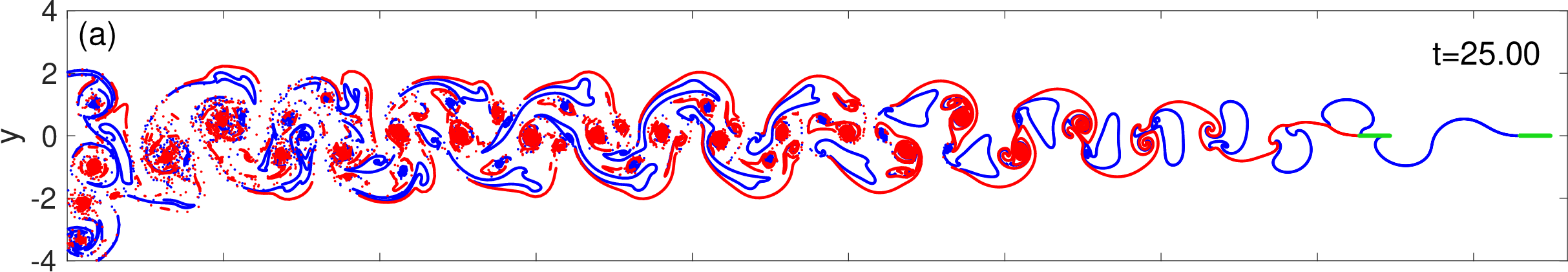}
\includegraphics[trim=0 0 0 0, clip, width=0.99\textwidth]{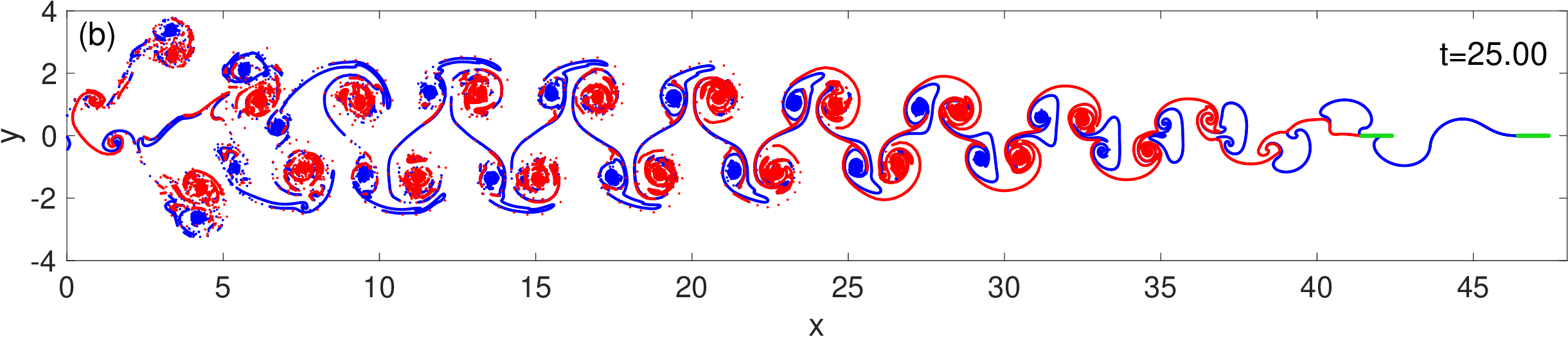}
\caption{Wake at $t=25$ behind a pair of plates ($n=2$) with heaving amplitude $A_0=0.2$ and initial separation distance $d_0 = 4.2$. The plots are (a) without stabilization, and (b) stabilized according to~\eqref{E:simplestab} with $\beta=2$.}
\label{F:stabilizedsheet}
\end{figure}

The stabilization has a significant impact on the vortex wake behind the plates, as an equilibrium state is reached more quickly. In figure \ref{F:stabilizedsheet} we compare snapshots of the vortex sheet behind two plates without (figure~\ref{F:stabilizedsheet}(a)) and with (figure~\ref{F:stabilizedsheet}(b)) stabilization. Figure~\ref{F:stabilizedsheet}(a) is the same as the one shown earlier in figure~\ref{F:allsheet25}(b), but at a different scale. The formation of a wake of quadrupoles is seen early on (near the plates) in figure~\ref{F:stabilizedsheet}(a), but no clear pattern emerges far from the plates. On the other hand, the results obtained with stabilization (figure~\ref{F:stabilizedsheet}(b)) clearly show the emergence of a stable pattern of quadrupoles behind the plates. A positive (negative) vortex is created behind each plate by the plate's upward (downward) motion. With stabilization, these four vortices per oscillation period are positioned in a regular stable configuration, which grows slightly in time and space as the plates move forward.
%
%figures made by 
%threeswimmersmove/datoutamp/plotcl15v2.m

\begin{figure}
\centering
\includegraphics[trim=100 115 98 105 , clip, width=0.99\textwidth]{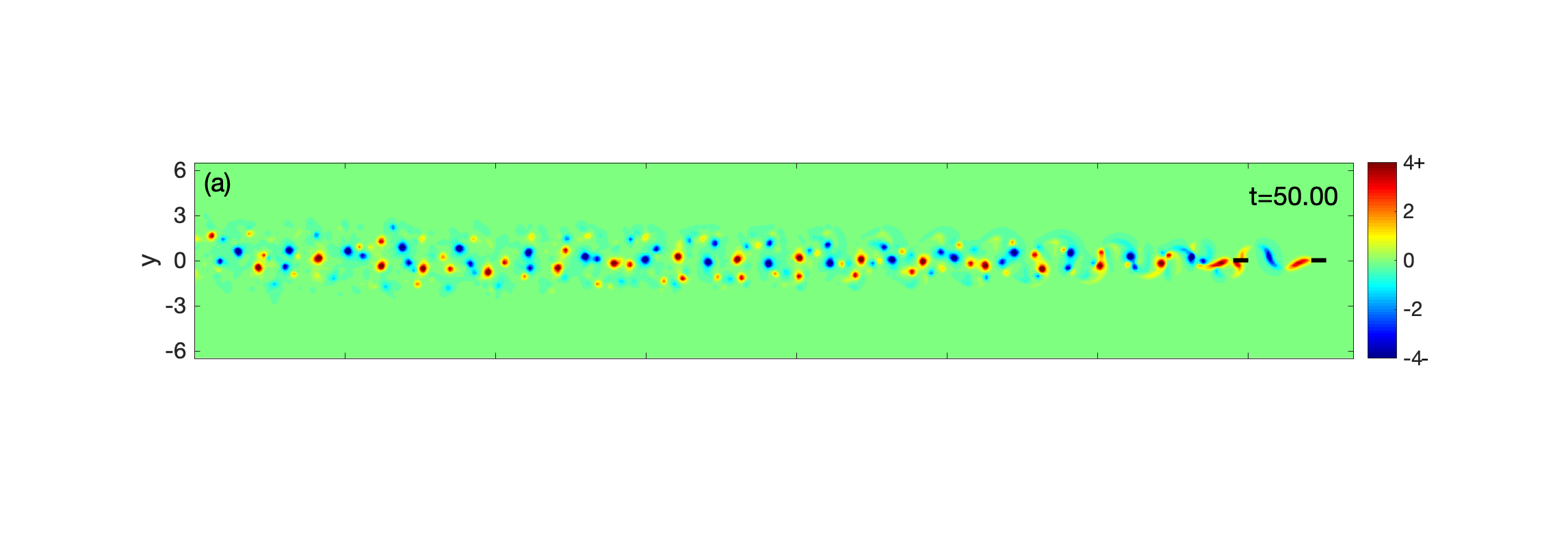}
\includegraphics[trim=100 83 98 110, clip, width=0.99\textwidth]{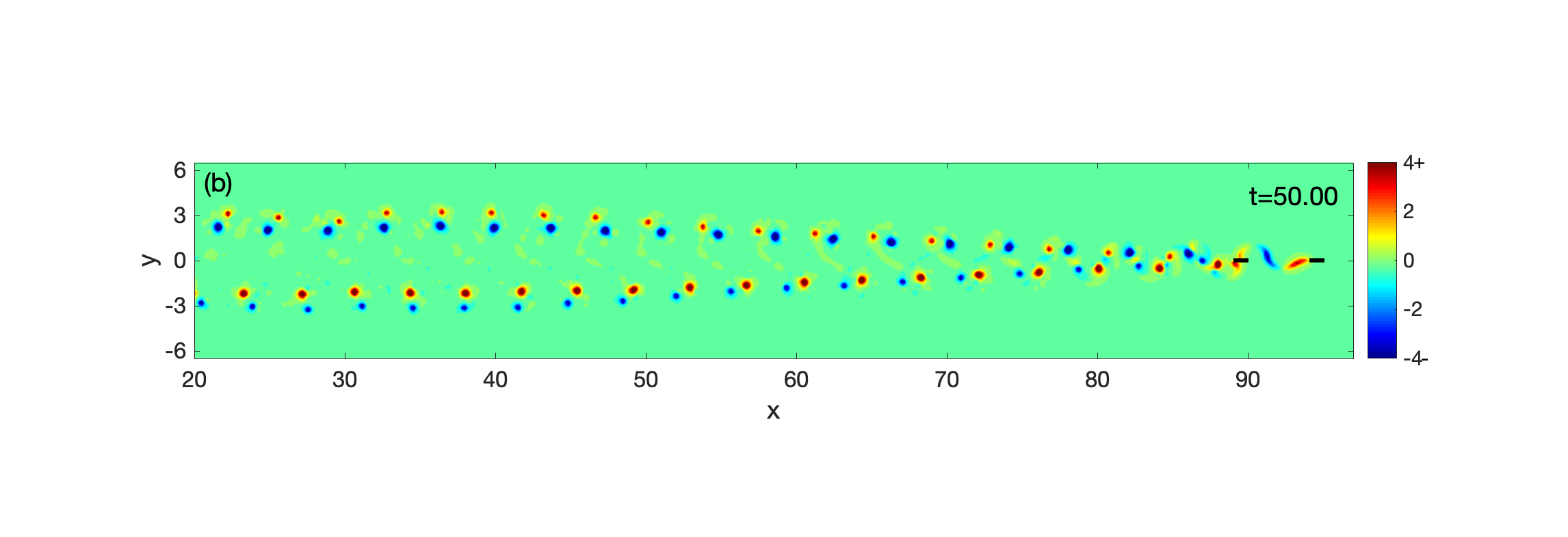}
\caption{Vorticity at $t=50$, with the same parameters as in figure~\ref{F:stabilizedsheet}.
} 
\label{F:stabilizedvorticity}
\end{figure}

\begin{figure}
\centering
\includegraphics[trim=100 115 98 105, clip, width=0.99\textwidth]{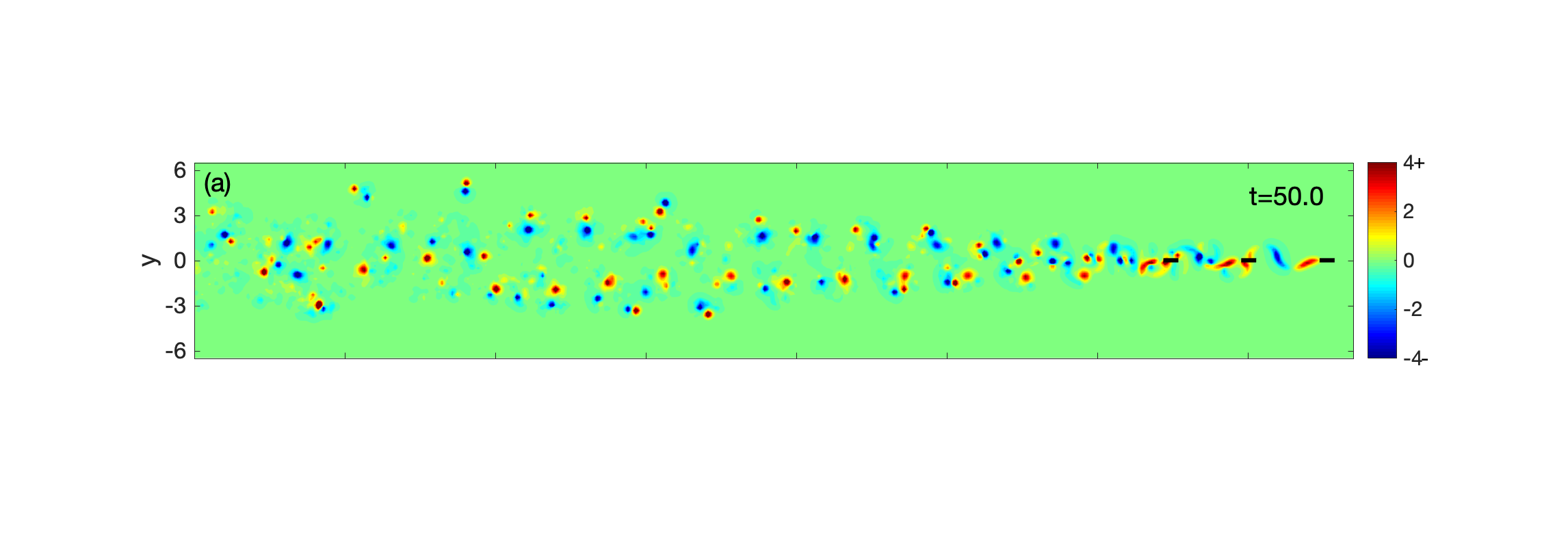}
\includegraphics[trim=100 83 98 110, clip, width=0.99\textwidth]{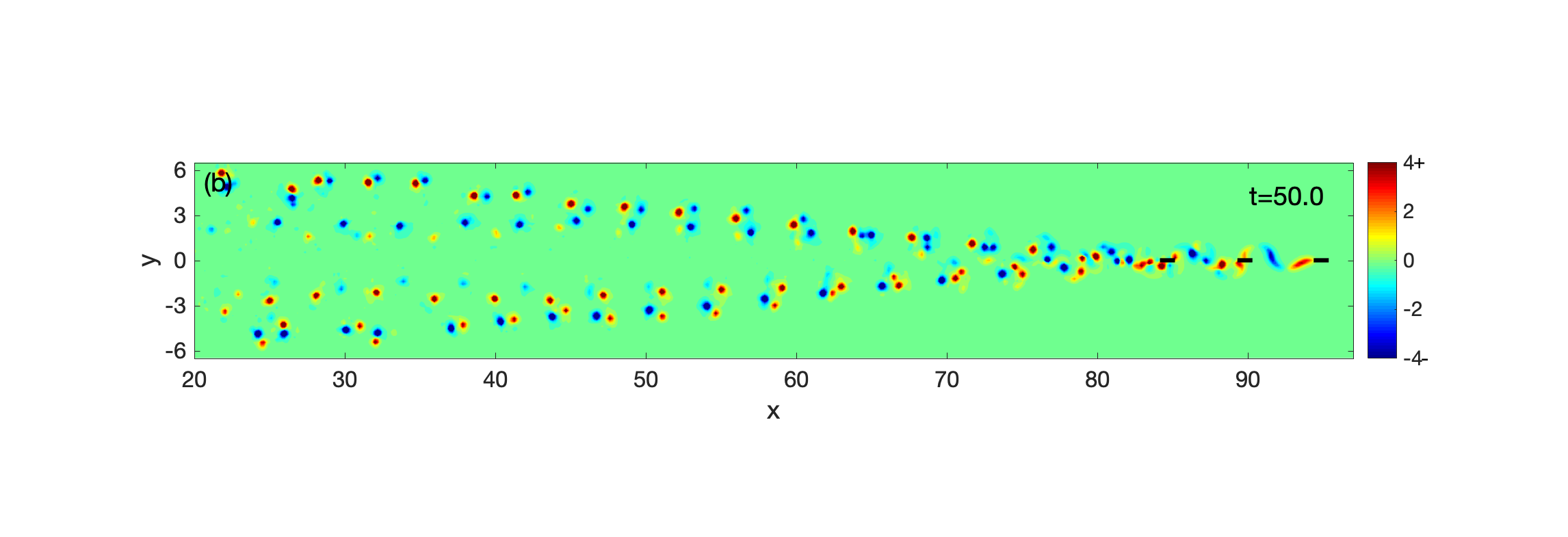}
\caption{Vorticity at $t=50$ behind three  plates ($n=3$) with heaving amplitude $A_0=0.2$ and initial separation distance $d_0 = 4.2$. The plots are (a) without stabilization, and (b) stabilized according to~\eqref{E:simplestab} with $\beta=2$.}
\label{F:stabilizedvorticityn3}
\end{figure}

Figure \ref{F:stabilizedvorticity}  shows the vorticity in the wake, corresponding to the regularized vortex sheet shown in figure~\ref{F:stabilizedsheet}. Figure~\ref{F:stabilizedvorticity}(b) clearly shows the stable pattern in the wake, growing in time and space. The contrast with figure~\ref{F:stabilizedvorticity}(a), which shows much less organized wake vorticity without any growth, is striking. 

Figure \ref{F:stabilizedvorticityn3} shows the vorticity in the wake of $n=3$ plates, with and without stabilization. Here, each heaving cycle of the three plates generates six vortices per period. Without stabilization (figure~\ref{F:stabilizedvorticityn3}(a)) the vortices exhibit a coherent pattern only at short times after separation, but they become less organized further downstream. By contrast, with stabilization (figure~\ref{F:stabilizedvorticityn3}(b)) a regular configuration of six vortices  per period clearly emerges. The width of the wake grows in time and space, as was the case for a pair of plates ($n=2$, figure~\ref{F:stabilizedvorticity}(b)).

We were unable to compute such stable patterns of eight vortices per period in the wake of $n=4$ plates due to the computational cost of doing so. While the stabilization mechanism allows stable schooling modes to be achieved relatively quickly, the vortex configurations take longer to reach equilibrium. %it should be possible to reach equilibrium vortex configurations %using stabilization to avoid collisions and irregularities, for $n=4$ as well. However, we suspect that these configurations are achieved after a longer time, a time for which simulations with $n=4$ plates would be computationally prohibitive.} %so provided the computation can be performed to sufficiently long times. 
Long time simulations using fast tree-algorithms~\citep{Greengard1987,WangKrasnyTlupova2020} can be performed in the future to study this phenomenon further. 

\subsection{Reduced model based on thin-airfoil theory}\label{ssec:reduced}

We now propose and analyze a reduced model to provide rationale for two of the phenomena observed in our simulations (\S~\ref{ssec:loss}), namely, that collectives of heaving plates in an in-line formation are increasingly stabilized as the flapping amplitude is increased progressively, and such collectives become less stable as the number of plates increases. To that end, we deploy the seminal thin-airfoil theory of \citet{Wu1961Swimmingwavingplate}, who considered the motion of a single flat plate executing a small amplitude heaving motion ($A \ll 1$) while translating horizontally at a velocity $U$ in an inviscid and incompressible 2D fluid. To simplify the problem, the plate is assumed to shed a flat vortex sheet from its trailing edge, and the sheet strength is determined by the Kutta condition. By solving the linearized Euler equations using conformal mapping, the (dimensionless) thrust $T_0$ on the plate is found to be
\begin{align}
T_0 = \frac{\pi}{2}(\pi A)^2|C(\sigma)|^2,\quad\mathrm{where}\quad C(\sigma)=\frac{\mathrm{K}_1(\rmi\sigma)}{\mathrm{K}_0(\rmi\sigma)+\mathrm{K}_1(\rmi\sigma)}
\end{align}
is the so-called Theodorsen function, $\mathrm{K}_{0,1}$ are the modified Bessel functions of the second kind of orders zero and one, and $\sigma = \pi/(2U)$ is the reduced frequency. An analogous theory has been developed to describe how a follower plate located downstream of the leader and executing the same heaving kinematics would be affected by the leader's wake~\citep{Wu1975Extractionflowenergy,Sophie}. In this theory, the no-penetration boundary condition is not satisfied simultaneously on both wings; instead, an approximation for the hydrodynamic thrust on the follower is found by assuming that the leader's wake is unaffected by the follower's presence. This assumption is consistent with our numerical simulations before the leader's wake has reached the follower, as discussed in \S~\ref{ssec:several}. This thrust is found to be $T = T_0\left[1+F(d)\right]$, where $d$ is the distance between the leader's trailing edge and follower's leading edge (as in figure~\ref{F:sketch}) and
\begin{align}
F(d) &= G(\sigma)^2+2G(\sigma)\cos\left[2\pi\left(\frac{d}{2U}-g(\sigma)+\frac{\sigma}{\pi}\right)\right].
\end{align}
Here, the real-valued functions $G(\sigma)$ and $g(\sigma)$ are defined through the relation
\begin{align}
G(\sigma)\mathrm{e}^{2\pi\rmi g(\sigma)}&= \frac{\mathrm{K}_1(-\rmi\sigma)\mathrm{K}_0(\rmi\sigma)+\mathrm{K}_0(-\rmi\sigma)\mathrm{K}_1(\rmi\sigma)}{\mathrm{K}_1(\rmi\sigma)\left[\mathrm{K}_0(\rmi\sigma)+\mathrm{K}_1(\rmi\sigma)\right]}.
 \end{align}

We thus propose the following model for a collective of $n$ plates that execute small-amplitude heaving kinematics in an in-line formation:
\begin{align}
M\ddot{x}_1+C_{\mathrm{d}}\dot{x}_1^{3/2}&=T_0,\nonumber \\
M\ddot{x}_m+C_{\mathrm{d}}\dot{x}_m^{3/2}&=T_0\left[1+F(x_{m-1}-x_{m}-1)\right], \quad 2\leq m\leq n,\label{WuEqnMot}
\end{align}
where $x_m$ is the horizontal position of the center of the $m$th plate. We have assumed that each plate is influenced only by its nearest upstream neighbor; that is, the leader wing ($x_1$) is unaffected by all of the followers, and moves at a steady speed $U$ determined by the balance between drag and thrust forces on it:
\begin{align}
C_{\mathrm{d}}U^{3/2}&=\frac{\pi }{2}\left(\pi A\right)^2|C(\sigma)|^2.\label{UBase}
\end{align}
Every other wing is influenced by its upstream neighbor through the function $F$, which depends on the {\it steady} speed $U$. That is, our theory is valid in the quasi-steady regime in which the wings' velocities are all close to $U$. The nearest-neighbor assumption may be justified using our simulations that employ stabilization (figure~\ref{F:distall+stabilization}); specifically, $d_1^{\infty,k}\approx d_2^{\infty,k}\approx d_3^{\infty,k}$, which suggests that the equilibrium distances are mostly insensitive to the number of upstream wings. The assumption is also supported by recent experiments on up to five flapping hydrofoils in an in-line formation~\citep{Newbolt_2024}. %We note that a nearest-neighbor model for flapping wing interactions was recently proposed by~\citet{Newbolt_2024}. 
Table~\ref{SteadySpeeds} shows a comparison between the steady speeds $U$ obtained in our vortex sheet simulations and those predicted from Eq.~\eqref{UBase}, with surprisingly good agreement given the simplified nature of our reduced model. 
\begin{table}
\begin{center}
\begin{tabular}{c||c|c}
%\hline
$A$ & simulations & thin-airfoil theory \\  \hline \hline
0.1 & 0.665 & 0.544 \\
0.2 & 1.85 & 1.506 \\
0.3 & 3.584 & 3.006  \\
0.4 & 5.826 & 5.147 \\
\hline
\end{tabular}
\caption{The values of the velocity $U$ of a single wing, as obtained in our vortex sheet simulations (\S~\ref{ssec:single}) and predicted by thin-airfoil theory from Eq.~\eqref{UBase}, are shown for different values of the flapping amplitude $A$. }\label{SteadySpeeds}
\end{center}
\end{table}

We proceed by seeking steadily translating solutions of Eq.~\eqref{WuEqnMot}, in which the wings move at a constant speed $U$ with a constant separation distance $d$ between them, $x_{m}(t)-x_{m+1}(t)=d+1$. By substituting the form $x_m(t)=Ut-(m-1)(d+1)$ into Eq.~\eqref{WuEqnMot}, we find that $d$ satisfies the algebraic equation
\begin{align}
F(d)=0.\label{FEquil}
\end{align}
This equation has infinitely many positive solutions $d^{\infty,k}$ for $k\in\mathbb{Z}_{> 0}$. For the parameter values adopted in this paper, $M = 1$, $C_{\mathrm{d}}=0.1$, and $A = 0.1-0.4$, we find that the corresponding schooling numbers $S_k=d^{\infty,k}/(2U)$ are equal to $S_1+k-1$ where $S_1$ is between 0.95 and 1.1, in good agreement with those obtained in our vortex sheet simulations (figure~\ref{F:distall}). We note that a similar result was obtained for two plates by~\citet{Sophie} using a similar thin-airfoil theory, and by~\citet{Baddoo_2023} using a more sophisticated thin-airfoil theory based on conformal maps of multiply-connected domains. Reduced models of flapping wings that approximate the hydrodynamic interactions using time-delay differential equations also recover near-integer values of the equilibrium schooling numbers~\citep{Heydari_2024,Newbolt_2024}. We also note that Eq.~\eqref{FEquil} is satisfied by a second family of solutions with schooling numbers $\tilde{S}_k=\tilde{S}_1+k-1$, where $\tilde{S}_1$ is between 0.6 and 0.7 for the parameter values adopted in this paper. We neglect these solutions in the forthcoming discussion because they are linearly unstable to perturbations and thus not observed in our simulations nor in laboratory experiments~\citep{Sophie}.

We now examine the linear stability of the state in which the distance between each pair of wings corresponds to the schooling number $S_1$, the analysis of the other schooling numbers $S_k$ being identical due to the periodicity of the function $F$. To that end, we substitute the form $x_m(t)=Ut-(m-1)\left(d^{\infty,1}+1\right)+\epsilon\tilde{x}_m(t)$ into Eq.~\eqref{WuEqnMot} and retain terms at leading order in $\epsilon$. The linearized equations are thus (dropping the tildes)
\begin{align}
M\ddot{x}_1+\frac{3C_{\mathrm{d}}}{2}U^{1/2}\dot{x}_1&=0,\nonumber \\
M\ddot{x}_m+\frac{3C_{\mathrm{d}}}{2}U^{1/2}\dot{x}_m&=T_0\xi\left(x_{m-1}-x_{m}\right),\quad 2\leq m\leq n,\label{WuEqnLin}
\end{align}
where $\xi = F^{\prime}\left(d^{\infty,1}\right)$. We assume an impulsive perturbation is applied to the leader:
\begin{align}
x_1(0)=1,\quad \dot{x}_1(0)=s_1\equiv-\frac{3C_{\mathrm{d}}U^{1/2}}{2M}\quad\mathrm{and} \quad x_m(0)=\dot{x}_m(0)=0\quad\text{for}\quad m\geq 2.\label{e:s1}
\end{align}
It is straightforward to obtain the solutions
\begin{align}
x_1(t)&=\rme^{s_1t}\quad\mathrm{and}\quad  x_2(t)=\rme^{s_1t} + \mathrm{Re}\left[B_0^{(2)}\rme^{s_2t}\right],\quad  \mathrm{where}\quad B_0^{(2)}=-1+\frac{\rmi s_1}{2\omega},\nonumber \\
\omega &= \frac{|s_1|}{2}\sqrt{\frac{16MU^{1/2}\xi}{9C_{\mathrm{d}}}-1}\quad\mathrm{and}\quad  s_2=\frac{s_1}{2}+\rmi\omega
\end{align} 
is a root of the quadratic polynomial $p(s)\equiv Ms^2+(3C_{\mathrm{d}}/2)U^{1/2}s+T_0\xi$. We find that $\omega\in\mathbb{R}$ for the parameter values adopted in this paper. Each wing after the second oscillates in resonance with its upstream neighbor, so the general solution to Eq.~\eqref{WuEqnLin} has the form
\begin{align}
x_m(t)=\rme^{s_1t}+\mathrm{Re}\left[\sum_{k=0}^{m-2}B_k^{(m)}t^k\rme^{s_2t}\right].\label{LinSoln}
\end{align}
The coefficients $B_0^{(m)}$ may be obtained using the initial conditions:
\begin{align}
B_0^{(m)}=-1+\frac{\rmi}{\omega}\left(\frac{s_1}{2}+\mathrm{Re}\left[B_1^{(m)}\right]\right),\quad m\geq 3.\label{B0n}
\end{align}
To obtain the remaining coefficients, we write~\eqref{WuEqnLin} in the compact form $\mathcal{L}x_m=T_0\xi x_{m-1}$ for $m\geq 2$, where $\mathcal{L}=p(\rmd/\rmd t)$. Using the facts that
\begin{align}
\mathcal{L}\rme^{s_1t}&=T_0\xi\rme^{s_1t},\quad \mathcal{L}\rme^{s_2t}=0,\quad \mathcal{L}t\rme^{s_2t}=2\rmi\omega M\rme^{s_2t}\nonumber \\
\mathrm{and}\quad \mathcal{L}t^k\rme^{s_2t}&=\left[2\rmi\omega Mkt^{k-1}+Mk(k-1)t^{k-2}\right]\rme^{s_2t}\quad\mathrm{for}\quad k\geq 2,
\end{align}
it can be shown that the coefficients $B_k^{(m)}$ satisfy the recursion relation
\begin{subequations}
\begin{align}
&B_{m-2}^{(m)}=-\frac{\rmi T_0\xi B_{m-3}^{(m-1)}}{2\omega M(m-2)}\quad\mathrm{for}\quad m\geq 4,\label{Recursion1} \\
&2\rmi\omega MB_{k+1}^{(m)}(k+1)+MB_{k+2}^{(m)} (k+1)(k+2)=T_0\xi B_k^{(m-1)}\quad\mathrm{for}\quad k=0,\dots,m-4.\label{Recursion2}
\end{align}
\label{Recursion}
\end{subequations}
Taken together, Eqs.~\eqref{B0n} and~\eqref{Recursion} determine all of the coefficients in the solution~\eqref{LinSoln}. 

We proceed by making two observations about this solution. First, the equilibrium schooling states are linearly stable because the perturbations decay in time, $x_m(t)\rightarrow 0$ as $t\rightarrow\infty$. Moreover, the decay rate $s_1/2$ increases with $U$, as is evident from~\eqref{e:s1}, and thus with the flapping amplitude $A$, providing a rationale for the observation from our vortex sheet simulations (\S~\ref{ssec:loss}) that a larger flapping amplitude stabilizes the schooling state. Second, by plotting the solution~\eqref{LinSoln} for the parameter values adopted in this paper, we observe that the perturbations are amplified downstream, in that at intermediate times they exhibit oscillations that grow in time and have progressively larger amplitude as $m$ increases. To see why this is the case, we observe that the exponential decay factors $\rme^{s_2t}$ in~\eqref{LinSoln} are multiplied by  increasingly large powers of $t$ as $m$ increases. More precisely, note that~\eqref{Recursion1} can be solved explicitly for $B_{m-2}^{(m)}$, from which the long-time asymptotic behavior of $x_m(t)$ can be deduced:
\begin{align}
x_m(t)\sim \text{Re}[X_m(t)]\quad\text{as}\quad t\rightarrow\infty,\quad\text{where}\quad X_m(t)=\frac{B_0^{(2)}}{(m-2)!}\left(\frac{-\rmi T_0\xi}{2\omega M}\right)^{m-2}t^{m-2}\rme^{s_2t}.\label{xnAsymp}
\end{align}
We find that, for the parameter values adopted in this paper, $|X_m(t)|$ provides an adequate approximation for the envelope of the oscillating solution $x_m(t)$ in~\eqref{LinSoln}. Moreover, it is easy to see that $|X_m(t)|\propto t^{m-2}\rme^{s_1t/2}$ attains its maximum value of
\begin{align}   X_m^*\equiv|X_m(t_m)|=\frac{\left|B_0^{(2)}\right|}{(m-2)!}\left(\frac{T_0\xi(m-2)}{\omega M|s_1|\mathrm{e}}\right)^{m-2}\quad\text{at the time}\quad t=t_m\equiv \frac{2(m-2)}{|s_1|}
.
\end{align}
By using Stirling's formula, $m!\sim \sqrt{2\pi m}(m/\mathrm{e})^m$ as $m\rightarrow\infty$, we conclude that the peak of the envelope $X_m^*$ increases with $m$ for sufficiently large $m$ provided that $T_0\xi/(\omega M|s_1|) > 1$, as is the case for the parameter values adopted in this paper. Moreover, the time $t_m$ at which the peak is attained increases with $m$, which further illustrates that the oscillations propagate downstream. %While the linear theory predicts that these oscillations eventually decay in time, nonlinear effects could presumably further amplify the perturbations at intermediate times and thus dislodge the trailing plates from their equilibrium positions. 
The linear theory thus provides a qualitative rationale for the observation from our vortex sheet simulations (\S~\ref{ssec:loss}) that the schooling states become less stable as the number of wings increases.

\begin{figure}
    \centering
    \includegraphics[width=1\columnwidth]{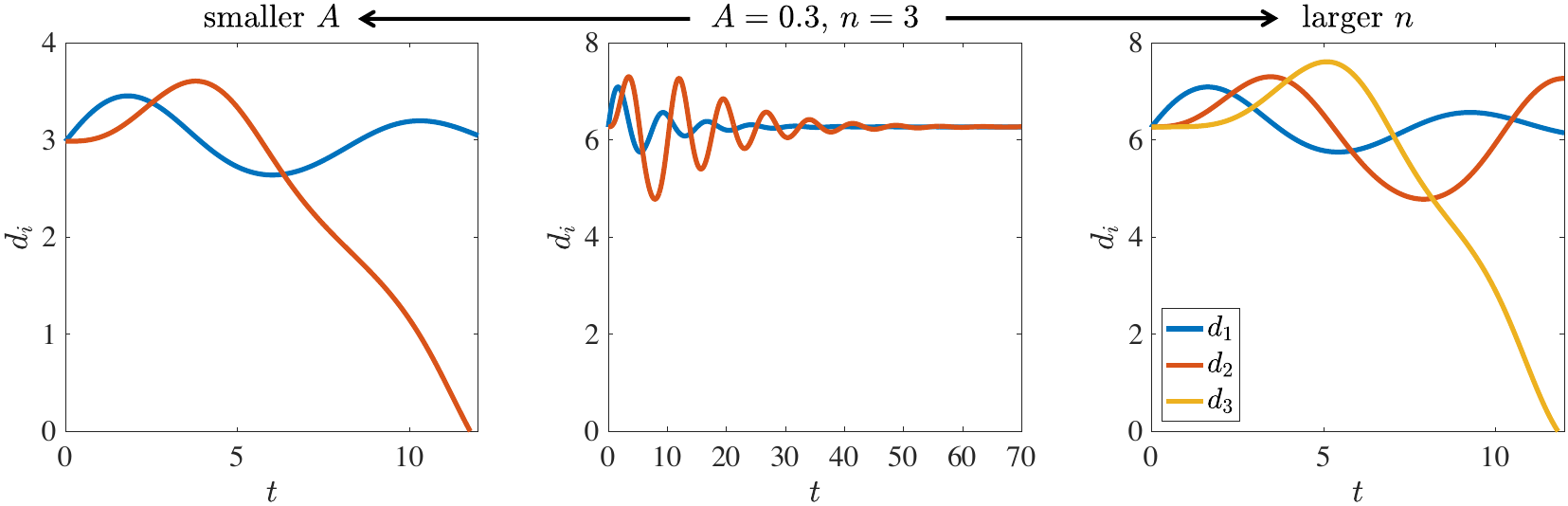}
    \caption{Results of numerical simulations of the (nonlinear) reduced model~\eqref{WuEqnMot}. The initial velocity of the leader is $1.3U$, and the remaining plates are initialized in an equilibrium schooling mode with velocity $U$ (given by Eq.~\eqref{UBase}) and inter-plate distance $d^{\infty,1}$ (given by Eq.~\eqref{FEquil}). The middle panel shows the inter-plate distances for $n=3$ wings flapping with amplitude $A=0.3$; after a transient, the system returns to the equilibrium schooling state. The right panel shows the effect of adding a plate ($A = 0.3$, $n=4$), and the left panel shows the effect of decreasing the flapping amplitude ($A=0.2$, $n=3$). It is evident that both lead to a collision between the trailing plates.}
    \label{fig:Wu}
\end{figure}

While the foregoing linear theory predicts that small-amplitude perturbations to the equilibrium schooling state eventually decay in time, we find that transient linear growth combined with nonlinear effects can cause finite-amplitude perturbations to dislodge the trailing plates from their equilibrium positions. Specifically, we performed numerical simulations of the nonlinear reduced model~\eqref{WuEqnMot}, initializing the plates with the positions $x_m(0)=-(m-1)\left(d^{\infty,1}+1\right)$ and velocities $\dot{x}_m(0)=U$ for $m > 1$ and $\dot{x}_1(0)=1.3U$; that is, we assume that the trailing wings initially move with the steady speed of a single wing, as determined by Eq.~\eqref{UBase}, and subject the leader to a finite-amplitude velocity perturbation. The results are shown in figure~\ref{fig:Wu}. The middle panel shows the inter-plate distances $d_i$  for $n = 3$ wings and flapping amplitude $A = 0.3$: after a transient, the trailing wings eventually settle back into their equilibrium positions. By contrast, increasing the number of wings to $n = 4$ (right panel) or decreasing the flapping amplitude to $A = 0.2$ (left panel) causes the trailing wings to become dislodged from their equilibrium positions and collide. A qualitatively similar behavior is observed for the other flapping amplitudes considered in this paper. We thus conclude that, despite its simplifications, the reduced model presented in the section captures the two salient features of our vortex sheet simulations, namely, that schooling modes become progressively less stable as the number of wings is increased or the flapping amplitude is decreased.

We note that a number of physical effects are neglected in our reduced model~\eqref{WuEqnMot} for the sake of analytical simplicity. Specifically, the assumptions are that the wake shed by each plate is flat and unaffected by modulations in the plate's horizontal speed, both of which are at odds with our vortex sheet simulations. The model also assumes nearest-upstream-neighbor interactions; while our simulation results seem to confirm this assumption (\S~\ref{ssec:several}), strictly speaking each plate feels the vortex sheets shed by {\it all} of the other plates.

\section{Conclusion}\label{sec:conc}
%A vortex sheet model is used to compute the motion of $n$ plates initially placed in-line and separated by distance $d_0$, that move vertically with a prescribed oscillatory motion of fixed frequency and amplitude $A$, and have an initial horizontal velocity $U_0$.  The plates move horizontally due to thrust and drag forces. The thrust forces are induced by the tip vortex sheet strength which in turn is induced by the prescribed vertical motion as well as the fluid vorticity. The drag forces depend on the plate velocity and are modelled using a simple $3/2$ power law.

%The long-term plate motion is found to be independent of the given initial velocity $U_0$. The plates move forward in an oscillatory manner, corresponding to the oscillatory tip sheet strength. The wake of each plate consists of a sequence of positive and negative vortices, one of each being formed during one oscillation period of the generating plate. The wake of $n$ plates consist of $2n$ vortices generated in each oscillation period, which interact and form a quasiperiodic pattern with wavelength $\uinf T$, where $T=1/\omega$ is the length of the oscillation period. 
%$\uinf$ decreases as $A$ decreases and is largely independent of $n$.

In this paper we have used a vortex sheet method to simulate the dynamics of $n$ flat plates in a 2D inviscid and incompressible fluid. The plates are in an in-line formation and execute a heaving kinematics with amplitude $A$ and frequency $f$ (figure~\ref{F:sketch}), but their horizontal velocities are determined by the balance of hydrodynamic thrust and drag forces. The thrust forces are induced by the vortex sheet strength at the leading edge, which in turn is induced by the prescribed vertical motion and the fluid vorticity. The drag forces depend on the plate velocity and are modelled using a simple $3/2$--power law. Particular care is taken to evaluate the near-singular integrals that arise when a vortex sheet is near a plate, which is required both to ensure that the vortex sheet does not cross the plate and to accurately compute the hydrodynamic thrust on the plate.

We find that the long-time motion of a single plate is independent of the imposed initial velocity $U_0$ (figure~\ref{F:uinf}). The plate moves forward in an oscillatory manner due to the oscillatory leading edge sheet strength, and leaves a wake of positive and negative vortices, one of each being formed during one oscillation period (figure~\ref{F:allvort25}(a)). The wake of $n$ plates consists of $2n$ vortices generated in each oscillation period, which interact and form a quasiperiodic pattern with wavelength $\uinf/f$ (figure~\ref{F:allvort25}(b--d)).

     Our simulations show that plates in the wake of prior ones move towards certain equilibrium positions (figure~\ref{F:equilibria}). These equilibria are quantized on the schooling number~\citep{Becker}: that is, the distance between a given plate and the one in front of it is approximately an integer multiple of the wavelength $\uinf/f$. This observation is consistent with prior experiments and vortex sheet simulations of both pairs~\citep{Sophie,Fang_Thesis,Heydari_2021} and larger collectives~\citep{Newbolt_2022,Newbolt_2024,Heydari_2024} of flapping wings. By varying the initial distance $d_0$ between each plate and its upstream neighbor, we find the first six equilibrium positions for $n=2$, 3 and 4 plates, for the three dimensionless flapping amplitudes $A=0.1$, 0.3 and 0.4 (figure~\ref{F:distall}). Which equilibrium is approached by a trailing plate depends on its initial distance from the preceding one. 

As more plates are added to the formation, they experience increasingly large oscillations in their horizontal motion. This is apparent for instance by comparing the three panels in the top row of figure~\ref{F:distall}, for which the flapping amplitude $A = 0.4$. The basins of attraction of the equilibrium positions evidently shrink, as evidenced by the trajectories that depart from the equilibria altogether. For smaller values of the flapping amplitude, the in-line formation becomes increasingly unstable, with a number of cases in which the plates not only leave their nearest equilibria but collide with their upstream neighbors. The loss of stability and %correspondingly increasing number of collisions increases drastically 
associated collisions are far more prevalent both
as the number of plates $n$ increases and as the amplitude $A$ decreases. For example, for $A=0.1$, $n=4$, less than a third of the initial conditions tested have not collided after thirty flapping periods. 

We find that a simple algorithm is sufficient to mitigate the flow-induced instability that we observe: specifically, if a plate approaches (separates from) its upstream neighbor, it slows down (speeds up) by adjusting its flapping amplitude $A$. This leads to out-of-sync oscillations until an equilibrium position is reached (figure~\ref{F:distall+stabilization}). 
The stabilized vortex wakes form beautiful organized vortex patterns that are not observed without stabilization (figures~\ref{F:stabilizedvorticity} and~\ref{F:stabilizedvorticityn3}). The stabilized wakes are expected to represent the long-time limit of unstabilized wakes that have not lost stability and reached an equilibrium. The cost of such long-time computations is prohibitive, and the stabilization scheme enables us compute these patterns in reasonable time, at least for $n=2$ and 3. 

We have also developed a reduced model for the dynamics of hydrodynamically interacting plates based on linear thin-airfoil theory~\citep{Wu1961Swimmingwavingplate,Wu1975Extractionflowenergy,Sophie}. In this model, we assume that each plate interacts only with its nearest upstream neighbor, and that each plate sheds a flat vortex sheet that is unperturbed by the other plates. The model is able to recover a number of phenomena observed in our simulations, namely, the quantization of stable equilibria on the schooling number, the dependence of the translation velocity on the flapping amplitude, and the decreasing stability of the formation as $A$ decreases and $n$ increases.

The flow-induced oscillatory instability and subsequent collisions within increasingly large in-line formations are reminiscent of the ``flonons" observed in recent experiments on flapping wings in a water tank~\citep{Newbolt_2024}. The authors of that experiment posited that the formation destabilized via a resonance cascade, a physical picture that is consistent with our reduced model in \S~\ref{ssec:reduced}. Such an instability was not observed in the vortex sheet simulations of in-line formations of pitching plates conducted by~\citet{Heydari_2024}, who instead observed that formations typically destabilized via a loss of cohesion in which trailing plates could not keep up with the rest of the school and fell behind. We note that one difference between our vortex sheet algorithm and that of~\citet{Heydari_2024} is that, in the latter, the point vortices shed by the plates are manually removed from the fluid after a prescribed dissipation time~\citep{Heydari_2021}. Another difference, which is potentially significant, is that~\citet{Heydari_2024} consider plates that undergo pitching oscillations about their leading edges, while our plates undergo heaving oscillations. As a consequence, the horizontal force on the plate is generated entirely by the leading-edge suction in our work, while in their work it is dominated by the pressure force normal to the plate~\citep{Heydari_2021}. We also note that we do not observe a ``compact" formation in which the plates are separated by less than a chord length, which has been observed in 2D Navier-Stokes simulations of elastic filaments~\citep{Zhu_2014} and  wings~\citep{Lin_2020} in flows with $\text{Re} = O(100)$, and also in 3D Navier-Stokes simulations of flexible plates~\citep{Arranz_2022}.

The in-line formation considered herein may be viewed as a minimal model for a fish school, in which only hydrodynamic interactions are considered at the expense of behavioral rules and sensing. Our results suggest that fish may need to employ active control mechanisms to mitigate the flow-induced instability experienced by increasingly large formations, but that relatively simple control mechanisms (like the one proposed in \S~\ref{ssec:stabilization}) could be sufficient. A worthwhile future direction would be to consider more sophisticated control mechanisms, such as those based on reinforcement learning~\citep{Novati_Learning,Verma_RL}. One could also explore biologically-relevant control mechanisms, such as those inspired by  the fish lateral line system, which detects the nearby flow velocity and pressure gradients and is thought to play an important role in mediating schooling behavior~\citep{Faucher2010,Mekdara2021}. While our study was focused on the formation's streamwise stability, another future direction would be to allow the plates to translate in both the streamwise and lateral directions and thus examine the two-dimensional stability of various formations, as has been done using models~\citep{Gazzola_JFM}, Navier-Stokes simulations~\citep{Gazzola_JCP,Lin_2021,Lin_2022} and experiments~\citep{Ormonde2023}. Moreover, the simulation framework presented herein could be straightforwardly extended to model more complex formations, such as the diamond lattice and so-called ``phalanx" formations exhibited by some fish species~\citep{Newlands_tuna,AshrafINT,AshrafPNAS}.
\\

\noindent{\bf Supplementary movie captions}: \\ \\ \noindent{\it Movie 1:} The movies show $n=2$ flapping plates (green) and their shed vortex sheets (blue and red). The plates' dimensionless heaving amplitude is $A = 0.2$, and their initial separation distances are $d_0 = 4.0$ (top), $d_0 = 6.0$ (middle) and $d_0 = 6.5$ (bottom). The arrow indicates the instantaneous tail-to-tip separation distance between the plates. Note that the top and middle movies converge to the first equilibrium schooling mode with separation distance $d_1^{\infty,1}=4.2$, whereas the bottom converges to the second with $d_1^{\infty,2}=8.0$. \\ \\ \noindent{\it Movie 2:} The movie shows $n=3$ flapping plates (green) and their shed vortex sheets (blue, red and pink). The plates' dimensionless heaving amplitude is $A = 0.2$, and their initial separation distance is $d_0 = 6.0$. The arrow indicates the instantaneous tail-to-tip separation distance between the plates. The simulation eventually converges to the first equilibrium schooling mode, $d_n(t)\rightarrow d_n^{\infty,1}$ for $n=1$ and $n=2$. \\ \\ \noindent{\it Movie 3:} The movie shows $n=4$ flapping plates (green) and their shed vortex sheets (blue, red, pink and cyan). The plates' dimensionless heaving amplitude is $A = 0.4$, and their initial separation distance is $d_0 = 14.0$. The arrow indicates the instantaneous tail-to-tip separation distance between the plates. The simulation eventually converges to the first equilibrium schooling mode, $d_n(t)\rightarrow d_n^{\infty,1}$ for $n=1$, 2 and 3. \\ \\ \noindent{\it Movie 4:} The movies show $n=3$ flapping plates (green) and their shed vortex sheets (blue, red and pink), without (top panel) and with (bottom panel) the stabilization algorithm implemented with parameter $\beta = 5.0$. The plates' dimensionless heaving amplitude is $A = 0.1$, and their initial separation distance is $d_0 = 2.25$. The arrow indicates the instantaneous tail-to-tip separation distance between the plates. Note that the collision between plates in the top panel is avoided because of the stabilization.
\\

\noindent{\bf Funding}: Support is acknowledged from NSF DMS-2108839 (AO)
and NSF DMS-1909407 (MS). 
\\

\noindent {\bf Declaration of interests}: The authors report no conflict of interest.
\\

\noindent {\bf Data availability statement}: The data that support the findings of this study are available upon request.
\\

\noindent{\bf Author ORCIDs}: 

\noindent M. Nitsche, https://orcid.org//0000-0003-1430-9968

\noindent A. U. Oza, https://orcid.org/0000-0002-9079-9172

\noindent M. Siegel, https://orcid.org/0009-0001-7265-5896.

\appendix

\section{Derivation of the leading-edge thrust}\label{App:Thrust}

We provide a simple derivation of the  thrust or `suction'  force on the leading edge of a wing in a general time-dependent flow.  The expression can also be found by combining equations (4.62) and (6.89) from~\citet{EldredgeBook}. 

We assume that at a fixed time $t$ an isolated wing (i.e., a flat plate) is located on the $x$-axis. For the sake of analytical convenience we adopt the convention that the plate's leading edge is at $x=0$ and its trailing edge at $x=1$, so that the plate moves to the left, which is opposite to the convention used in the main text. Let the plate  be parameterized in terms of an arclength variable $0 \leq s \leq 1$ by $\bx(s)=(s,0)$,  and  assume it has bound vortex sheet strength $\gamma(s)$ (we omit representing the dependence on time, with the understanding that all quantities are time dependent).  Let $\gamtil(s)=\gamma(s) \sqrt{s(1-s)}$ be a desingularized vortex sheet strength, which is assumed to be a regular function of $s$~\citet[page 120]{EldredgeBook}.
The fluid  velocity induced by the  vorticity is given by the Birkhoff-Rott integral
\begin{align}
\bu(x,y)=\frac{1}{2 \pi} \int_0^1 \frac{(- y,x-s)}{(x-s)^2+y^2} \frac{\gamtil(s)}{\sqrt{(1-s)s}} \ \rmd s + \boldsymbol{w},\label{BR}
\end{align}
where $\boldsymbol{w}$ is a smooth velocity due to free and other bound vorticity in the fluid.

Following~\citet[page 98]{Saffman}, the suction force is calculated by integrating the Euler momentum equation in a small disk $B_\epsilon$ of radius $\epsilon \ll 1$ about the leading edge  $(x,y)=(0,0)$: 
\begin{align}
 \frac{\partial}{\partial t} \int_{B_\epsilon} \bu \  \rmd V + \int_{S_\epsilon} [p  \  \bn + \bu(\bu \cdot \bn)] \   \rmd S  &= \int_{B_\epsilon} \boldsymbol{F} \ \rmd V \\
&= -T \boldsymbol{\hat{i}}, \label{eq:T}
\end{align}
%where we used the divergence theorem to convert  the second volume integral to a surface integral.
 where $p$ is the pressure, $\bn$ is the outward unit normal to the surface $S_\epsilon$ of $B_\epsilon$,  and $\boldsymbol{F}=-T  \delta(\bx) \boldsymbol{\hat{i}} $ is the external force, i.e., the force that the  leading edge of the wing exerts  on the fluid.  The  suction force of the fluid  on the wing  is  then $T \boldsymbol{\hat{i}}$.   Here we have taken  the fluid density  to be one. 
%The pressure and velocity are singular at the leading edge. 
The pressure can be eliminated using Bernoulli's law~\citep[page 19]{Saffman}
\begin{align}
\frac{\partial \phi}{\partial t} +p +\frac{1}{2} |\bu|^2=H(t),\label{eq:Bernoulli}
\end{align}
where $\phi$ is the velocity potential, and $H(t)$ is an arbitrary function of time.  Multiply (\ref{eq:Bernoulli}) by $\boldsymbol{n}$ and take the surface integral
over $S_\epsilon$  to obtain at leading order 
\begin{align}
\int_{S_\epsilon} p \bn \ \rmd S= - \frac{1}{2} \int_{S_\epsilon} |\bu|^2 \bn \ \rmd S,\label{pressure_int}
\end{align}
where we have neglected terms that are less singular at the plate's leading edge and thus give higher-order contributions in $\epsilon$. Substitute (\ref{pressure_int}) into (\ref{eq:T}) and note that the volume integral  $\frac{\partial}{\partial t} \int_{B_\epsilon} \bu \ dV$ gives   a  higher-order contribution 
in $\epsilon$
%(in view of being less singular at the plate's leading edge) 
to obtain  
\begin{align}
 T \boldsymbol{\hat{i}} =   \lim_{\epsilon \rightarrow 0}  \int_{S_\epsilon} \left[  \frac{1}{2} |\bu|^2 \bn -  \bu(\bu \cdot \bn) \right]  \ \rmd {S}.  \label{suction_force}
\end{align}

We now calculate the velocity at leading order in  $\epsilon$ for $(x,y) \in B_\epsilon$.
  In (\ref{BR}), make the change of variable $-v=x-s$ and expand $\tilde{\gamma}(x+v)=\tilde{\gamma}(x)+\tilde{\gamma}'(x) v + \ldots$ to obtain for $\boldsymbol{u}=(u_1,u_2)$
\begin{align}
\begin{split}
\label{u1u2}
u_1 (x,y)&=- \frac{y}{2 \pi} \left[\gamtil(x)   W_0(x,y)  +\gamtil'(x) W_1(x,y) \right]+ O(1), \\
u_2(x,y)  &= -\frac{\gamtil(x)}{2 \pi} W_1(x,y)+O(1),
\end{split}
\end{align}
where
\begin{align}
\begin{split}
\label{W0W1}
W_0(x,y)&= \int_{-x}^{1-x} \frac{1}{(v^2+y^2) \sqrt{(1-(v+x))(v+x)}} \  \rmd v,\\
W_1(x,y)&= \int_{-x}^{1-x} \frac{v}{(v^2+y^2) \sqrt{(1-(v+x))(v+x)}}  \ \rmd v,\\
\end{split}
\end{align}
and the $O(1)$ remainder  in (\ref{u1u2}) refers to terms for which the Birkhoff-Rott kernel  is integrable in  $v$ as $y \rightarrow 0 $. The integrals in (\ref{W0W1}) can be evaluated in closed form. Let $F(x,y)=\sqrt{(1-x)^2+y^2}$ and $G(x,y)=\sqrt{x^2+y^2}$. Then~\citep{Geer1968}
\begin{align}
\begin{split}
\label{Weval}
W_0 (x,y)&=\pi \left( F+G \right) \left\{ (FG)^2 \left[ (F+G)^2-1 \right] \right\}^{-1/2},\\
W_1 (x,y)&=\pi  \left( -x F+(1-x) G  \right)  \left\{ (FG)^2 \left[ (F+G)^2-1 \right] \right\}^{-1/2}.
\end{split}
\end{align}
Evaluation of~\eqref{Weval} for $x=\epsilon \cos \theta, ~y=\epsilon \sin \theta$ on $S_\epsilon$ gives to leading order in $\epsilon$, 
\begin{align}
\begin{split}
\label{Weval1}
%W_0 &=\frac{\pi}{\epsilon^{3/2}} \frac{1}{\sqrt{2(1-\cos\theta)}}+O(\epsilon^{-1/2}),\\
W_0 &= \frac{\pi}{\sqrt{2(1-\cos\theta)}} \epsilon^{-3/2} +O(\epsilon^{-1/2}),\\
W_1 &=\pi   \sqrt{\frac{1-\cos \theta}{2}}  \epsilon^{-1/2}+O(\epsilon^{1/2}).
\end{split}
\end{align}
From (\ref{u1u2}) and (\ref{Weval1}), the leading order velocity components  on $S_\epsilon$ are 
\begin{align}
\begin{split}
\label{ueval}
%u_1&= \frac{1}{\epsilon^{1/2}} \frac{\gamtil(0)}{2 \sqrt{2}} \frac{\sin \theta}{(1-\cos \theta)^{1/2}} + O(1),\\
%u_2&=  \frac{1}{\epsilon^{1/2}}  \frac{\gamtil(0)}{2 \sqrt{2}} (1-\cos \theta)^{1/2} + O(1).
u_1&=- \frac{\gamtil(0)}{2 \sqrt{2}} \frac{\sin \theta}{(1-\cos \theta)^{1/2}}  \epsilon^{-1/2}+ O(1),\\
u_2&= -  \frac{\gamtil(0)}{2 \sqrt{2}} (1-\cos \theta)^{1/2}  \epsilon^{-1/2}+ O(1).
\end{split}
\end{align}

The surface integral in the expression for the suction force    can now be computed by substituting (\ref{ueval}) and $\boldsymbol{n}=(\cos \theta, \sin \theta)$  into  (\ref{suction_force}).  We find  that the $y$-component of the suction force is zero as expected,  while the $x$-component satisfies
\begin{align}
T \boldsymbol{\hat{i}} = -\frac{\pi}{4} \gamtil^2(0) \boldsymbol{\hat{i}}.\label{T_formula}
\end{align}
More generally  the magnitude of the suction force  for a horizontal plate with leading edge located at $\boldsymbol{x}$ is
\begin{align}
|T| = \frac{\pi}{4} \gamtil^2(\boldsymbol{x}),\label{E:UnsteadyForceMag}
\end{align}
with the force oriented  in the negative-$x$ direction when the leading edge of the wing is located on the left, and in the positive-$x$ direction when the leading edge is on the right. 

%In the special case of a  steady uniform flow with speed $U$ directed  at angle $\alpha$ with respect to the plate,  \citet[page 153]{Acheson} gives  the suction force  as $T  \boldsymbol{\hat{i}}=- \pi (U \sin \alpha)^2  \boldsymbol{\hat{i}}=-\pi V^2  \boldsymbol{\hat{i}}$.  This  agrees with  (\ref{T_formula}) upon setting $\gamtil(0)=-2V$, which is the leading-edge normalized sheet strength  for a steady flow. \colr{I would vote to kill this paragraph, since it is all said in Appendix B.} \colg{I'm fine with this.} % \cite{} (Monika has a reference). 

\begin{figure}
\centering
\includegraphics[trim= 0 0 0 0, clip, height=2.9truein]{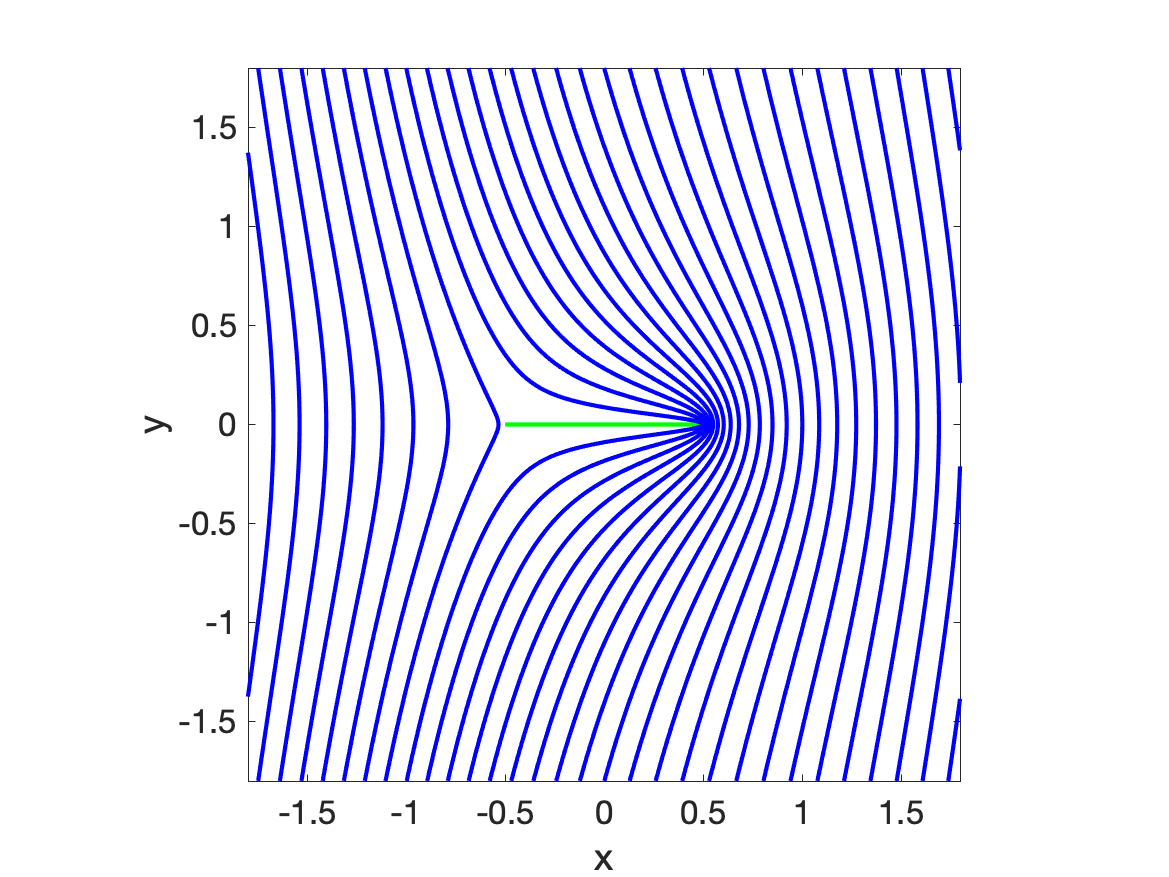}
\caption{Potential flow past a plate of unit length translating vertically with a velocity $V$, as obtained from~\eqref{E:compw} and~\eqref{E:GammaCirc}. The circulation around the plate is $\Gamma=-\pi V$, which cancels the singularity at the left edge. %in the presence of a far-field flow $\boldsymbol{U}_{\inf}=(0,1)$, with circulation $\Gamma = \pi V$ that cancels the singularity at the left edge.
} \label{F:steadyplate}
% Computed with streamlines/plotstream.m
\end{figure}

\section{Comparison between unsteady and steady flow around a vertically translating flat plate}\label{App:Steady} 

A simple argument for steady flow past a plate, as considered by~\citet[page 153]{Acheson},  gives results consistent with those of Appendix~\ref{App:Thrust}. Consider a plate of length $4a$ placed at $x\in[-2a,2a]$, aligned horizontally with the $x$-axis, moving with velocity $(U,V)$ in an inviscid fluid at rest at infinity. Since the horizontal component does not impact thrust, we set $U=0$. Here we derive the leading-edge normalized sheet strength and leading edge suction for steady plate motion with singular flow at the leading edge (at $z=2a$), and smooth flow satisfying the Kutta condition at the trailing edge (at $z=-2a$). The steady flow in a reference frame moving with the plate is obtained by the conformal map from the unit circle $|\zeta|=a$ onto the plate on the $x$-axis with $x\in[-2a,2a]$:
\begin{equation}
z=J(\zeta)=\zeta+\frac{a^2}{\zeta}~, \hbox{ with } \zeta=J^{-1}(z)= \frac{1}{2}\left(z+\sqrt{z^2-4a^2}\right)~.
\end{equation}
Potential flow around the circle with circulation $\Gamma$ is given by the complex potential
\begin{equation}
w(\zeta)=\rmi V\left(\zeta-\frac{a^2}{\zeta}\right)-\rmi\frac{\Gamma}{2\pi}\log \zeta~.\label{E:compw}
\end{equation}
The complex potential in the $z$-plane is $W(z)=w(J^{-1}(z))$, with complex velocity
\begin{equation}
\frac{\rmd W}{\rmd z}=\frac{\rmd w/\rmd \zeta}{\rmd z/\rmd \zeta}
=\rmi V\left(1+\frac{a^2}{\zeta^2}-\frac{\Gamma}{2\pi V\zeta}\right)
\frac{1}{1-\frac{a^2}{\zeta^2}}
=\frac{\rmi V}{\zeta^2-a^2}\left(\zeta^2 -\frac{\Gamma}{2\pi V}\zeta + a^2\right).
\end{equation}
The singularity at the trailing edge $z=-2a$ ($\zeta=-a$) is removed by setting 
\begin{equation}
\Gamma=-4\pi V a~,\label{E:GammaCirc}
\end{equation}
so that 
\begin{equation}
\frac{\rmd W}{\rmd z}= \rmi V\frac{\zeta+a}{\zeta-a}~.
\end{equation}
The resulting velocity on the plate is obtained by setting 
$\zeta=ae^{\rmi \theta} (x=2a\cos\theta)$,
where $\theta\in[0,\pi]$ corresponds to the top of the plate, and $\theta\in[\pi, 2\pi]$ to the bottom:
\begin{equation}
\begin{split}
\frac{\rmd W}{\rmd z} 
&=V\frac {\sin\theta} {1-\cos\theta}
=V\frac {\pm\sqrt{1-\cos^2\theta}} {1-\cos\theta}
=V\frac {\pm\sqrt{1+\cos\theta}} {\sqrt{1-\cos\theta}}\\
&=\pm V\frac{\sqrt{2a+x}}{\sqrt{2a-x}}~ 
=\pm V\frac{2a+x}{\sqrt{4a^2-x^2}}~. 
\end{split}
\end{equation}
For positive $V$, the top velocity is positive and the bottom is negative. This yields the vortex sheet strength
\begin{equation}
\gamma(x)=u_--u_+=-2V\frac{2a+x}{\sqrt{4a^2-x^2}} =-2V\frac{2a+x}{\sqrt{4a^2-x^2}} =\frac{\gamtil(x)}{\sqrt{4a^2-x^2}}
\end{equation}
where $\gamtil(x)=\gamma(x) {\sqrt{4a^2-x^2}}$. This definition of $\gamtil$ is consistent with~\eqref{BR}  with $s=x+1/2, a=1/4$, and~\eqref{E:totalvel} with $s=x, a=1/4$. For $a=1/4$, the leading-edge normalized sheet strength is $\gamtil(1/2)=-2V$.

\begin{figure}
\centering
\includegraphics[trim= 0 0 0 0, clip, height=3.2truein]{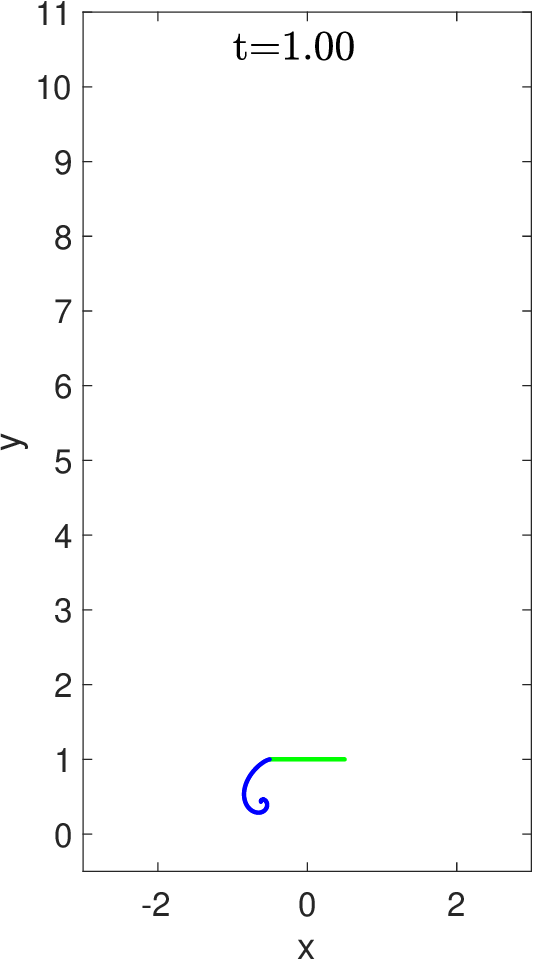}
\includegraphics[trim= 35 0 0 0, clip, height=3.2truein]{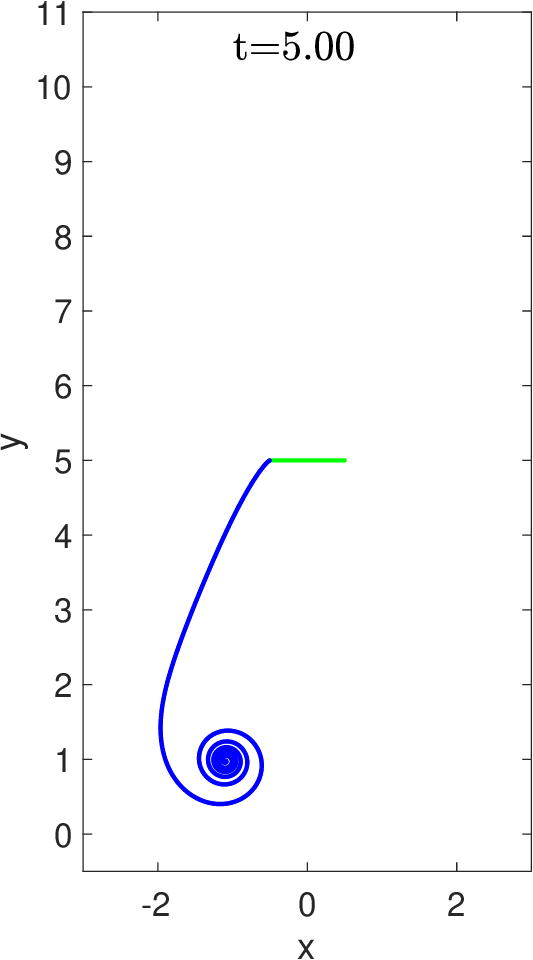}
\includegraphics[trim= 35 0 0 0, clip, height=3.2truein]{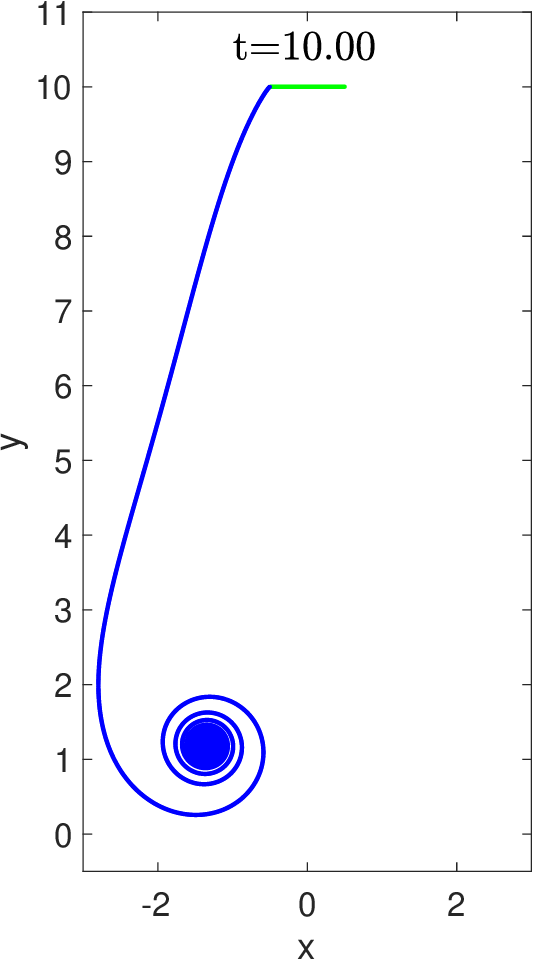}
\caption{Impulsively started plate that translates vertically with constant velocity $V=1$, and the vortex sheet that separates from its left endpoint, shown at the indicated times.}
\label{F:positionconstantvelo}
\end{figure}

Using Blasius' theorem we find that the total force on the plate is
\begin{equation}
\begin{split}
F_x-\rmi F_y
&=\frac{i\rho}{2}\oint_{C_z}{\left(\frac{\rmd W}{\rmd z}\right)}^2\,\rmd z
%&=\frac{i\rho}{2}\oint{\left(\frac{dW}{dz}\right)}^2J'(\zeta)d\zeta
%&=\frac{\rmi \rho}{2}\oint_{C_z}{\left(\frac{\rmd W}{\rmd z}\right)}^2\frac{\rmd z}{\rmd\zeta}\,\rmd\zeta
=\frac{\rmi\rho}{2}\oint_{C_\zeta}{\left(\frac{\rmd W}{\rmd z}\right)}^2\frac{\rmd z}{\rmd\zeta}\,\rmd\zeta\\
%=\frac{i\rho}{2}\oint\frac{(dw/d\zeta)^2}{J'(\zeta)}d\zeta
%=\frac{i\rho}{2}\oint\frac{(dw/d\zeta)^2}{J'(\zeta)}d\zeta
&=-V^2\frac{\rmi\rho}{2}\oint_{C_\zeta}\frac{(\zeta+a)^3}{(\zeta-a)\zeta^2}\,\rmd\zeta
=4\pi\rho V^2a 
=-\rho V\Gamma
=\pi\rho a [\gamtil(1/2)]^2,\label{E:SteadyForce}
\end{split}
\end{equation}
%\colr{I think the first line above should read
%\begin{align}
%F_x-\rmi F_y=\frac{\rmi \rho}{2}\oint_{C_z}{\left(\frac{\rmd W}{\rmd z}\right)}^2\,\rmd z
%=\frac{\rmi\rho}{2}\oint_{C_\zeta}\frac{\left(\rmd W/\rmd\zeta\right)^2}{\rmd z/\rmd\zeta}\,\rmd\zeta
%\end{align}
%}
picking up the residues at $\zeta=0,a$. Expression~\eqref{E:SteadyForce} is consistent with~\eqref{E:UnsteadyForceMag} by setting $\rho=1$, $a=1/4$.

Figure \ref{F:steadyplate} shows the streamlines of the resulting steady flow past the plate with $V=1$ and $a=1/4$, in a reference frame moving with the plate, where the circulation given by~\eqref{E:GammaCirc} cancels the singularity at the trailing (left) edge. Note the resulting stagnation point at the trailing edge and the symmetry across $y=0$.

We now consider an impulsively started plate with $a=1/4$, initially on the $x$-axis and centered at the origin,  moving vertically upward with constant velocity $V=1$ for $t>0$. The singularity at the left edge is resolved by satisfying the Kutta condition through vortex sheet generation and separation via the method described in \S~\ref{sec:model}.  Figure \ref{F:positionconstantvelo} shows the position of the plate and the separated vortex sheet at a sequence of times. 

\begin{figure}
\centering
\includegraphics[trim=  0 0  0 0, clip, height=2.29truein]{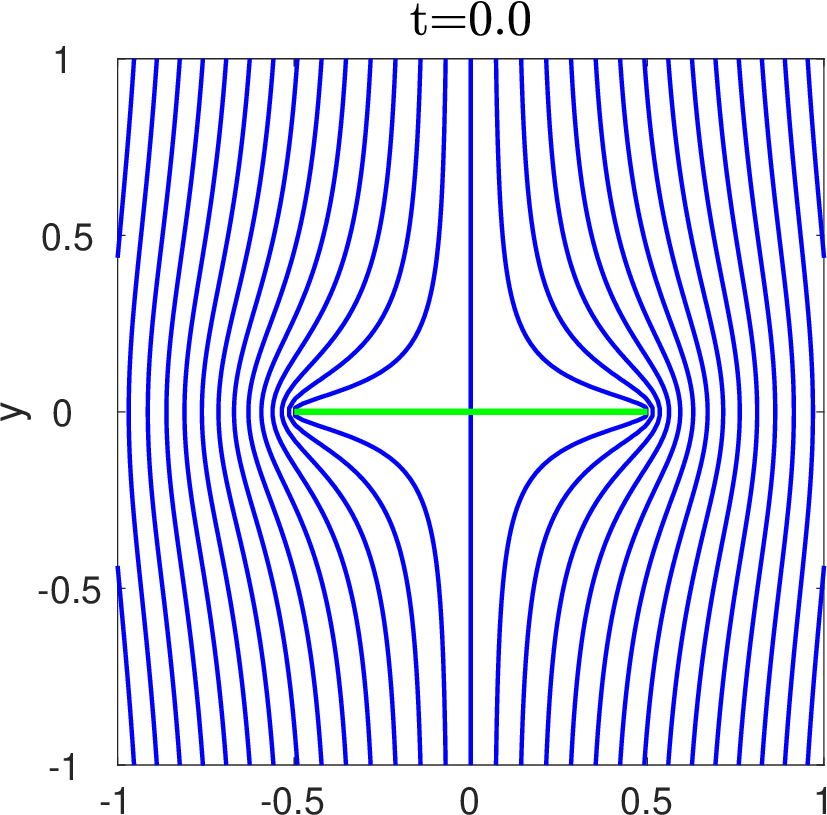}\hskip15pt
\includegraphics[trim=  0 0  0 0, clip, height=2.29truein]{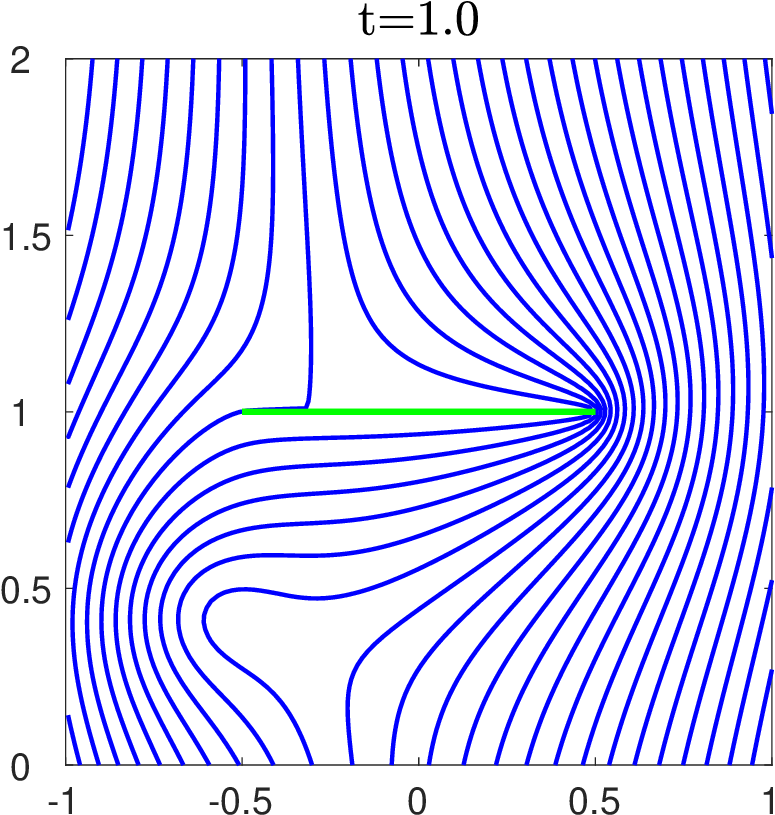}\\
\vskip10pt
\includegraphics[trim=  0 0  0 0, clip, height=2.4truein]{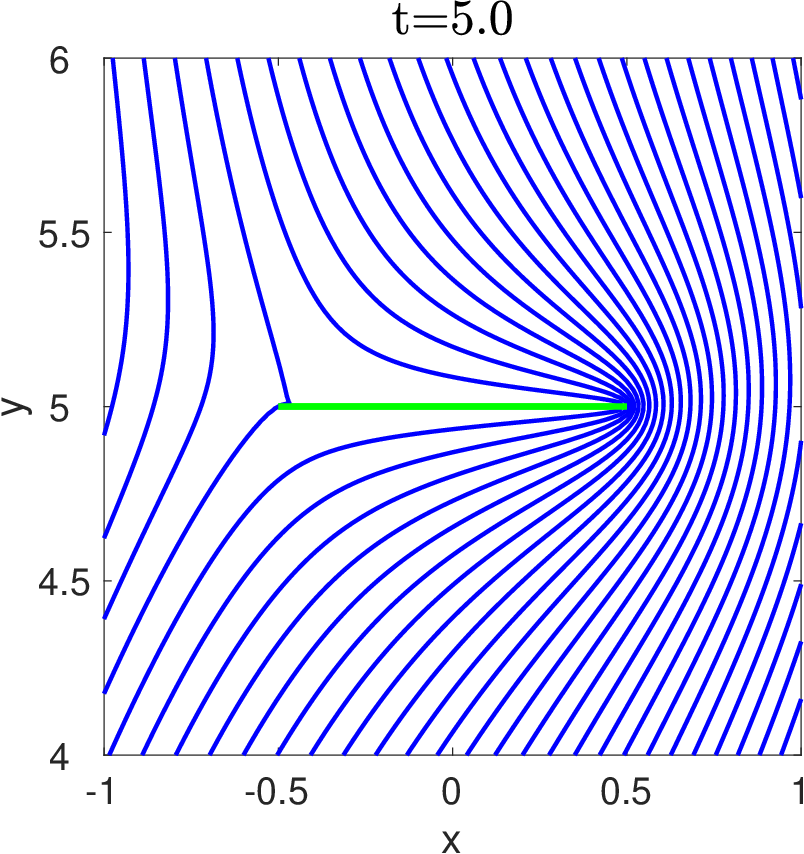}\hskip15pt
\includegraphics[trim=  0 0  0 0, clip, height=2.4truein]{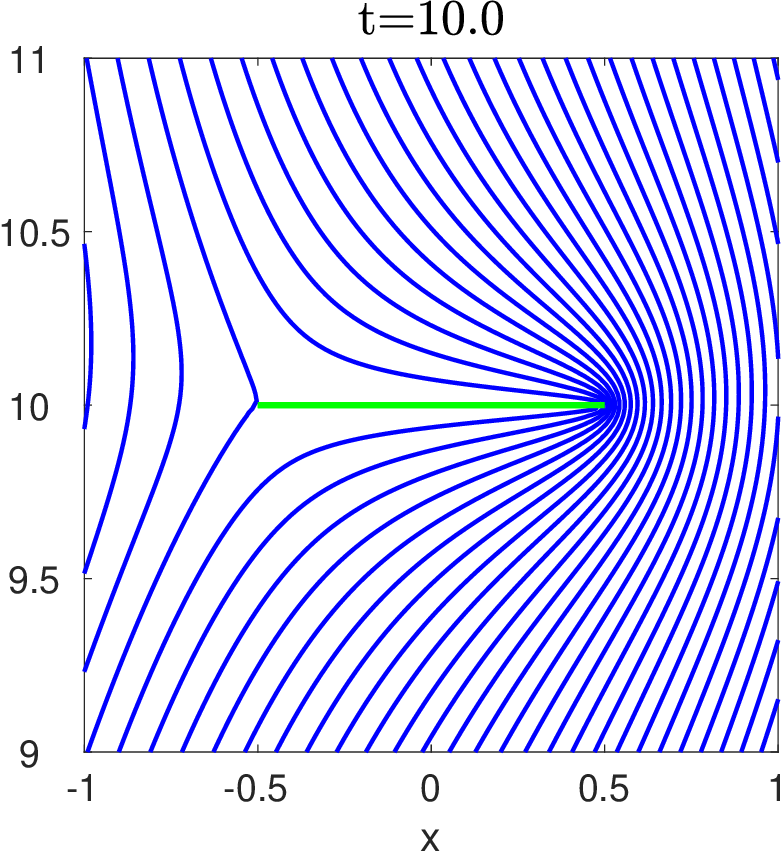}
\caption{Streamlines of the flow in figure~\ref{F:positionconstantvelo}, shown in a reference frame moving with the plate.}
\label{F:streamlinesvx}
\end{figure}

Figure \ref{F:streamlinesvx} shows the corresponding streamlines, in a reference frame moving with the plate. 
%created with streamlines/plotstreamfcn.m
It shows that as $t\to\infty$ the flow approaches the steady symmetric potential flow discussed above and shown in figure \ref{F:steadyplate}. Furthermore, note the position of the stagnation point on the upper plate side. Initially, the impulsively started flow is symmetric, with a stagnation point at the middle of the plate at $x=0$.  As time evolves, the unsteady flow approaches the steady flow configuration, and the stagnation point moves moves progressively closer to the trailing edge.

\begin{figure}
\center
\includegraphics[trim= 0 0 0 0, clip, width=2.4truein]{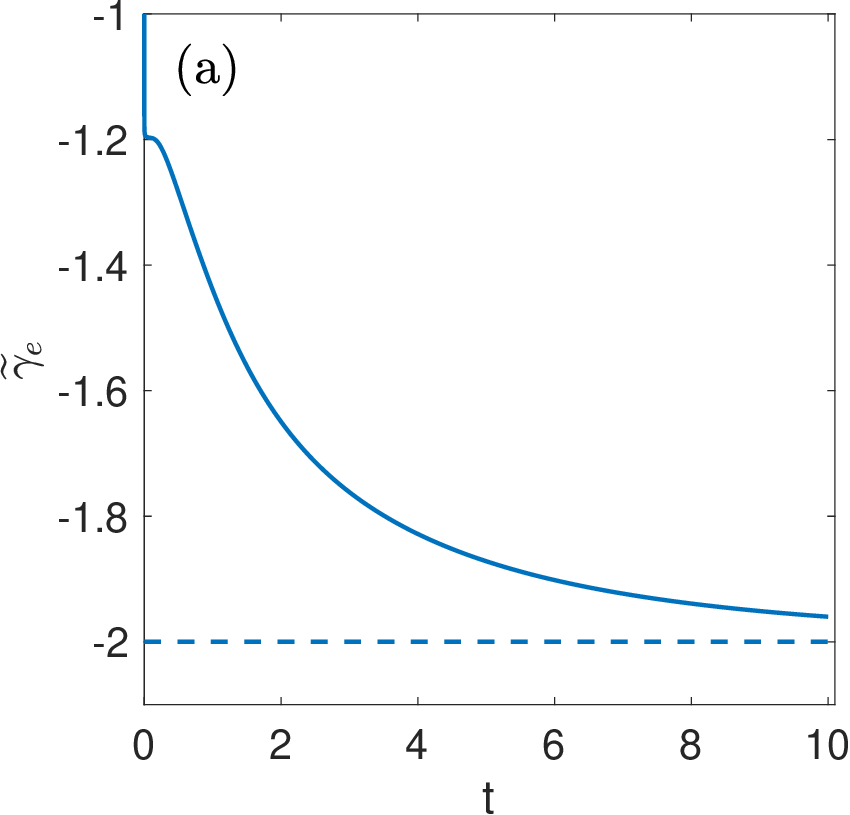}
\includegraphics[trim= 0 0 0 0, clip, width=2.4truein]{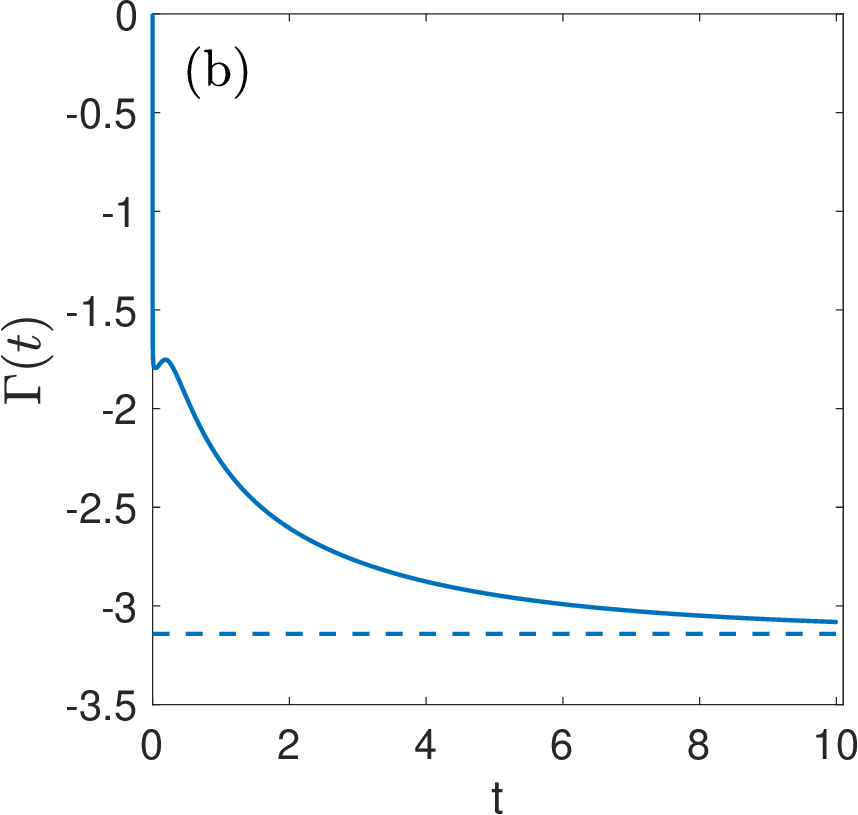}
\caption{
(a) Desingularized leading edge sheet strength $\gamtil(1/2)$, and (b) total circulation around the plate $\Gamma(t)$, corresponding to the flow in figure~\ref{F:positionconstantvelo}.
}
%computed with U=0, delta=0.2
\label{F:tipstrengthconstantvelo}
\end{figure}

Figure \ref{F:tipstrengthconstantvelo} shows the desingularized leading edge sheet strength and the circulation about the plate for the impulsively started plate.  Note that the total circulation about the plate and the shed vortex sheet is enforced to stay constant at zero. The plate circulation is thus the negative of the circulation in the shed vortex sheet. The figure shows that both the shed sheet strength and the plate circulation approach the steady state values as $t\to\infty$, with $\gamtil\to -2V$ and  $\Gamma\to -\pi V$. 

These numerical results illustrate the difference between steady and unsteady flow, and provide further consistency between the steady state results derived here and the unsteady results derived in Appendix~\ref{App:Thrust}.

%created in oneswimmermove/datout020/plotshedgam
%created in oneswimmermove/datout020/plottipgamconst

\section{Alternative model for the drag force acting on the plates}\label{App:Drag}

To compute the drag on plate $j$, we simply use the plate velocity $U^j$, which induces drag relative to the stagnant background flow.  An alternative model that better accounts for the horizontal velocity above and below the plate, used by \cite{Fang_Thesis} and \cite{Heydari_2021}, is obtained by replacing
\begin{equation}
C_d(U^j)^{3/2}
\label{E:origdrag}
\end{equation}
in Equation~\eqref{E:vel} by
\begin{equation}
\frac{C_d}{2} [ (U^j-\overline{u}_+^j)^{3/2} + (U^j-\overline{u}_-^j)^{3/2}]
\label{E:reviseddrag}
\end{equation}
where 
\begin{equation}
\overline{u}_{\pm}^j=\int_0^1 u_{\pm}^j(s,t) \,\mathrm{d}s
\end{equation}
are the average horizontal velocities just above and below the plate, and $u_{\pm}^j(s,t)$ are the limiting horizontal velocities from above and below the plate, respectively. Equation~\eqref{E:reviseddrag} assumes that $U^j-\overline{u}_{\pm}^j>0$, which is the case in the present paper. 

The velocities $u^j_{\pm}(s)$ have two components, one stemming from the $j$th plate and one stemming from the far field vorticity in all the other $n-1$ plates and all the $n$ wakes,
\begin{equation}
u^j_{\pm}= u^j_{\pm \mathrm{plate}} +u^j_{\mathrm{far}}~. 
\end{equation}
The average of this sum is the sum of the average of each component.

The first component is discontinuous across the plate. Note that the jump in the horizontal velocity across the plate equals the negative sheet strength, and the average of the horizontal velocities above and below the plate, induced by itself, is zero,  
\begin{subequations}
\begin{align}
 u^j_{+\mathrm{plate}} -u^j_{-\mathrm{plate}}&=-\gamma(s,t)~,\\
\frac{1}{2}\left(u^j_{+\mathrm{plate}} +u^j_{-\mathrm{plate}}\right)&=0~,
\end{align}
\end{subequations}
and thus,
\begin{equation}
 u^j_{+\mathrm{plate}} =-\gamma(s)/2~,\quad u^j_{-\mathrm{plate}} =+\gamma(s)/2~.
\end{equation}
As a result,
\begin{equation}
\overline{u}^j_{\pm \mathrm{plate}} =\mp\frac{1}{2}\int_0^1\gamma(s)\,\mathrm{d}s%=\mp\frac{1}{2}[-\Gamma_T^j]
=\pm\frac{1}{2}\Gamma_T^j
\end{equation}
The last equality follows since the total circulation about each plate is the negative of the shed vorticity $\Gamma_T^j$.

Furthermore, note that if $\Gamma_T^j/2<U^j$, the contribution of this self-induced term to the difference between (\ref{E:reviseddrag}) and (\ref{E:origdrag}) is of order
\begin{equation}
C_d\frac{3}{4}\left(\frac{\Gamma_T^j}{2U^j}\right)^2\label{E:dragdiff}
\end{equation}
so is excepted to be small. Indeed, we observe in our simulations that $\Gamma^T/2U\le 0.4$ for the amplitudes $A=0.1-0.4$ we considered, so the contribution~\eqref{E:dragdiff} is expected to be $\le 2\%$ for $C_d=0.1$.

The second component, the far field contribution $u^j_{\mathrm{far}}$ to $u^j_{\pm}(s,t)$, is continuous across the $j$th plate. It consists only of the contribution of the free vorticity in all $n$ wakes, and not of any bound plate vorticity. The reason the other $n-1$ plates do not contribute to the horizontal velocity on the $j$th plate is that they are positioned in line with the $j$th plate.
We compute $u^j_{\mathrm{far}}$ by evaluating the horizontal component of
\begin{equation}
\ub(\xo,t) = \frac{1}{2\pi}\sum_{l=1}^n 
\int\limits_0^{\Gamma_T^l}
\frac{(\xo-\xb_f^l(\Gamma,t))^{\perp}} {|\xo-\xb_f^l(\Gamma,t)|^2+\delta^2}\rm d\Gamma 
\label{E:wakevel}
\end{equation}
at all midpoints $\xo=\xb^j(\alf_k^m)$ and then using the midpoint rule to obtain
$$\overline{u}^j_{\mathrm{far}}=\sum_{k=0}^n u(\xb^j(\alf_k^m))s'(\alf_k^m)(\alf_{k}-\alf_{k-1})~.$$
This second component is expected to be small for a single plate ($n=1$), since the wake is behind the plate. It is expected to be less-negligible for $n>1$ plates, when the $j$th plate is in the wake of the upstream plates.

As a result, we expect the change introduced by the alternative model (\ref{E:reviseddrag}) over the simple model (\ref{E:origdrag}) to be small if $n=1$, and thus not significantly affect the steady state velocity if $n=1$. For $n>1$, the rear plates have been shown to have small impact on the first plate.
Since at steady state all plates travel at the same speed, the steady state velocity for $n>1$ computed with the alternative model is expected to be similar to that in the original model.

These conjectures were confirmed by numerical results for sample cases using $A=0.2$ and $C_d=0.1$. We computed the translation velocity $U^j$ for $n=1$ and $n=2$, and the equilibrium separation distance $d_1^{\infty,1}$ for $n=2$, and found that the translation velocity computed with the alternative drag model decreased by 4\% over the simple model, and the separation distance decreased by 3\%. This shows that there is little difference between the two drag models for the parameter regime considered herein, and changing from one to the other is not expected to change the main conclusions of this paper.

\bibliographystyle{jfm}
\bibliography{SchoolingBib}

@article{Faucher2010,
title = {Fish lateral system is required for accurate control of shoaling behaviour},
journal = {Animal Behaviour},
volume = {79},
number = {3},
pages = {679-687},
year = {2010},
issn = {0003-3472},
doi = {https://doi.org/10.1016/j.anbehav.2009.12.020},
url = {https://www.sciencedirect.com/science/article/pii/S0003347209005788},
author = {Karine Faucher and Eric Parmentier and Christophe Becco and Nicolas Vandewalle and Pierre Vandewalle}
}

@article{Mekdara2021,
    author = {Mekdara, Prasong J and Nasimi, Fazila and Schwalbe, Margot A B and Tytell, Eric D},
    title = {Tail Beat Synchronization during Schooling Requires a Functional Posterior Lateral Line System in Giant Danios, \emph{{D}evario aequipinnatus}},
    journal = {Integrative and Comparative Biology},
    volume = {61},
    number = {2},
    pages = {427-441},
    year = {2021},
    month = {05},
    issn = {1540-7063},
    doi = {10.1093/icb/icab071},
    url = {https://doi.org/10.1093/icb/icab071}
}

@article{Greengard1987,
title = {A fast algorithm for particle simulations},
journal = {Journal of Computational Physics},
volume = {73},
number = {2},
pages = {325-348},
year = {1987},
doi = {https://doi.org/10.1016/0021-9991(87)90140-9},
url = {https://www.sciencedirect.com/science/article/pii/0021999187901409},
author = {L Greengard and V Rokhlin},
}

@InProceedings{Sheng2012,
author="Sheng, J. X.
and Ysasi, A.
and Kolomenskiy, D.
and Kanso, E.
and Nitsche, M.
and Schneider, K.",
editor="Childress, Stephen
and Hosoi, Anette
and Schultz, William W.
and Wang, Jane",
title="Simulating Vortex Wakes of Flapping Plates",
booktitle="Natural Locomotion in Fluids and on Surfaces",
year="2012",
publisher="Springer New York",
address="New York, NY",
pages="255--262"
}

@Article{Baddoo_2023,
	author = {P. J. Baddoo and N. J. Moore and A. U. Oza and D. G. Crowdy},
	title = {Generalization of waving-plate theory to multiple interacting swimmers},
	journal = {Communications on Pure and Applied Mathematics},
	year = {2023},
volume = {76},
number = {12},
pages = {3811--3851},
	doi ={https://doi.org/10.1002/cpa.22113},
	url = {https://onlinelibrary.wiley.com/doi/abs/10.1002/cpa.22113}
}

@article{WangKrasnyTlupova2020,
  title={A kernel-independent treecode based on barycentric Lagrange interpolation},
  author={L. Wang and R. Krasny and S. Tlupova},
  journal={Commun. Comput. Phys.},
  volume={28},
  number={4},
  pages={1415--1436},
  year={2020}
}

@article{Krasny1986,
  title={Desingularization of periodic vortex sheet roll-up},
  author={Krasny, R.},
  journal={J. Comput. Phys.},
  volume={65},
  pages={292--313},
  year={1986}
}

@article{Nitsche1994,
  title={A numerical study of vortex ring formation at the edge of a circular tube},
  author={Nitsche, M. and Krasny, R.},
  journal={J. Fluid Mech.},
  volume={276},
  pages={139--161},
  year={1994}
}

@article{Shelley1992,
  title={A study of singularity formation in vortex-sheet motion by a spectrally accurate vortex method},
  author={M. J. Shelley},
  journal={J. Fluid Mech.},
  volume={244},
  pages={493--526},
  year={1992}
}

@book{EldredgeBook,
author = {Eldredge, J. D.},
publisher = {Springer Cham},
title = {Mathematical Modeling of Unsteady Inviscid Flows},
year = {2019}
}

@book{Saffman,
  title={Vortex dynamics},
  author={Saffman, P.G.},
  year={1995},
  publisher={Cambridge University Press}
}

@book{Acheson,
  title={Elementary fluid dynamics},
  author={Acheson, D.J.},
  year={1990},
  publisher={Oxford University Press}
}

@article{Newbolt_2024,
  title = {Flow interactions lead to self-organized flight formations disrupted by self-amplifying waves},
  author = {Newbolt, Joel W. and Lewis, Nickolas and Bleu, Mathilde and Wu, Jiajie and Mavroyiakoumou, Christiana and Ramananarivo, Sophie and Ristroph, Leif},
  journal = {Nature Communications},
  volume = {15},
  issue = {1},
  pages = {3462},
  year = {2024},
  doi = {10.1038/s41467-024-47525-9},
  url = {https://doi.org/10.1038/s41467-024-47525-9}
}

@phdthesis{Fang_Thesis,
  author       = {Fang Fang}, 
  title        = {Hydrodynamic interactions between self-propelled flapping wings},
  school       = {New York University},
  year         = {2016},
}

@article{Newbolt_2022,
  title = {Lateral flow interactions enhance speed and stabilize formations of flapping swimmers},
  author = {Newbolt, Joel W. and Zhang, Jun and Ristroph, Leif},
  journal = {Phys. Rev. Fluids},
  volume = {7},
  issue = {6},
  pages = {L061101},
  numpages = {8},
  year = {2022},
  month = {Jun},
  publisher = {American Physical Society},
  doi = {10.1103/PhysRevFluids.7.L061101},
  url = {https://link.aps.org/doi/10.1103/PhysRevFluids.7.L061101}
}

@article{Heydari_2024, 
 title={Mapping Spatial Patterns to Energetic Benefits in Groups of Flow-coupled Swimmers}, 
journal = {eLife},
volume = {13},
number = {RP96129},
 url={http://dx.doi.org/10.1101/2024.02.15.580536}, 
 DOI={10.1101/2024.02.15.580536}, 
 publisher={Cold Spring Harbor Laboratory}, 
 author={Heydari, Sina and Hang, Haotian and Kanso, Eva}, 
 year={2024}
 }

@article{Heydari_2021, 
title={School cohesion, speed and efficiency are modulated by the swimmers flapping motion}, 
volume={922}, 
DOI={10.1017/jfm.2021.551}, 
journal={Journal of Fluid Mechanics}, 
author={Heydari, Sina and Kanso, Eva}, 
year={2021}, 
pages={A27}
}

@article{Lin_2022, 
title={Two-dimensional hydrodynamic schooling of two flapping swimmers initially in tandem formation},
volume={941}, 
DOI={10.1017/jfm.2022.315},
journal={Journal of Fluid Mechanics}, 
author={Lin, Xingjian and Wu, Jie and Yang, Liming and Dong, Hao}, 
year={2022}, 
pages={A29}
}

@article{Lin_2021, 
title={Flow-mediated organization of two freely flapping swimmers}, 
volume={912}, 
DOI={10.1017/jfm.2020.1143}, 
journal={Journal of Fluid Mechanics}, 
author={Lin, Xingjian and Wu, Jie and Zhang, Tongwei and Yang, Liming}, 
year={2021}, 
pages={A37}
}

@article{Park_2018,
title={Hydrodynamics of flexible fins propelled in tandem, diagonal, triangular and diamond configurations}, 
volume={840}, 
DOI={10.1017/jfm.2018.64}, 
journal={Journal of Fluid Mechanics}, 
author={Park, Sung Goon and Sung, Hyung Jin},
year={2018}, 
pages={154–189}
}

@article{Dai_2018, 
title={Stable formations of self-propelled fish-like swimmers induced by hydrodynamic interactions}, 
volume={15}, 
journal={Journal of the Royal Society Interface}, 
author={Dai, Longzhen and He, Guowei and Zhang, Xiang and Zhang, Xing}, 
year={2018}, 
number = {20180490},
DOI={http://dx.doi.org/10.1098/rsif.2018.0490}
}

@article{Peng_2018, 
title={Hydrodynamic schooling of multiple self-propelled flapping plates}, 
volume={853}, 
DOI={10.1017/jfm.2018.634}, 
journal={Journal of Fluid Mechanics}, 
author={Peng, Ze-Rui and Huang, Haibo and Lu, Xi-Yun}, 
year={2018}, 
pages={587–600}
}

@article{Lin_2020, 
title={Self-organization of multiple self-propelling flapping foils: energy saving and increased speed}, 
volume={884},
DOI={10.1017/jfm.2019.954}, 
journal={Journal of Fluid Mechanics}, 
author={Lin, Xingjian and Wu, Jie and Zhang, Tongwei and Yang, Liming}, 
year={2020}, 
pages={R1}
}

@article{Zhu_2014,
  title={Flow-mediated interactions between two self-propelled flapping filaments in tandem configuration},
  author={Zhu, Xiaojue and He, Guowei and Zhang, Xing},
  journal={Phys. Rev. Lett.},
  volume={113},
  number={23},
  pages={238105},
  year={2014},
  publisher={APS}
}

@article{Arranz_2022, 
title={Flow interaction of three-dimensional self-propelled flexible plates in tandem}, 
volume={931}, 
DOI={10.1017/jfm.2021.918},
journal={Journal of Fluid Mechanics}, 
author={Arranz, G. and Flores, O. and García-Villalba, M.}, 
year={2022}, 
pages={A5}
}

@Article{Nitsche_2021,
	author = {M. Nitsche},
	title = {Evaluation of near-singular integrals with application to vortex sheet flow},
	journal = {Theoretical and Computational Fluid Dynamics},
	year = {2021},
	volume = {35},
	number = {5},
	pages = {581-608}
}

@Article{Li2020, 
  Title                    = {Vortex phase matching as a strategy for schooling in robots and in fish}, 
  Author                   = {L. Li and M. Nagy and J. M. Graving and J. Bak-{C}oleman and G. Xie and I. D. Couzin},
  Journal                  = {Nat. Comm.},
  Year                     = {2020},
  Volume = {11},
  Number = {5408}
}

@Article{Huang2016, 
  Title                    = {Hovering in oscillatory flows}, 
  Author                   = {Y. Huang and M. Nitsche and E. Kanso},
  Journal                  = {J. Fluid Mech.},
  Year                     = {2016},
  Volume = {804},
  Pages = {531--549}
}

@Article{Fang2017, 
  Title                    = {A computational model of the flight dynamics and aerodynamics of a jellyfish-like flying machine}, 
  Author                   = {F. Fang and K. L. Ho and L. Ristroph and M. J. Shelley},
  Journal                  = {J. Fluid Mech.},
  Year                     = {2017},
  Volume = {819},
  Pages = {621--655}
}

@Article{Ormonde2023, 
title={Two-dimensionally stable self-organisation arises in simple schooling swimmers through hydrodynamic interactions}, 
volume={1000}, 
DOI={10.1017/jfm.2024.1086}, 
journal={Journal of Fluid Mechanics}, 
author={Ormonde, Pedro C. and Kurt, Melike and Mivehchi, Amin and Moored, Keith W.}, 
year={2024}, 
pages={A90}
}

@article{Geer1968,
 title={Uniform asymptotic solutions for potential flow around a thin airfoil and the electrostatic potential about a thin conductor},
  author={J. F. Geer and J. B. Keller},
  journal={{S}{I}{A}{M} Journal on Applied Mathematics},
  volume={16},
  number = {1},
  pages = {75-101},
    year={1968}
}

@article{Geder_2017,
 title={Development of an unmanned hybrid vehicle using artificial pectoral fins},
  author={J. D. Geder and R. Ramamurti and D. Edwards and T. Young and M. Pruessner},
  journal={Marine Technology Society Journal},
  volume={51},
  number = {56},
    year={2017}
}

@Book{Lighthill_Book,
  Title                    = {Mathematical Biofluiddynamics},
  Author                   = {J. Lighthill},
  Publisher                = {SIAM},
  Year                     = {1975},
  Volume                   = {17},
  address = {Philadelphia}
}

@Article{Gazzola_JCP, 
  Title                    = {Simulations of single and multiple swimmers with non-divergence free deforming geometries}, 
  Author                   = {Mattia Gazzola and Philippe Chatelain and Wim M. van Rees and Petros Koumoutsakos},
  Journal                  = {Journal of Computational Physics},
  Year                     = {2011},
  Volume                   = {230},
  Pages = {7093--7114},
}

@Article{Gazzola_JFM, 
  Title                    = {Learning to school in the presence of hydrodynamic interactions}, 
  Author                   = {M. Gazzola and A. A. Tchieu and D. Alexeev and A. de Brauer and P. Koumoutsakos},
  Journal                  = {Journal of Fluid Mechanics},
  Year                     = {2016},
  Volume                   = {789},
  Pages = {726--749},
}

@Article{Novati_Learning, 
  Title                    = {Synchronisation through learning for two self-propelled swimmers}, 
  Author                   = {Guido Novati and Siddhartha Verma and Dmitry Alexeev and Diego Rossinelli and Wim M van Rees and Petros Koumoutsakos},
  Journal                  = {Bioinspiration and Biomimetics},
  Year                     = {2017},
  Volume                   = {12},
  Number = {036001},
}

@Article{Verma_RL, 
  Title                    = {Efficient collective swimming by harnessing vortices through deep reinforcement learning}, 
  Author                   = {Siddhartha Verma and Guido Novati and Petros Koumoutsakos},
  Journal                  = {Proceedings of the National Academy of Sciences},
  Year                     = {2018},
  Volume                   = {115},
  Number = {23},
  Pages = {5849--5854},
}

@Article{Alben_Street, 
  Title = {Passive and active bodies in vortex-street wakes}, 
  Author = {Silas Alben},
  Journal = {Journal of Fluid Mechanics},
  Year = {2010},
  Volume = {2642},
  Pages = {95--125},
}

@Article{Alben_Flexible, 
  Title = {Simulating the dynamics of flexible bodies and vortex sheets}, 
  Author = {Silas Alben},
  Journal = {Journal of Computational Physics},
  Year = {2009},
  Volume = {228},
  Pages = {2587--2603},
}

@Article{Triantafyllou_Review, 
  Title                    = { Hydrodynamics of fishlike swimming}, 
  Author                   = {M. S. Triantafyllou and G. S. Triantafyllou and D. K. P. Yue},
  Journal                  = {Annual Review of Fluid Mechanics},
  Year                     = {2000},
  Volume                   = {32},
  Pages = {33---53},
}

@Article{Bajec_Review,
  Title                    = {Organized flight in birds}, 
  Author                   = {Iztok Lebar Bajec and Frank H. Heppner},
  Journal                  = {Animal Behaviour},
  Year                     = {2009},
  Volume                   = {78},
  Pages = {777--789},
}

@Article{Pavlov_Review, 
  Title                    = {Patterns and Mechanisms of Schooling Behavior in Fish: A Review}, 
  Author                   = {D. S. Pavlov and A. O. Kasumyan},
  Journal                  = {J. Ichthyology},
  Year                     = {2000},
  Volume                   = {40},
  Pages = {S163--S231},
}

@Article{Portugal2014,
  Title                    = {Upwash exploitation and downwash avoidance by flap phasing in ibis formation flight}, 
  Author                   = {S. Portugal and T. Hubel and J. Fritz and S. Heese and D. Trobe and B. Voelkl and S. Hailes and A. M. Wilson and J. R. Usherwood},
  Journal                  = {Nature},
  Year                     = {2014},
  Volume                   = {505},
  Pages = {399-402},
}

@Article{Tunstrom1,
  Title                    = {Collective States, Multistability and Transitional Behavior in Schooling Fish},
  Author                   = {Kolbj{\o}rn Tunstr{\o}m and Yael Katz and Christos C. Ioannou and Cristi\'{a}n Huepe and Matthew J. Lutz and Iain D. Couzin},
  Journal                  = {PLOS Computational Biology},
  Year                     = {2013},
  Volume                   = {9},
  Number = {2},
  Pages = {e1002915},
}

@Article{Jolles1,
  Title                    = {Consistent Individual Differences Drive Collective
Behavior and Group Functioning of Schooling Fish},
  Author                   = {Jolle W. Jolles and Neeltje J. Boogert and Vivek H. Sridhar and Iain D. Couzin and Andrea Manica},
  Journal                  = {Current Biology},
  Year                     = {2017},
  Volume                   = {27},
  Pages = {2862--2868},
}

@Article{Swain1,
  Title                    = {Coordinated Speed Oscillations in Schooling Killifish Enrich Social Communication},
  Author                   = {Daniel T. Swain and Iain D. Couzin and Naomi Ehrich Leonard},
  Journal                  = {Current Biology},
  Year                     = {2015},
  Volume                   = {25},
  Number = {5},
  Pages = {1077--1109},
}

@Article{Gudger,
  Title                    = {Fishes that swim heads to tails in single file},
  Author                   = {E. W. Gudger},
  Journal                  = {Copeia},
  Year                     = {1944},
  Volume                   = {1944},
  Number = {3},
  Pages = {152--154},
}

@Article{Pitcher_Minnow,
  Title                    = {THE THREE-DIMENSIONAL STRUCTURE OF SCHOOLS IN THE
MINNOW, {\it {P}HOXINUS PHOXINUS} (L.)},
  Author                   = {T. J. Pitcher},
  Journal                  = {Animal Behaviour},
  Year                     = {1973},
  Volume                   = {21},
  Pages = {673--686},
}

@Article{Partridge_3D,
  Title                    = {The Three-Dimensional Structure of Fish Schools},
  Author                   = {B. L. Partridge and T. Pitcher and J. M. Cullen and J. Wilson},
  Journal                  = {Behav. Ecol. Sociobiol.},
  Year                     = {1980},
  Volume                   = {6},
  Pages = {277--288},
}

@Article{Newlands_tuna,
  Title                    = {Measurement of the size, shape and structure of {A}tlantic bluefin tuna schools in the open ocean},
  Author                   = {K. Newlands and T. A. Porcelli},
  Journal                  = {Fisheries Research},
  Year                     = {2008},
  Volume                   = {91},
  Pages = {42--55},
}

@Article{AshrafPNAS,
  Title                    = {Simple phalanx pattern leads to energy saving in cohesive fish schooling},
  Author                   = {Intesaaf Ashraf and Hana\'{e} Bradshaw and Thanh-Tung Ha and Jos\'{e} Halloy and Ramiro Godoy-Diana and Benjamin Thiria},
  Journal                  = {Proceedings of the National Academy of Sciences},
  Year                     = {2017},
  Volume                   = {114},
  Number = {36},
  Pages = {9599--9604},
}

@Article{AshrafINT,
  Title                    = {Synchronization and collective swimming patterns in fish ({\it Hemigrammus bleheri})},
  Author                   = {I. Ashraf and R. Godoy-Diana and J. Halloy and B. Collignon and B. Thiria},
  Journal                  = {J. R. Soc. Interface},
  Year                     = {2016},
  Volume                   = {13},
  Number = {20160734}
}

@Article{Sophie,
  Title                    = {Flow interactions lead to orderly formations of flapping
wings in forward flight},
  Author                   = {S. Ramananarivo and F. Fang and A. Oza and J. Zhang and L. Ristroph},
  Journal                  = {Phys. Rev. Fluids},
  Year                     = {2016},
  Number                    = {071201(R)},
  Volume                   = {1},
}

@Article{Wu1961Swimmingwavingplate,
  Title                    = {Swimming of a waving plate},
  Author                   = {T. Y. Wu},
  Journal                  = {J.~Fluid Mech.},
  Year                     = {1961},
  Number                   = {03},
  Pages                    = {321--344},
  Volume                   = {10},
  Publisher                = {Cambridge Univ Press}
}

@InCollection{Wu1975Extractionflowenergy,
  Title                    = {Extraction of flow energy by fish and birds in a wavy stream},
  Author                   = {T. Y. Wu and A. T. Chwang},
  Booktitle                = {Swimming and flying in nature},
  Publisher                = {Springer},
  Year                     = {1975},
  Pages                    = {687--702},
}

@Article{Becker,
Title                    = {Hydrodynamic schooling of flapping swimmers},
  Author                   = {Alexander D. Becker and Hassan Masoud and Joel W. Newbolt and Michael Shelley and Leif Ristroph},
  Journal                  = {Nat. Commun.},
  Year                     = {2015},
  Number = {8514},
  Volume                   = {6},
}

\end{document}